\documentclass{cambridge6A}

\usepackage[numbers,square,comma,sort&compress]{natbib}
\usepackage{rotating}
\usepackage{floatpag}
\usepackage{epsfig}
\usepackage{array}

\rotfloatpagestyle{empty}

\usepackage{amsthm}
\usepackage{graphicx}
\usepackage{breakurl}
\usepackage{amsmath, amssymb, amsfonts}
\usepackage{xspace} % Sensible space treatment at end of simple macros
\usepackage{dcolumn}% Align table columns on decimal point
\usepackage{bm}% bold math
\usepackage{multirow} % Multiple rows in tables

\makeatletter
\renewcommand\@makefntext[1]%
    {\noindent\makebox[0pt][r]{\textsuperscript{\@thefnmark}\,}#1}
\makeatother

\usepackage[usenames, dvipsnames]{color}
\usepackage[colorlinks, pdfborder={0 0 0}, plainpages=false]{hyperref}

\usepackage{ulem}
\def\gsim{\mathrel{\rlap{\raise 0.511ex \hbox{$>$}}{\lower 0.511ex\hbox{$\sim$}}}}
\def\lsim{\mathrel{\rlap{\raise 0.511ex \hbox{$<$}}{\lower 0.511ex\hbox{$\sim$}}}}
\def\laq{\raise 0.4ex\hbox{$<$}\kern -0.8em\lower 0.62ex\hbox{$\sim$}}
\def\gaq{\raise 0.4ex\hbox{$>$}\kern -0.7em\lower 0.62ex\hbox{$\sim$}}

\newcommand{\blankout}[1]{}

\begin{document}

\title[{Chapter 6: \bf Sources of Gravitational Waves: Theory and 
Observations}]{{\bf General Relativity and Gravitation:\\ A Centennial 
Perspective}
%{\small\footnote{Please note that the references are updated through Spring 2014, which is when the article was finalized for the Centenary volume.}}
}

\author[Alessandra Buonanno and B.S.~Sathyaprakash]
{\it Alessandra Buonanno$^{1,2}$ and B.S.~Sathyaprakash$^3$\\[10pt]
$^1$Max Planck Institute for Gravitational Physics (Albert Einstein Institute), 
Am M\"uhlenberg 1, Potsdam-Golm, 14476, Germany\\[10pt]
$^2$Maryland Center for Fundamental Physics \& Joint Space-Science Institute, Department of Physics,
University of Maryland, College Park, MD 20742, USA\\[10pt]
$^3$School of Physics and Astronomy, Cardiff University, 5, The Parade, Cardiff, UK, CF24 3YB}

\maketitle
\tableofcontents

\addtocounter{chapter}{5}

\chapter[Sources of Gravitational Waves: Theory and Observations]
{Sources of Gravitational Waves: Theory and Observations
}

\begin{abstract}
Gravitational-wave astronomy will soon become a new tool for observing the
Universe. Detecting and interpreting gravitational waves will require
deep theoretical insights into astronomical sources. The past 
three decades have seen remarkable progress in analytical and numerical 
computations of the source dynamics, development of search algorithms and 
analysis of data from detectors with unprecedented sensitivity. This 
Chapter is devoted to examine the advances and future challenges in 
understanding the dynamics of binary and isolated compact-object systems, 
expected cosmological sources, their amplitudes and rates, and highlights 
of results from gravitational-wave observations. All of this is a 
testament to the readiness of the community to open a new window for 
observing the cosmos, a century after gravitational waves were first predicted 
by Albert Einstein.
\end{abstract}

\section{Historical perspective}
\label{sec:history}

James Clerk Maxwell discovered in 1865 that electromagnetic phenomena 
satisfied wave equations and found that the velocity of these waves in 
vacuum was numerically the same as the speed of light~\cite{Maxwell:1865}. 
Maxwell was puzzled at this coincidence between the speed of light 
and his theoretical prediction for the speed of electromagnetic phenomena 
and proposed that ``light is electromagnetic disturbance propagated through
the field according to electromagnetic laws"~\cite{Maxwell:1865}. 

Because any theory of gravitation consistent with special relativity 
cannot be an action-at-a-distance theory, in many ways, Maxwell's theory, 
being the first relativistic physical theory, implied the existence 
of gravitational waves (GWs) in general relativity (GR). Indeed, years before 
Einstein derived the wave equation in the linearised 
version of his field equations and discussed the generation of GWs as one of the first 
consequences of his new theory of gravity~\cite{Einstein:1916,Einstein:1918}, 
Henri Poincar\'e proposed the existence of {\em les ondes gravifiques} purely 
based on consistency of gravity with special relativity~\cite{Poincare:1905}. 
However, for many years GWs caused much controversy and a lot of doubts 
were cast on their existence~\cite{Einstein:1937,Bondi:1959,Kennefick:2007,Saulson:2011}. 
The year 1959 was, in many ways, the turning point ---
it was the year of publication of a seminal paper by 
Bondi, Pirani and Robinson~\cite{Bondi:1959} on the exact plane wave solution 
with cylindrical symmetry and the energy carried by the waves~\cite{Bondi:1962VII}. 
This paper proved that wave solutions exist not just in the weak-field approximation and 
that GWs in GR carry energy and angular momentum away from their sources.    
These results cleared the way for Joseph Weber~\cite{Weber1959} to start pioneering 
experimental efforts. The discovery of the 
Hulse-Taylor binary~\cite{HulseTaylor75}, a system of two neutron stars in orbit 
around each other, led to the first observational evidence for the existence of gravitational 
radiation~\cite{TFMc79,WeisTay05,WeisbergTaylor10}. The loss of energy and angular momentum to GWs causes
the two stars in this system to slowly spiral in towards each other.
Since 1974, a few other pulsar binaries have been discovered. For the most relativistic binary 
the observed rate of change of the period agrees with the GR prediction to better than $0.03\%$~
\cite{kramer2013}.

Once the most obvious theoretical impediments in defining gravitational 
radiation were tackled and observational evidence of the existence of the radiation 
was firmly established, serious research on the modeling of astrophysical and 
cosmological sources of GWs began along side experimental and data-analysis
efforts to detect GWs.  This chapter examines the last thirty years of endeavours that have brought the field to 
the dawn of the first, direct detections of GWs, hopefully 
marking the one-hundred year anniversary of the first articles by Einstein on 
gravitational radiation~\cite{Einstein:1916,Einstein:1918}. 

Impressive theoretical advances and a few breakthroughs have occurred since the publication of the two notable \index{stellar-mass binaries}
reviews: {\it Gravitational Radiation} in 1982~\cite{LesHouches82} and {\it Three-Hundred 
Years of Gravitation} in 1987~\cite{300years}. Since then, a network of ground-based, GW 
laser interferometers has been built and has taken data\footnote{See the chapter on 
Receiving Gravitational Waves in this volume.}. Sophisticated and robust analytical 
techniques have been developed to predict highly-accurate gravitational waveforms 
emitted by compact-object binary systems (compact binaries, for short) during the 
inspiral, but also plunge, merger and ringdown stages (see Secs.~\ref{sec:approx} and 
\ref{sec:compact object binaries}). After many years of attempts, today 
numerical-relativity (NR)  simulations\footnote{See also the chapter 
on Numerical Relativity in this volume.} are routinely employed to predict merger 
waveforms and validate analytical models of binary systems composed of black holes 
(BHs) and/or neutron stars (NSs). Simulations of binary systems containing NSs are 
becoming more realistic, and robust connections to astrophysical, observable phenomena (i.e., 
electromagnetic counterparts) are under development \index{numerical relativity (NR)}
(see Sec.~\ref{sec:compact object binaries}). The internal structure of NSs has still
remained a puzzle, but there are hints of superfluidity of neutrons in the core.
Neutron-star normal modes are now understood in far more detail and a 
new relativistic instability, that could potentially explain NSs in X-ray
binaries, was discovered (see Sec.~\ref{sec:sources:neutron stars}). Supernova simulations are 
getting more sophisticated and are able to include a variety of micro- and macro-physics
(full GR, neutrino transport, weak interaction physics, magnetohydrodynamics),
but not all in a generic 3-dimensional simulation with realistic equations of state. 
Many challenges remain, including understanding the basic supernova mechanism 
\index{supernovae}
of core collapse and bounce (see Sec.~\ref{sec:sources:neutron stars}). 
Thirty years ago, only a few rough predictions of GW signals 
from the primordial dark age of the Universe existed. Today we know a plethora 
of physical mechanisms in the early Universe that could generate GWs (see Sec.~\ref{sec:primordial}).
Because of the weakness of GW signals, scientists working closely with the experiments have 
established strong collaborations with theorists, astrophysicists and cosmologists, 
so that searches for GWs are fully optimised (see Secs.~\ref{sec:compact object binaries}, 
 \ref{sec:sources:neutron stars} and \ref{sec:primordial}). 
Concurrently with the construction of initial interferometers, geometrical approaches to optimizing 
data analysis were developed, which now form the backbone of all GW data-analysis quests. 
As we shall discuss, searches with initial LIGO and Virgo 
detectors have already produced astrophysically and cosmologically significant upper limits. 
Pulsar Timing Arrays have reached unprecedented sensitivity levels and could detect
a stochastic background from a population of supermassive black holes over the next five years. \index{supermassive black hole binaries} \index{stochastic background} \index{pulsar timing array (PTA)}

Although the progress has been tremendous and the field has reached an unprecedented 
degree of maturity, several challenges still need to be tackled to take full advantage 
of the discovery potential of ground- and space-based detectors, and pulsar timing 
arrays. Part of this chapter is also devoted to highlight those challenges.

Later in this chapter we will discuss sources that can be observed in different
types of detectors. These include ground-based interferometers such as initial 
and Advanced LIGO (iLIGO and aLIGO), Virgo and Advanced Virgo (AdV), KAGRA and 
Einstein Telescope (ET) (see Fig.~\ref{fig:sources-g}), space-based detectors 
LISA and eLISA (see Fig.~\ref{fig:sources-s}) and Pulsar Timing Arrays 
(PTA) and the Square Kilometre Array (SKA)\footnote{ET is a third generation detector concept being
studied in Europe whose conceptual design study was completed in 2011 
\cite{ET-DSD}. The European Space Agency has selected GW Observatory \index{pulsar timing array (PTA)}
as the science theme for the 3rd large mission (L3) in its future science 
program, scheduled for a launch in 2034. eLISA is the current straw man design 
for L3.}. 

Lastly, while this Chapter was being finalised, the BICEP2 \index{BICEP2} experiment claimed to have observed 
the polarisation of the cosmic microwave background (CMB) photons (the so-called B-modes), caused 
by primordial GWs~\cite{Ade:2014xna}. If confirmed, this result will constitute a landmark discovery 
in cosmology and GW science, enabling us to probe epochs very close to the Big Bang.  
However, at this stage it is not clear if the observed signal is truly primordial in nature 
and not due to astrophysical foregrounds and synchrotron emission by intervening dust 
\cite{Mortonson:2014bja,Flauger:2014qra}. Measuring the imprint of primordial GWs on the CMB is different from, 
but equally relevant to, the detection of GWs with ground- and space-based detectors or PTAs. 
Indeed, detectors like BICEP2 infer the presence of GWs through their interaction 
with the CMB radiation at the time the latter was produced. They do not detect 
GWs passing by the detector on the Earth today. LIGO, Virgo, KAGRA, eLISA, PTA, etc., will probe contemporary 
GWs, yielding spectral and sky position data, and a plethora of new and unique information about 
our Universe and its contents.  The focus of this Chapter is on the new window on the Universe 
that those detectors will enable us to open.

\section{Analytical approximation methods}
\label{sec:approx}

\paragraph{Progress over the past three decades:} 
There is no doubt that the field of analytical relativity has matured considerably and has made 
tremendous progress since the notable Les Houches school on {\it Gravitational Radiation} in 1982~\cite{LesHouches82}. \index{quadrupole formula}
During the last thirty years there have been significant advances on the problems emphasised 
in the historic discussion organised and moderated by A. Ashtekar at 
the end of the school. The validity of the quadrupole formula for gravitational radiation far away 
from a binary source was questioned by some of the participants at the Les Houches school, appealing to 
the ongoing debate at that time~\cite{Wagoner:1976am,Ehlers:1976ji,Ehlers:1980,
Walker:1980tm,Walker:1980zz,Damour:1981bh,Futamase:1983,Futamase:1984dh} 
relating to the difficulties in describing nonlinearities in GR within a precise mathematical framework. 
At the Les Houches school, the Paris group (Damour, Deruelle, ...) proposed a consistent 
framework in terms of which to formulate questions about the two-body dynamics and GW emission. 
The Paris group was motivated by the impressive observational work related to the discovery of 
the Hulse-Taylor binary pulsar~\cite{HulseTaylor75} and wanted to develop a mathematical framework 
that could match the standards set by the ever more accurate pulsar-timing data. Eventually, 
as we shall discuss below, a precise mathematical framework was also needed for {\it direct} 
detection of gravitational waves on the ground, because in this case nonlinearities play a role that 
is much more crucial than in binary pulsar observations. \index{pulsar timing array (PTA)}

Between 1980 and 1992 important theoretical foundations in gravitational radiation and 
post-Newtonian (PN) \index{post-Newtonian (PN) formalism} theory were carried out by a number of researchers~\cite{thorne80,
DamourDeruelle1985,Damour-Schafer85,DamourDeruelle1986,BlDa.86,Blanchet1987,Damour:1988mr,
Blanchet88,Blanchet:1989cu,BD89,Damour:1990ji,Lincoln-Will:1990,BD92,Thorne:1992prd}. 
However, during those years, the analytical work on the two-body problem was 
considered mostly academic. It was not clear how relevant it would be to push calculations beyond 
the quadrupole formula for the direct observation of GWs. \index{quadrupole formula}
The first important turning point was in 1993 when~\cite{Cutler:1992tc} 
pointed out the importance of computing the GW phasing beyond the leading order. 
Many crucial developments took place in the subsequent 
years~\cite{Blanchet93,Iyer:1993xi,Blanchet95a,Blanchet:1995ez,Blanchet96a,Jaranowski:1996nv,Jaranowski:1997ky,Kidder:1992fr,
kidder95,Will:1996zj, Poisson:1993vp,Tagoshi:1993dm,Poisson:1993zr,Shibata:1994jx,Tagoshi:1994sm,
Tagoshi:1996gh}. The second important turning point, which brought theory and observations closer, occurred in the mid and late 1990s when the construction of LIGO, Virgo, GEO600 and TAMA 300 
detectors started~\cite{Abramovici:1992ah}. The TAMA 300 and LIGO detectors took the first 
data in 1999~\cite{Tagoshi:2000bz} and 2002~\cite{Abbott:2003pj}, respectively. (The first
ever coincident operation of a pair of interferometers was between the Glasgow 10 m and 
Garching 30 m prototypes~\cite{Nicholson:1996ys}.) The third turning point happened in the late 1990's  
and early 2000's when, pressed by the construction of GW interferometers, 
the analytical effective-one-body (EOB) \index{effective-one-body (EOB)} approach~\cite{Buonanno:1998gg,Buonanno:2000ef} made a bold prediction 
for the late inspiral, merger and ringdown waveform emitted by comparable-mass binary BHs. The 
EOB formalism builds on PN and perturbation theory results, and it 
is guided by the notion that non-perturbative effects can be captured analytically 
if the key ingredients that enter the two-body 
dynamics and GW emission are properly resummed about the (exact) test-particle limit results. 
Moreover, in the early 2000's, a pragmatic, numerical and analytical, hybrid approach aimed at 
predicting the plunge and merger waveform was bravely carried out~\cite{Baker:2002qf,Baker:2001nu}. \index{effective-one-body (EOB)} \index{gravitational waveform}
This approach, called the Lazarus project, consisted of evolving the binary system in full
numerical relativity (NR) \index{numerical relativity (NR)} for less than an orbit just prior to merger before 
stopping the evolution, extracting the spacetime metric from the results of the simulation 
of a deformed BH, and using perturbation theory calculations to complete the evolution 
during ringdown. The fourth relevant turning point occurred in 2005, when after more than thirty years of attempts, the first numerical-relativity simulations of binary BHs at last unveiled the merger waveforms~\cite{Pretorius2005a,Campanelli2006a,Baker2006a}. Since then, synergies and interplays between different analytical 
and numerical techniques to solve the two-body problem in GR have grown considerably. 
A few paradigms were broken, in particular the nature of the binary BH merger waveform, which 
turned out to be much simpler than what most people had expected or predicted. Finally, recent years have seen 
remarkable interactions between GW data analysts, astrophysicists
and theorists to construct templates to be used for the searches, making analytical 
relativity a crucial research area for experiments that will soon revolutionise our 
understanding of the Universe. 

In the rest of this section we shall discuss the main approximation methods that have been developed 
to study the two-body problem in GR, highlighting some of the key 
theoretical ideas that have marked the last thirty years. As we shall see, in all the approximation schemes 
one needs to develop a conceptual and consistent framework and solve difficult technical problems. 
Since Einstein did not conceive the theory of GR starting from approximations to it, 
we shall see in relation to GWs what we gain and what we lose from the original theory 
when investigating it through approximation methods. 

\begin{figure}
\begin{center}
\includegraphics[width=0.80\textwidth, angle=0]{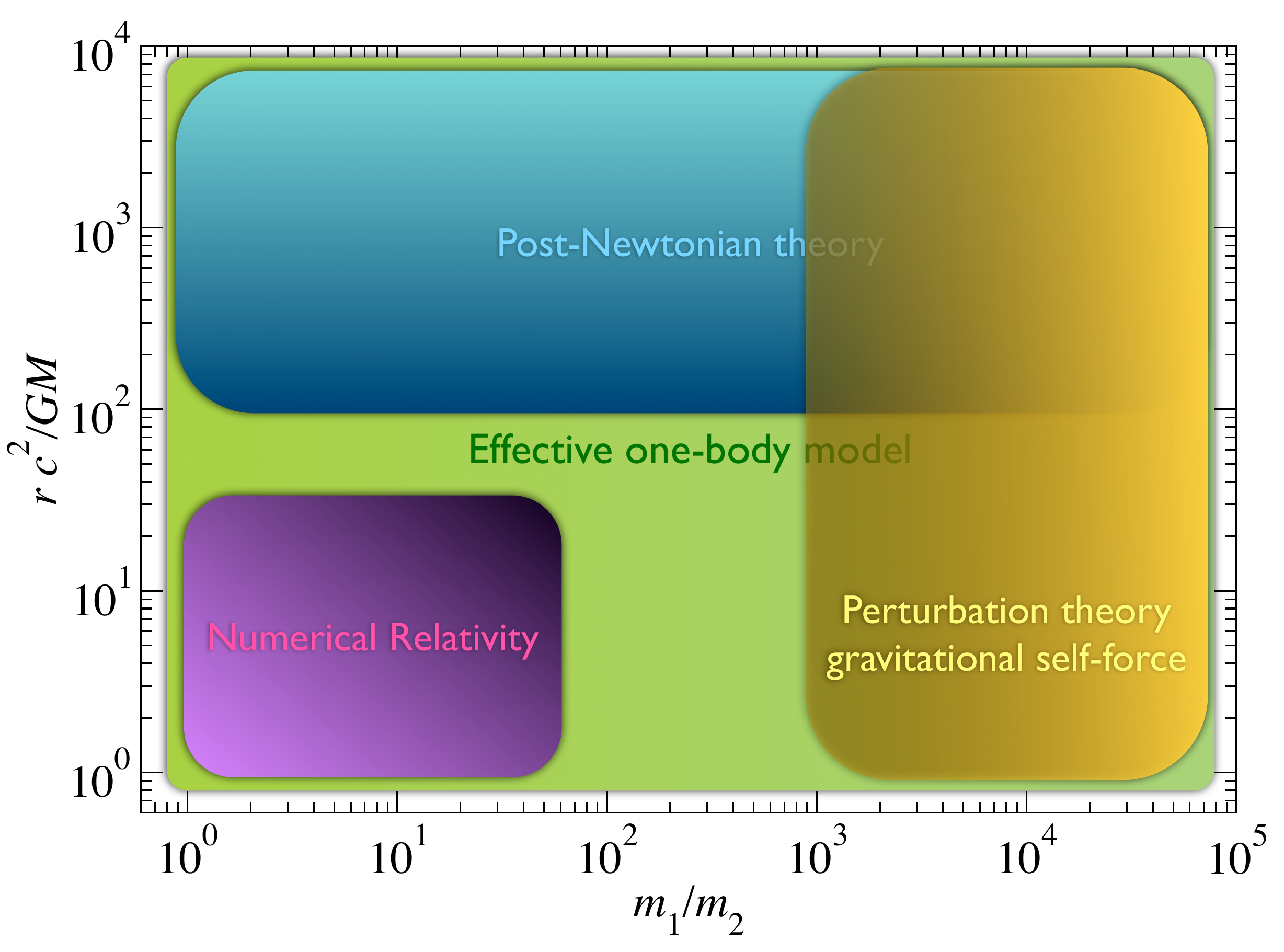}
\caption [Range of validity of analytical and numerical methods.] {\label{Fig:2bodymethods} Current range of validity of the main analytical and numerical methods to 
solve the two-body problem.}
\end{center}
\end{figure}
\paragraph{Physical scales and methods in the two-body problem:}
Three main methods have been proposed to tackle the two-body problem in GR: 
the PN approach, perturbation theory (and the gravitational self-force formalism), 
\index{gravitational self-force (GSF) formalism} \index{perturbation theory}
and NR. \index{numerical relativity (NR)} The first step towards setting up those approximation methods consists 
in identifying some small (dimensionless) parameters. In Fig.~\ref{Fig:2bodymethods} we show the range of 
validity of each method using the parameters $r c^2/(G M)$ and $m_1/m_2 \geq 1$, where $m_1$ and $m_2$ are 
the binary's component masses, $M=m_1+m_2$ is its total mass 
and $r$ is the separation between the two bodies. If we consider astrophysical sources that are held 
together by gravitational interactions, then, as a consequence of the virial theorem,  
$v^2/c^2 \sim G M/(r c^2)$, where $v$ is the characteristic velocity of the bodies 
in the binary. 

The PN formalism expands the dynamics and gravitational waveforms in powers of $v/c$. 
\index{post-Newtonian (PN) formalism} \index{gravitational waveform}
It is valid for any mass ratio but, in principle, only for slow motion, which, 
for self-gravitating objects, also implies  large separations. Perturbation theory is suitable to describe the motion and radiation 
of a small body moving around a large body. It expands the Einstein equations around the BH metric 
in powers of the mass ratio $m_2/m_1$. At leading order the small-body moves along geodesics of the background 
spacetime and can reach any speed $v \,\laq\, c$. When taking into account the back-reaction 
of the gravitational field of the small body on its motion, one needs to develop a consistent 
framework free of divergences. This is done within the gravitational self-force (GSF) formalism. 
Finally, NR solves the Einstein equations on a computer.  \index{radiation reaction}
In principle, it could be used for any mass ratio, binary separation and velocity. However, 
the computational cost and the requirements on the accuracy of the numerical solutions limit 
its range of validity. For the past thirty years the two-body problem has been tackled 
keeping in mind that each method has a domain of validity displayed in Fig.~\ref{Fig:2bodymethods}. 
As we shall see below, by proceeding without blinkers, recent work at the interfaces 
between the different methods has demonstrated that 
the limits of validity of those approaches are more blurred than expected --- for example PN 
calculations can be pushed into the mildly relativistic regime $v/c \,\laq \, 0.1$ and GSF 
predictions could be used also for intermediate (or perhaps even comparable) mass binary systems. 
Moreover, an analytical formalism, namely, the EOB approach, exists that can incorporate the results of the different \index{intermediate mass black hole binaries}
methods in such a way as to span the entire parameter space and provide highly-accurate templates to search
for BH binaries in GW data.

Because of space limitations, the presentation of the different approximation methods will be sketchy and incomplete. 
The reader is referred to the original articles, reviews~\cite{Sasaki:2003xr,Blanchet2006,Futamase:2007zz,
Barack:2009ux,Poisson:2011nh,Damour:2013hea} and books~\cite{Landau-Lifshitz,Schutz:Book,Maggiore2008,PNmethods} for more 
details. 

\subsection{Post-Newtonian formalism}
\label{sec:PN}

\begin{figure}
\begin{center}
\includegraphics[width=0.66\textwidth, angle=0]{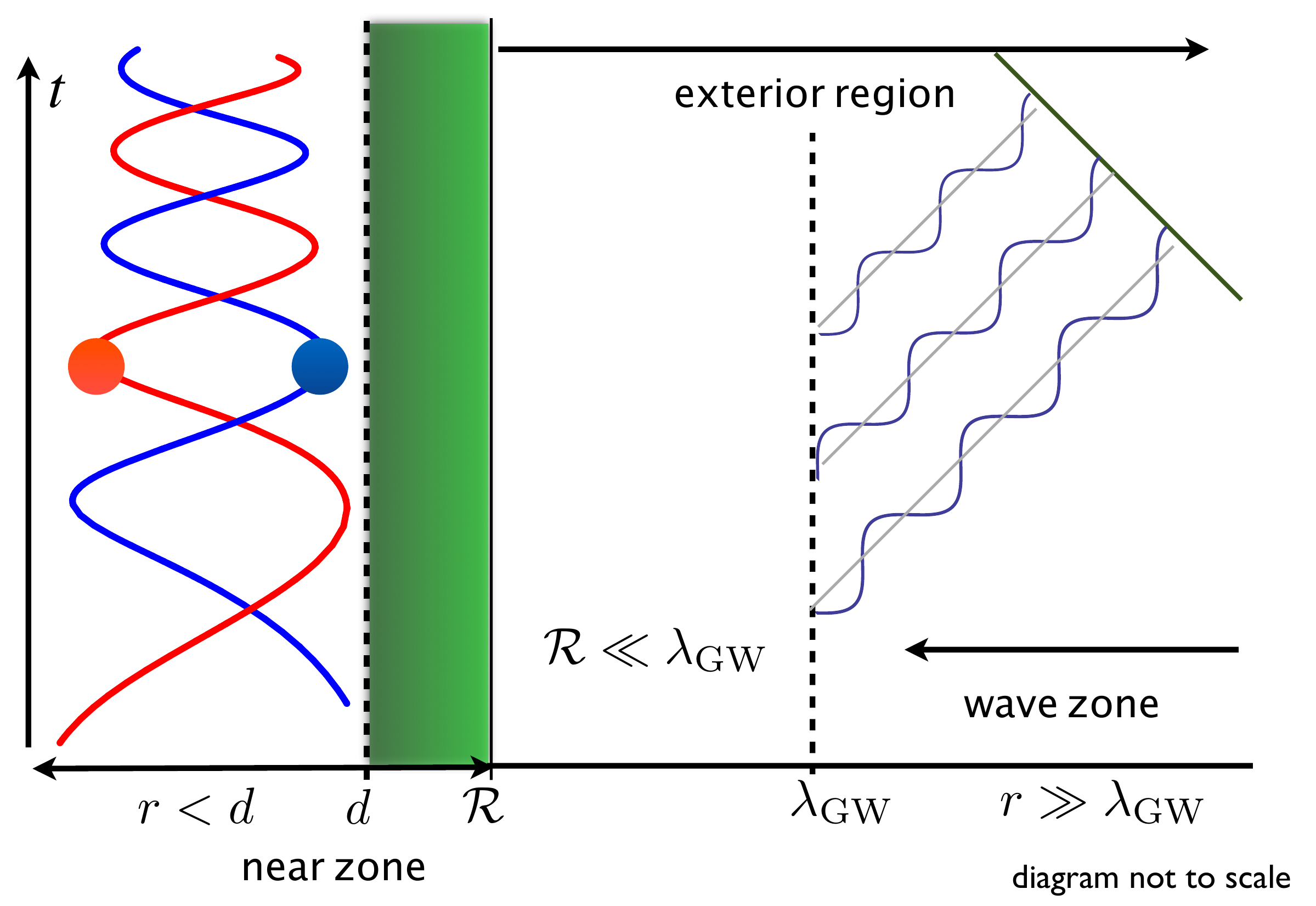}
\caption [Schematic diagram of an in spiraling binary] {\label{Fig:inspiral} Schematic diagram of an inspiraling binary 
showing various scales used in the PN formalism.}
\end{center}
\end{figure}

\index{post-Newtonian (PN) formalism} The Einstein field equations $R_{\alpha \beta} - g_{\alpha \beta} R/2 = 8 \pi G T_{\alpha \beta}/c^4$ 
can be recast in a convenient form introducing the field $h^{\alpha \beta} = \sqrt{-g}\,g^{\alpha \beta} - 
\eta^{\alpha \beta},$ which is a measure of the deviation of the background from Minkowskian metric $\eta_{\alpha\beta}$, and imposing the harmonic gauge condition $\partial_\beta h^{\alpha \beta}=0$~\cite{Landau-Lifshitz},
\begin{equation}
\Box h^{\alpha \beta} = \frac{16 \pi G}{c^4}|g| T^{\alpha \beta} + \Lambda^{\alpha \beta}\equiv 
\frac{16 \pi G}{c^4} \tau^{\alpha \beta}\,,
\label{EE}
\end{equation}
where $\Box$ is the D'Alambertian operator in flat spacetime, $g \equiv {\rm det}{(g_{\alpha \beta})}$, 
$T^{\alpha \beta}$ is the matter stress-energy tensor and $\Lambda^{\alpha \beta}$ 
depends on non-linear terms in $h^{\mu \nu}$ and $g^{\mu \nu}$ and their derivatives. 
By imposing no-incoming-radiation boundary conditions, one can formally solve Eq.~(\ref{EE}) 
in terms of retarded Green functions 
\begin{equation}
{h}^{\alpha\beta}(t,{\bm r}) = \frac{16 \pi G}{c^4} \Box_{\rm ret}^{-1} \tau^{\alpha \beta}= -\frac{4G}{c^4}
\int \frac{\tau^{\alpha\beta}(t-|{\bm r}-{\bm r}'|/{c}, 
{\bm r}')}{|{\bm r} - {\bm r}'|} {\rm d}^3 r'.
\label{hformal}
\end{equation}
Limiting to leading order in $G$ and considering $r\equiv |\bm{r}| \gg d$ 
(i.e., the field point is at a far greater distance compared to the size $d$ of the source), we can expand the 
integrand in powers of $1/r$ and find at leading order
\begin{equation}
{h}^{\alpha \beta}(t,{\bm r}) = -\frac{4G}{c^4\,r}
\int T^{\alpha \beta}(t- r/c+ \bm{n}\cdot\bm{r'}/{c},\bm{r'}) {\rm d}^3 r',
\label{hfar}
\end{equation}
where $\bm{n} = \bm{r}/r$. 
Let us assume that the source is a PN source (i.e., it is slowly moving, weakly stressed and 
weakly self-gravitating). This means that $|{T^{0i}}/{T^{00}}|\ \sim \sqrt{|{T^{ij}}/{T^{00}}|} 
\sim \sqrt{| {U}/{c^2}|} \ll 1,$  where $U$ is the source's Newtonian potential. It is customary 
to indicate the magnitude of the above small quantities with a small parameter $\epsilon$, which is 
essentially $v/c$ where $v$ is the characteristic, internal velocity of the source. 
Assuming that the source's size is $d$ and it oscillates at frequency $\omega$, the characteristic 
speed of the source is $v \sim \omega d$. From analogy to the electromagnetic 
case, we expect 
$\lambda_{\rm GW} \sim (c/v)\,d$. For slow motion $v/c \ll 1$, thus $\lambda_{\rm GW} \gg d$ 
and the source is located well within one wavelength. Historically, the region 
at a distance $r \ll \lambda_{\rm GW}$ from the source, extending to ${\cal R}$ with 
${\cal R} \ll \lambda_{\rm GW}$, has been denoted the {\it near zone}, whereas 
the region which extends to $r \gg \lambda_{\rm GW}$ is denoted the {\it wave zone} 
(see Fig.~\ref{Fig:inspiral}). 

Using the conservation of the energy-momentum tensor at linear order in $G$, that is 
$\partial_\alpha T^{\alpha \beta} =0$, and expanding the integral (\ref{hfar}) in powers of $v/c$, one can obtain the  
gravitational field at linear order in $G$ as a function of the derivatives of the 
source multipole moments ~\cite{thorne80}. As originally derived by Einstein~\cite{Einstein:1918} 
and then by Landau and Lifshitz, at lowest order in the wave-generation formalism, 
the gravitational field in the transverse-traceless (TT) gauge and in a suitable 
radiative coordinate system  $X^\mu=(c\,T,\bm{X})$ reads (``far-field quadrupole formula'') \index{quadrupole formula}
\begin{equation}
\label{quadr_form}
h^{\rm TT}_{ij} = \frac{2 G}{c^4 R}\,\sum_{k, l}{\cal P}_{ij kl}(\bm{N}) 
\left [\frac{d^2}{d T^2} Q_{ k l}\left (T - \frac{R}{c}\right ) + {\cal O}\left (\frac{1}{c} \right ) \right ] 
+ {\cal O}\left (\frac{1}{R^2} \right ),  
\end{equation}
where $R = \sqrt{\sum_i X^2_i}$ is the distance to the source, 
$\bm{N} = \bm{X}/R$ is the unit vector from the source to the observer, and ${\cal P}_{ij kl} = {\cal P}_{ik}\,
{\cal P}_{jl} - {\cal P}_{ij}\,{\cal P}_{kl}/2$ is the TT projection operator, where
${\cal P}_{ij} = \delta_{i j} - N_i\,N_j$ is the operator that projects onto the plane orthogonal to $\bm{N}$. 
The radiative coordinate system $X^\mu$ can be related to the source-rooted coordinate system 
$x^\mu =(c\,t,\bm{x})$~\cite{Blanchet2006}.  The source quadrupole moment at Newtonian order is 
\begin{equation}
Q_{ij}(t) = \int_{\rm source} d^3 \bm{x'}\,\rho(t, \bm{x}')\, \left (x_i'\,x_j' - \frac{1}{3} 
\delta_{i j}\,{x'}^2 \right )\,,
\end{equation}
where $\rho$ is the Newtonian mass density. The gravitational field (\ref{quadr_form}) is 
generally referred to as Newtonian because the 
evolution of the quadrupole moment of the source is computed using Newton's law of gravity. 

Several methods have been proposed for going beyond the leading-order result to 
solve Eq.~(\ref{hformal}) approximately. In the near zone the solution $h_{\mu \nu}$ 
can be written in terms of instantaneous potentials, while in the wave zone retardation 
effects must be taken into account. Note that at higher orders
gravitational waves can themselves act as sources; so $\tau^{\alpha \beta}$ 
in Eq.~(\ref{hformal}) contains both compact and non-compact support terms. 

\paragraph{The multipolar post-Minkowskian--post-Newtonian formalism:}
\index{multipolar post-Minkowskian-post-Newtonian formalism (MPM)}
\index{post-Newtonian (PN) formalism} Early on Epstein and Wagoner~\cite{Epstein75}, and Thorne~\cite{thorne80} proposed a PN extension 
of the Landau-Lifshitz derivation and computed the ${\cal O}(v^2/c^2)$ corrections to the quadrupole 
formula for binary systems. However, in their approach the formal application of the PN 
expansion leads to divergent integrals. Building on pioneering work by 
Bonnor et al.~\cite{Bonnor61,Bonnor66} and Thorne~\cite{thorne80}, Blanchet 
and Damour~\cite{BlDa.86,Bl.87,BD89} introduced a GW generation 
formalism in which the corrections to the leading quadrupolar formalism are 
obtained in a mathematically well-defined way. In this approach 
the wave-zone expansion, in which the external 
vacuum metric is expanded as a multipolar post-Minkowskian (MPM) series (i.e., 
a non-linear expansion in $G$), is matched to the near zone expansion, in which a PN 
expansion (i.e., a non-linear expansion in 
$1/c$) is applied, the coefficients being in the form of a multipole expansion. 
The multipolar expansions employ a multipole decomposition in irreducible 
representations of the rotation group. The gravitational field all over space 
can be obtained  by matching the near-zone to the wave-zone fields and the matching 
can be accomplished at all orders, as shown by Blanchet and 
collaborators~\cite{Blanchet95,Blanchet98,Poujade:2001ie,Blanchet:2005ft}. We shall denote 
this approach the MPM-PN formalism.

\paragraph{Wave-zone multipolar post-Minkowskian approximation:}
\index{wave-zone multipolar post-Minkowskian approximation}
In the multipolar post-Minkowskian expansion, as one moves away from a weakly self-gravitating source, the spacetime 
quickly approaches the Minkowskian spacetime.  At a distance $r$ from the source with mass $M$,  
the deviations from the metric scale like $r_s/r$ with $r_s = 2 G M/c^2$. 
Thus, in the region $ d < r < +\infty$, one can solve the vacuum Einstein 
equations iteratively in powers of $G$. Indeed, one sets $\sqrt{-g} g^{\alpha \beta} 
= \eta^{\alpha \beta} + G h^{\alpha \beta}_1 + G^2 h^{\alpha \beta}_2 + \cdots $, that is  $h^{\alpha \beta} = \sum_{n=1} 
G^n h_{n}^{\alpha \beta}$, substitutes this expression in the Einstein's equations (\ref{EE}) in vacuum, obtaining  
\begin{equation}
\Box h^{\alpha \beta} = \Lambda^{\alpha \beta} 
= N^{\alpha \beta}(h,h) + M^{\alpha \beta}(h,h,h) + {\cal O}(h^4)\,,
\end{equation}
and equates terms of the same order in $G$. 

At linear order in $G$, the most general homogeneous 
solution satisfying the harmonic gauge can be written in terms of symmetric trace-free tensors (STF), 
which are a complete basis of the rotation-group's 
representation, that is $ h_1^{\alpha \beta} = \sum_{\ell=0} \partial_L[K_L^{\alpha \beta}(t-r/c)/r]$ where 
$L=i_1\cdots i_\ell$ denotes a multi-index composed of $\ell$ STF indices 
$i_1, \cdots, i_\ell,$ ranging from 1 to 3. More explicitly, one finds that the most general solution 
satisfying the gauge condition is given by $h_1^{\alpha \beta} =k_1^{\alpha \beta} + \partial^\alpha \varphi_1^\beta + 
\partial^\beta \varphi_1^\alpha - \eta^{\alpha \beta} \partial_\rho \varphi_1^\rho$, where $k_1^{\alpha \beta}$ and 
$\varphi_1^\alpha$ can be expressed in terms of two {\it source} multipole moments,  
$\{I_L, J_L\}$, that encode properties of the source and four multipole 
functions, $\{W_L, X_L,Y_L,Z_L\}$, that parameterise gauge transformations. Eventually, these 
six moments can be reduced to two gauge-inequivalent  
canonical ones $\{M_L,S_L\}$. At quadratic order in $G$, one needs to invert the equation 
$\Box h_2^{\alpha \beta} = N^{\alpha \beta}(h_1,h_1)$. Because $h_1^{\alpha \beta}$ is known only 
in the exterior region ($r > d$), one cannot employ retarded or advanced Green functions 
for which it is necessary to know the solution everywhere in space. However, at each finite PN 
order only a finite number of multipole moments contribute. So, one applies a multipolar post-Minkowskian \index{multipolar post-Minkowskian-post-Newtonian formalism (MPM)}
expansion outside the source for $d/r < 1$ and introduces a regularisation to extend  
$N^{\alpha \beta}(h_1,h_1)$ at the origin. Blanchet and Damour proposed that at each order $G^n$ the function 
$\Lambda_n^{\alpha \beta}$ be multiplied by $r^B$, $B$ being a complex number whose real part is positive and 
sufficiently large, so that $r^B \Lambda_n^{\alpha \beta}$ is regular at the origin. Then, the solution of $\Box h_n^{\alpha \beta} = 
r^B \Lambda_n^{\alpha \beta}$ is obtained by using retarded Green functions, analytic continuation in the complex 
$B$-plane and extracting the coefficient of the zeroth power of $r$ ($B=0$), that is $u^{\alpha \beta}_n={\rm FP}_{B=0}\left \{\Box_{\rm ret}^{-1}[r^B \Lambda_n^{\alpha \beta}]\right \}$, where ${\rm FP}$ stands for finite part~\cite{BlDa.86}.  
The most general solution is obtained by adding to the inhomogeneous solution $u_n^{\alpha \beta}$  
the homogeneous solution $v_n^{\alpha \beta},$ such that the harmonic gauge condition $\partial_\alpha h_{n}^{\alpha \beta} 
=0$ is satisfied. Thus, $h_n^{\alpha \beta} = u_n^{\alpha \beta} + v_n^{\alpha \beta}$. 
The solution $h_n^{\alpha \beta}$ depends on the canonical multipole moments $M_L$ 
and $S_L$, which at this stage do not know anything about the matter 
source. They only parameterise the most general solution of the Einstein equations in vacuum. 

\paragraph{Near-zone post-Newtonian approximation:}
\index{near-zone post-Newtonian approximation}
\index{post-Newtonian (PN) formalism} In the near zone one wants to obtain the solution of Eq.~(\ref{hformal}) in a multipolar PN 
expansion. Expanding $h^{\alpha \beta}$ and $\tau^{\alpha \beta}$ in powers of $1/c$, that is  
$h^{\alpha \beta}=\sum_{n = 2} {}^{(n)} h^{\alpha \beta}/c^n$  and $\tau^{\alpha \beta}=
\sum_{n = -2} {}^{(n)} \tau^{\alpha \beta}/c^n$,  substituting them in the 
Einstein equations (\ref{EE}) and equating terms of the same order in $1/c$, one finds 
the following relation~\cite{Maggiore2008} $\nabla^2 [{}^{(n)}h^{\alpha \beta}] = 16 \pi G \,[{}^{(n-4)}\tau^{\alpha \beta}] + 
\partial_t^2 [{}^{(n-2)}h^{\alpha \beta}] \equiv {}^{(n)} f^{\alpha \beta}$. If one were solving 
the above differential equation via Poisson integrals, the presence of non-compact--support 
terms at some order in the PN expansion would prevent the integral from 
converging at spatial infinity. The convergence problem becomes even more severe 
when the multipolar expansion 
$1/|\bm{r}-\bm{r}'| = 1/r + \bm{r}\cdot \bm{r}'/r^3 + \cdots$ 
is applied. These were the problems that affected the original method of 
Epstein, Wagoner and Thorne. They were overcome by applying a more sophisticated 
mathematical method to invert the Laplacian and find the correct solution. 
This method is a variant of the analytic continuation technique developed 
by Blanchet and Damour in the wave zone, and it was carried out in~\cite{Blanchet95,Blanchet96,Blanchet98,Poujade:2001ie,Blanchet:2005ft}. Basically, one multiplies the function ${}^{(n)}f^{\alpha \beta}$ by $r^B$, where
$B$ is a negative real number whose modulus is sufficiently large that the integral 
is regular at spatial infinity. Then, the inhomogeneous solution is derived by analytic continuation in the complex 
$B$-plane and extracting the coefficient of the pole at $B=0$, that is 
${}^{(n)}u^{\alpha \beta}={\rm FP}_{B=0}\left \{(\nabla^2)^{-1}[{}^{(n)}f^{\alpha \beta}\,r^B ]\right \}$. 
The most general solution is obtained by adding the homogeneous solution 
${}^{(n)}v^{\alpha \beta}$ that is regular at the origin $r=0$, to the 
inhomogeneous solution, that is ${}^{(n)}h^{\alpha \beta} = 
{}^{(n)}u^{\alpha \beta} + {}^{(n)}v^{\alpha \beta}$. 
As derived in~\cite{Blanchet:1993ng,Blanchet95,Blanchet96,Blanchet98,Poujade:2001ie,Blanchet:2005ft}, 
the solution in the near zone that matches the external field and satisfies correct boundary 
conditions at infinity  involves a specific homogenous solution which can be expressed 
in terms of STF tensors as $ \sum_{\ell =0} \partial_L [F_L^{\alpha \beta}(t-r/c)/r 
- F_L^{\alpha \beta}(t+r/c)/r]$ and it is fixed by 
matching it to the post-Minkowskian solution. In the region $d < r < {\cal R}$ the multipolar PN 
and post-Minkowskian series are both valid and one can resort to the standard method 
of matched asymptotic expansion to relate them, obtaining the solution all over space, \index{multipolar post-Minkowskian-post-Newtonian formalism (MPM)}
$0 < r < + \infty$.  In particular, the matching allows expression of the canonical multipole 
moments $M_L$ and $S_L$  (or the source multipole moments $\{I_L, J_L, W_L, X_L,Y_L,Z_L\}$) in 
terms of integrals that extend over the matter and gravitational fields described by the 
PN--expanded $\tau^{\alpha \beta}$.  

\paragraph{Direct integration of the relaxed Einstein equations:} 
\index{direct integration of relaxed Einstein equation (DIRE)}
A different formalism that also cures the convergence issues that plagued previous brute-force, slow-motion 
approaches to gravitational radiation from isolated sources, was developed by Will, Wiseman and 
collaborators~\cite{PW00,Pati:2002ux,Will:1996zj}. It is called the Direct Integration of the Relaxed Einstein Equation (DIRE). 
It differs from the MPM-PN approach in the definition of the source multipole moments. In both 
formalisms, the moments are generated by the PN expansion of $\tau^{\alpha \beta}$ in Eq.~(\ref{hformal}).  
However, in the DIRE formalism they are defined by compact-support integrals terminating at the 
radius ${\cal R}$ enclosing the near zone, while in the MPM-PN approach, as we have discussed, 
the moments are defined by integrals covering the entire space and are regularised using the 
finite-part procedure. \index{post-Newtonian (PN) formalism}

\paragraph{Gravitational waveform in the wave zone:}
\index{gravitational waveform}
When neglecting terms of order $1/R^2$ or higher, the general expression of the 
TT waveform that goes beyond the leading-order term (\ref{quadr_form}) reads  
\begin{equation}\label{eq:hijTT}
h^\mathrm{TT}_{ij} = \frac{4G}{c^2R} \sum _{k,q} \mathcal{P}_{ijkq}(\bm{N}) 
\sum^{+\infty}_{\ell=2}\frac{1}{c^\ell\ell !} \biggl[
N_{L-2} U_{kqL-2} - \frac{2\ell\,N_{m L-2}\, \varepsilon_{mn(k}\, V_{q)nL-2}}{c(\ell+1)} \biggr]\,,
\end{equation}
where the integer $\ell$ refers to the multipolar order, $\varepsilon_{ijk}$ is the 
Levi-Civita antisymmetric symbol, parentheses denote symmetrisation and $U_L$ and $V_L$ 
are the multipole moments at infinity (called {\it radiative} multipole moments), 
which are functions of the retarded time $T-R/c$. The radiative multipoles $U_L$ and 
$V_L$ can be expressed in terms of the canonical multipole moments $M_L$ and $S_L$ as 
$U_L(T) = {d^\ell M_L}/{d T^\ell} + {\cal F}_L[M(T^\prime),S(T^\prime)]$ and 
$V_L(T) = {d^\ell S_L}/{d T^\ell} + {\cal G}_L[M(T^\prime),S(T^\prime)]$ 
where ${\cal F}_L$ and ${\cal G}_L$ are multi-linear retarded functionals of 
the full past behaviour, with $T^\prime < T$. For example, the radiative mass-type multipole reads 
\begin{equation}
U_L = \frac{d^\ell M_L}{d T^\ell} + \frac{2 GM}{c^3}\int_0^{\infty} d \tau \,M_L^{(\ell +2)}(T_R - \tau)\,
\left [\log \left (\frac{c \tau}{2 r_0} \right ) + \kappa_{\ell} \right ] + {\cal O}\left ( \frac{1}{c^5} \right )\,,
\end{equation}
where $T_R=T-R/c$, $r_0$ is an arbitrary length scale and $\kappa_\ell$ are 
constants. The term at order $1/c^3$ in the equation above describes the effect of back scattering of 
the gravitational waves on the Schwarzschild-like curvature associated with the total mass $M$ of 
the source, the so-called {\it tail terms}~\cite{Blanchet88,Wiseman:1993aj}. 
At order $1/c^5$, one has another 
hereditary term called the {\it memory term}, which is generated by non-linear interactions between 
multipole moments~\cite{BD92}. At order $1/c^6$, tails back scatter again with the 
Schwarzschild-like curvature generating {\it tail-of-tail terms}~\cite{Blanchet:1997jj}.

\paragraph{Gravitational-wave flux at infinity:}
The GW energy flux (or luminosity) ${\cal L}$ can be expressed 
in terms of the radiative multipole moments. It reads: 
\begin{align}
{\cal L} &= \sum_{\ell=2} \frac{G}{c^{2\ell +1}}\,\left 
[ \frac{(\ell +1)(\ell +2)}{(\ell -1)\ell \ell!(2\ell+1)!!}\,\left (\frac{d U_L}{dT} \right)^2 
\right . \nonumber \\&\left. + \frac{4\ell (\ell +2)}{c^2(\ell -1)(\ell +1)!(2\ell+1)!!}\,\left (\frac{d V_L}{dT} \right)^2 \right ]\,.
%{\cal G}_i &= \epsilon_{iab}\sum_{\ell=2} \frac{G}{c^{2\ell +1}}\,\left 
%[ \frac{(\ell +1)(\ell +2)}{(\ell -1)\ell!(2\ell+1)!!}\, U_{a L-1}\,\frac{d U_{b L-1}}{dT} \right . \nonumber \\
%&\left. + \frac{4\ell^2 (\ell +2)}{c^2(\ell -1)(\ell +1)!(2\ell+1)!!}\, V_{a L-1}\,\frac{d V_{b L-1}}{dT} \right ]\,,
\end{align}
and at leading order, using $U_{i j} = d^2{Q}_{ij}/dT^2 + {\cal O}(1/c^3)$, the luminosity reduces to the famous
``Einstein quadrupole formula'' \index{quadrupole formula}
\begin{equation}
{\cal L} = \frac{G}{5 c^5}\,\left [\frac{d^3 Q_{ij}}{d T^3}\,\frac{d^3 Q_{ij}}{d T^3} 
+ {\cal O}\left (\frac{1}{c^2}\right) \right ]\,.
\label{eq:leading order flux}
\end{equation}
\paragraph{Gravitational radiation reaction:}
The gravitational radiation acts back on the motion of the binary through a radiation-reaction force. To have \index{gravitational waveform} \index{radiation reaction}
an explicit temporal representation of the waveform $h_{ij}^{\rm TT}$ and the fluxes, one needs to 
solve the problem of motion of the source, including radiation-reaction effects. The first 
relativistic terms in the equations of motion, at the 1PN order, were derived by Lorentz and 
Droste~\cite{Lorentz17}. Then Einstein, Infeld 
and Hoffmann obtained the full 1PN corrections using the surface-integral method~\cite{EiInHof.38}, in which the 
equations of motions are deduced from the vacuum field equations and they are valid for any compact object (NS, 
BH, etc.). Petrova~\cite{Petrova}, Fock~\cite{Fock1939} and Papapetrou~\cite{Papapetrou1951} 
also obtained the equations of motion for the centres of extended bodies at 1PN order. Kimura, 
Ohta and collaborators introduced the ADM Hamiltonian formalism for doing PN computations~\cite{Kimura61,Ohtaetal74} 
and started the computation of the equations of motion for nonspinning bodies at 2PN order~\cite{Ohta1973,Ohta1974}. \index{post-Newtonian (PN) formalism} The equations of 
motion for nonspinning point masses through 2.5PN order in harmonic coordinates were obtained by Damour and 
Deruelle~\cite{DamourDeruelle1985,DamourDeruelle1986}, who built on the non-linear iteration 
of the metric proposed by Bel et al.~\cite{Bel1981}. The gravitational radiation-reaction 
force in the equations of motion at 2.5PN order made it possible to unambiguously test 
general  relativity through the observation of the secular 
acceleration in the orbital motion of binary pulsars~\cite{Damour:1991rd}. Quite importantly, because of the 
{\it effacement principle}, the 2.5PN equations of motion are independent of the internal 
structure of the bodies~\cite{Damour:1987af}.  In fact, the latter effect appears only 
at 5PN order for compact bodies. The 2.5PN equations of motion for point masses were also 
derived using extended compact objects by Kopeikin~\cite{kopeikin1985}. Itoh, Futamase and 
Asada~\cite{Itoh:2001np} also computed the equations of motion through 2.5PN order, using a 
variant of the surface-integral method. In Table~\ref{Tab:PNcm} we summarise the current status 
of the computation of the two-body equations of motion. 

\paragraph{Advances in regularisation method:} In the MPM-PN approach the two bodies are treated as point
particles using delta functions. \index{post-Newtonian (PN) formalism} As a consequence, singularities
appear when computing the near-zone metric and the equations of
motion, because the gravitational field needs to be computed at the
location of the particles. Thus, it is necessary to introduce a
regularisation. Until the early 2000's the Hadamard
regularisation was employed~\cite{Blanchet:2000nu}, but it does not provide unambiguous results 
at 3PN order, and so dimensional regularisation, which 
is a well-known regularisation scheme in particle physics, was
adopted~\cite{Damour:2001bu}. 

\paragraph{Canonical Hamiltonian approach:}
Another analytical approach that has been very effective in computing the two-body equations of motion 
at high PN orders is the canonical Hamiltonian approach. \index{post-Newtonian (PN) formalism}
The canonical Hamiltonian 
formulation of GR was developed in 1958-1963 
by Dirac~\cite{Dirac58,Dirac59a,Dirac59b}, Arnowitt, Deser, Misner (ADM)~\cite{ADM60a,ADM60b}, 
and Schwinger~\cite{Schwinger63}, with important contributions 
by deWitt~\cite{deWitt67}, and Regge and Teitelboim~\cite{Regge-Teit74} in 
the 60s and 70s. The original motivation of the formulation 
was the quantisation of GR. 

Assuming asymptotically flat spacetime and an asymptotically Minkowskian reference frame, 
one considers the 4-dimensional metric $g_{\alpha \beta}$ and the 
extrinsic curvature $K_{ij}$ of the spacelike hypersurface $x^0 = {\rm const}$. The 
Lagrangian describing non-spinning point particles interacting gravitationally is 
\begin{equation}
L = \sum_A \bm{p}_A \frac{d \bm{x}_A}{d t} + \int d^3 x (\pi^{ij} g_{ij\,,0} - N_\alpha {\cal H}^{\alpha}) 
- \oint d^2 s_i \,\partial_j(g_{ij} - g_{kk}\delta_{ij})\,,
\label{Lagr}
\end{equation}
where the surface integral is computed at infinity on the spacelike hypersurface $x^0 = {\rm const}$, 
$\pi^{ij} = -\sqrt{g}(K^{ij} - g^{ij} K)$ is the variable conjugate to $g_{ij}$, $K = g^{ij} K_{ij}$,  
$N_0 = (- g^{00})^{-1/2}$ and $N_i = g_{0 i}$ are the shift and lapse functions, respectively, and 
${\cal H}_\alpha$ are the four-momentum densities, which depend on the intrinsic curvature of 
the spacelike hypersurface $x^0 = {\rm const}$ and on the matter energy-momentum tensor. 
The Hamiltonian can be derived from the Lagrangian through a Legendre transformation. It reads
\begin{equation}
H = \int d^3 x N_\alpha \, {\cal H}^\alpha + \oint d^2 s_i \,\partial_j(g_{ij} - g_{kk}\delta_{ij})\,,
\label{Ham}
\end{equation}
where the second term on the right hand side is the usual ADM energy. 
Variations of the Lagrangian (\ref{Lagr}) with respect to $g_{ij}$ and $\pi^{ij}$ give the field equations, 
while variations with respect to $N_\alpha$ yields the constraint equations ${\cal H}^\alpha =0$.
ADM introduced suitable coordinate conditions that allow solution of the field and constraint equations 
for the metric coefficients such that they become Minkowskian asymptotically. 
By adopting such a coordinate system and imposing the constraint equations, 
one obtains the {\it reduced} Hamiltonian $H_{\rm reduced} = E[h_{ij}^{\rm TT},\pi^{ij\,{\rm TT}},\bm{x}_A,\bm{p}_A]$,  
which contains the full information for the dynamical evolution of the canonical field variables $h_{ij}^{\rm TT}$ and 
$\pi^{ij\,\rm TT}$ and the canonical particle variables $\bm{x}_A$ and $\bm{p}_A$. The reduced Hamiltonian has been  
explicitly computed by solving the constraint equations in a PN expansion and 
by adopting a suitable regularisation procedure. It contains a matter piece, an interaction piece that yields 
the radiation-reaction force, and a radiation piece. The energy and angular \index{radiation reaction}
momentum losses can be computed through surface integrals in the wave zone. 
The ADM Hamiltonian has been successfully extended to gravitationally interacting spinning particles using a 
tetrad generalisation of the ADM canonical formalism. This allowed the computation of spin-orbit 
and spin-spin couplings at quite high PN orders. \index{post-Newtonian (PN) formalism} So far, 
the ADM Hamiltonian approach has been developed mostly for the conservative dynamics through 
high PN orders~\cite{Kimura61,Ohtaetal74,Damour-Schafer85,Schafer85,Jaranowski:1996nv,Jaranowski:1997ky,
Jaranowski:1999ye,Damour:2000kk,Damour:2001bu,Damour:2000ni,Ledvinkaetal08,Jaranowski:2012eb,
Jaranowski:2013lca}. 

\paragraph{Effective-field theory approach:}
The two-body equations of motion and gravitational radiation have also been computed using Feynman diagrams, \index{effective field theory (EFT)}
in GR by Bertotti and Plebanski~\cite{Bertotti1960}, Hari Dass and Soni~\cite{HariDass:1980tq}, 
and in scalar-tensor theories by Damour and Esposito-Farese~\cite{Damour:1995kt}. An important turning point occurred in 2006, 
when Goldberger and Rothstein proposed a more systematic use of Feynman diagrams within effective field theory (EFT)  
to describe non-relativistic extended objects coupled to gravity~\cite{Goldberger:2004jt}. As discussed above, 
there are three relevant scales in the two-body problem: $r_s$, the internal structure scale or the size of the 
compact object; $r$, the orbital separation, and $\lambda_{\rm GW} \sim r(c/v)$, the radiation wavelength, where
$v$ is the typical velocity  of the body in the binary. To carry out calculations 
in a systematic manner at high orders in $v,$ Goldberger and Rothstein took advantage of 
the separation among those scales to set up a tower of effective field theories that 
account for effects at each scale~\cite{Goldberger:2006bd}.

When using the EFT approach to describe the object's size $r_s$, one does not need to resort 
to a specific model of the short-scale physics to resolve the point-particle singularity. Instead, 
one integrates out the internal structure of the object by matching onto an effective theory 
that captures the relevant degrees of freedom. Thus, one 
systematically parameterises the ignorance of the internal structure by building an  
effective point-particle Lagrangian that includes the most general set of operators consistent 
with the symmetry of GR (i.e., with general coordinate invariance). The operators 
in the point-particle Lagrangian have coefficients which encapsulate the properties 
of the internal structure of the extended objects. Moreover, short-distance divergences can be regularised and 
renormalised using standard methods in quantum field theory. Given a model for the internal structure, 
the value of the coefficients in the point-particle Lagrangian can be adjusted by a short-distance 
matching calculation so that they reproduce the observables of the isolated object. 
The resulting effective point-particle Lagrangian correctly describes length scales all the way to the 
orbital separation $r$. Then, to describe the binary problem at the scale $r$, it would be necessary 
to go beyond the point-particle effective Lagrangian and integrate out all modes of the graviton with wavelengths 
between the scales $r_s$ and $r$, so that one obtains an effective Lagrangian of composite particles 
interacting with long wavelength modes of the gravitational field. 

So far, the EFT approach has recovered and confirmed the PN results that were previously obtained with \index{direct integration of relaxed Einstein equation (DIRE)} \index{effective field theory (EFT)}
the MPM-PN, DIRE and ADM canonical Hamiltonian methods for the two-body equations of motion 
up to 3.5PN order when spins are neglected~\cite{Gilmore:2008gq,Foffa:2011ub}. When spin effects 
are included, the EFT approach has extended the knowledge of the conservative dynamics and 
multipole moments to high PN orders~\cite{Porto:2005ac,Porto:2006bt,Kol:2007bc,Porto:2008jj,Porto:2008tb,
Porto:2010zg,Porto:2010tr,Levi:2010zu,Levi:2011eq,Hergt:2011ik,Hergt:2012zx,Porto:2012as}. \index{post-Newtonian (PN) formalism}

\paragraph{Summary of results and final remarks:} 
In Table~\ref{Tab:PNcm} we summarise
the impressive current status of PN calculations for the conservative
dynamics, equations of motion and waveforms, for compact objects with and without intrinsic rotation 
(or spin) and tidal effects. We indicate with $n/2$-PN the PN term of formal order ${\cal O}(1/c^n)$ relative to
the leading non-spinning term\footnote{The spin of a rotating body is 
on the order $S \sim m l v_{\rm rot}$, where $m$ and $l$ denote the mass and 
typical size of the body, respectively, and where $v_{\rm rot}$ represents the 
velocity of the body's surface. Here, we consider bodies which are both compact, 
$l \sim Gm/c^2$, and maximally rotating, $v_{\rm rot} \sim c$. For such objects 
the magnitude of the spin is roughly $S \sim G m^2/c$.}. (For PN 
results of highly eccentric systems the reader may consult \index{post-Newtonian (PN) formalism}
\cite{Blanchet2006} and references therein.) 

\begin{table}[!t]
\addtolength\tabcolsep{0pt}% to stretch columns, if required
\caption [State-of-the-art of {\rm PN} calculations for compact binaries 
with comparable masses] {\label{Tab:PNcm} State-of-the-art of {\rm PN} calculations for compact binaries 
with comparable masses. We list main references  that contributed to the current accuracy. Unless otherwise specified, 
$n/2$-{\rm PN} refers to 
the {\rm PN} term of formal order ${\cal O}(1/c^n)$ relative to
the leading non-spinning term.}
\begin{tabular}{@{} c @{} c c c c@{}}
        \hline \hline \\[-5pt]
& No Spin & Spin-Linear & Spin-Squared& Tidal\\[5pt]\hline\\[-5pt]
{} & {4PN}\footnote{Partial higher-order PN terms in 
the two-body energy for circular orbits have been computed both analytically and numerically~\cite{Blanchet:2010zd,Blanchet:2012at,LeTiec:2011dp,Shah:2013uya,Blanchet:2013txa,Bini:2013rfa}. Bini-Damour work built on Refs.~\cite{Mano:1996vt,Mano:1996gn,Mano:1996mf}.}  & {3.5PN} & {3PN} & 7PN\footnote{2PN tidal effects in the conservative dynamics are known explicitly only for circular orbits.}\\[2pt]
{Conservative} & \cite{Jaranowski:1999ye,Damour:2000kk,Foffa:2011ub} & \cite{Kidder:1992fr,Will:1996zj,Levi:2010zu} & \cite{Kidder:1992fr,
Will:1996zj,Porto:2008tb} & \cite{Vines:2010ca,Damour:2009wj,Bini:2012gu}\\[2pt]
Dynamics & \cite{Jaranowski:2013lca,Blanchet:2000ub,deAndrade:2000gf,Blanchet:2003gy,Itoh:2003fy,Foffa:2012rn,Bini:2013zaa,Damour:2014jta} 
& \cite{Porto:2010tr,Tagoshi:2000zg,Faye:2006gx,Damour:2007nc,Hartung:2011te,Marsat:2012fn} & 
\cite{Levi:2008nh,Steinhoff:2007mb,Steinhoff:2008ji,Porto:2008jj} & 
\\[5pt] \hline\\[-5pt]
{Energy Flux}  & {3.5PN} & {4PN} & {2PN} & 6PN\\[2pt]
at Infinity & \cite{Blanchet:1997jj,Blanchet:2004ek,Blanchet:2005tk} & \cite{Blanchet:2006gy,Blanchet:2011zv,Bohe:2013cla,Marsat:2013caa} & \cite{kidder95,Will:1996zj,
Gergely:1999pd,Gergely:2000jz,Mikoczi:2005dn} & \cite{Vines:2011ud} \\[5pt]\hline\\[-5pt]

{} & 4.5PN & {4PN} & {4.5PN} & 6PN\\[2pt]
{RR Force} & \cite{Pati:2002ux,Konigsdorffer:2003ue,Nissanke:2004er,Blanchet88,Gopakumar1997} & \cite{Will:2005sn,Zeng:2007bq,Wang:2011bt} 
& \cite{Wang:2007ntb} & \cite{Vines:2010ca} \\[5pt]\hline\\[-5pt] 

{Waveform} &  {3.5PN} &  {4PN} & {2PN} & 6PN\\[2pt]
{Phase}\footnote{We refer to quasi-circular orbits only.} & \cite{Blanchet:2001ax} & \cite{Blanchet:2006gy,Bohe:2013cla,Marsat:2013caa}& \cite{Will:1996zj,Gergely:1999pd,Gergely:2000jz,Mikoczi:2005dn,Arun:2009} & \cite{Flanagan:2007ix,Vines:2011ud} \\[5pt]\hline\\[-5pt] 

{Waveform}
&  {3PN}\footnote{The $-2$ spin-weighted $(2,2)$ mode is known through 3.5PN order \cite{Faye:2012we}
.} &  {2PN} & {2PN} & 6PN\\[2pt]
{Amplitude}\footnote{We refer to quasi-circular orbits only.}  & \cite{Blanchet:1996pi,Arun:2004ff,Kidder:2007rt,Blanchet:2008je} & \cite{Arun:2009,Buonanno:2012rv} & \cite{kidder95,Will:1996zj,Buonanno:2012rv,Arun:2009} & 
\cite{Damour:2009wj,Vines:2011ud} \\[5pt]\hline\\[-5pt]

{BH Horizon}&  5PN & 3.5PN & {4PN}\footnote{Spin couplings beyond the squared ones have also been computed.} & $-$\\[2pt]
Energy Flux\footnote{We count the PN order with respect to the leading-order luminosity at infinity. BH horizon flux terms start at 4PN and 2.5PN orders in the non-spinning and spinning case, respectively.}  & \cite{Taylor:2008xy} & \cite{Alvi:2001mx,Chatziioannou:2012gq} & \cite{Alvi:2001mx,Chatziioannou:2012gq} & $-$ \\[5pt]
\hline \hline
\end{tabular}
\end{table}

In summary, during the last thirty years there has been tremendous progress in the PN computation 
of the two-body equations of motion and gravitational radiation. Results obtained using different analytical 
techniques (i.e., point particles described by Dirac delta-functions, surface-integral methods, post-Minkowskian and PN expansions, 
canonical Hamiltonian formalism, and EFT approach) have been compared with each other 
and agree. We stressed at the beginning of Sec.~\ref{sec:approx} that Einstein did not conceive the theory of GR starting  \index{effective field theory (EFT)}
from approximations to it. In spite of this, the PN approximation to the two-body dynamics is 
able to capture all relevant features of the weak-field, slow-motion dynamics and it can approximate quite 
well the theory of GR as long as the motion of comparable-mass compact objects is not highly relativistic 
and PN corrections at the highest order currently known are included. 
Future work at the interface between numerical and 
analytical relativity will be able to determine more precisely the PN region of accuracy, but this will 
imply doing very long NR simulations (i.e., simulations over hundreds of orbits) 
for generic binary configurations. \index{post-Newtonian (PN) formalism} \index{numerical relativity (NR)}

As we shall discuss in Sec.~\ref{sec:interface}, current PN results allow us 
to compute the GW phasing with sufficient accuracy to 
detect quasi-circular, neutron-star inspirals and extract physical parameters if neutron stars carry mild spins. 
Moreover, the knowledge of higher-order PN calculations has made it possible to test the 
reliability of the PN expansion as the two bodies 
approach each other and also understand the practicability of extending those 
calculations at any PN order. It turns out that as one approaches the last stages 
of inspiral crucial quantities \index{gravitational waveform}
that enter the computation of the waveforms start being very sensitive to 
the PN truncation error, leading to unreliable results. As we shall see below, by wisely combining 
different analytical techniques one can avoid such shortcomings 
and further improve the accuracy of the two-body dynamics and 
GW emission up to coalescence. 

\subsection{Perturbation theory and  gravitational self force}
\label{sec:GSF}

\index{gravitational self-force (GSF) formalism} Extreme--mass-ratio inspirals composed of a stellar-mass compact object orbiting \index{perturbation theory} \index{extreme mass ratio inspiral (EMRI)}
a supermassive or massive BH are promising sources for space-based and (future) ground-based 
detectors. The orbits are expected to be highly eccentric, non-equatorial and 
relativistic. To detect such binary systems and extract the strong-field information encoded 
in the spacetime around the larger body, one needs to model very accurately the equations of motion of 
the smaller body orbiting the BH and develop a consistent, appropriate wave-generation formalism. 
The PN framework, which is limited to slow velocities, 
is not suitable in this case. One needs to include relativistic effects at all orders and 
expand the field equations in the binary mass ratio (see Fig.~\ref{Fig:2bodymethods}). 
Henceforth, we let $m$ denote the mass of the small object, $M$ the mass of the central object and $q=m/M$    
the mass ratio. Typically, extreme mass-ratio binaries have $q \,\laq\, 10^{-5}$. 
If $g_{\mu \nu}$ is the metric of the background spacetime, 
the perturbation produced by the particle is $h_{\mu \nu} = \texttt{g}_{\mu \nu} - g_{\mu \nu}$, 
where $\texttt{g}_{\mu \nu}$ is the metric of the perturbed spacetime. The metric perturbation can be 
written as $h_{\alpha \beta} = \sum_{n \geq 1} h_{\alpha \beta}^{(n)}$ with $h_{\alpha \beta}^{(n)} \propto q^n$.   \index{post-Newtonian (PN) formalism}
In the limiting case of a very small test mass orbiting a heavy central mass, 
one works at first order in $q$ and obtains equations  
for the linear perturbations of the background geometry roughly of the kind 
$\Box_g h_{\alpha \beta}^{(1)} \sim T_{\alpha \beta}[z] \sim {\cal O}(m)$, where $T_{\alpha \beta}$ is the 
energy-momentum tensor of the small body and $z^\mu$ its worldline. [Essentially one obtains 
Eq.~(\ref{EE}) with $\eta_{\mu \nu}$ replaced by the background metric $g_{\mu \nu}$ and the 
higher-order terms $\Lambda_{\mu \nu}$ neglected.] Those equations were derived in 1950's and 1970's 
for the metric perturbations by Regge-Wheeler and Zerilli (RWZ)~\cite{Regge:1957td,Zerilli:1971wd} in the 
Schwarzschild case, and for the curvature perturbations by Teukolsky~\cite{Teukolsky:1973ha} in the Kerr case. 
Using suitable gauges, those equations (or variants of them~\cite{Sasaki:1981sx}) can be integrated analytically, for \index{post-Newtonian (PN) formalism}
quasi-circular orbits, by PN expansion in powers of $v/c$, $v$ being the velocity of the small body, 
obtaining the gravitational radiation and luminosity at very high PN orders. 
(Strictly speaking one computes analytically the Green function associated 
with the master equations.) In Table~\ref{Tab:PNem} we summarise the current status of PN 
calculations in BH perturbation theory. Furthermore, at leading order in the computation of 
the radiation field, one can assume that the small test mass moves along an adiabatic 
sequence of geodesics of the fixed background spacetime and \index{gravitational waveform}
compute the gravitational radiation numerically solving the RWZ and Teukolsky 
equations~\cite{Cutler:1993vq,Tagoshi:1993dm,Poisson:1993vp,Apostolatos:1993nu,Cutler:1994pb}. 
Much progress has been made in the last twenty years to evolve those equations 
in a robust, accurate and fast way~\cite{Hughes:1999bq,Hughes:2001jr,Fujita:2004rb,Sundararajan:2007jg,
Sundararajan:2008zm,Fujita:2009uz,Zenginoglu:2011zz,Bernuzzi:2011aj}, and compute the gravitational 
waveform $h_{\alpha \beta}^{(1)}$ in the wave zone. Today, time-domain RWZ and Teukolsky equations 
can compute not only the waveform emitted during the very long inspiral stage, but also the plunge, merger 
and ringdown stages~\cite{Damour:2007xr,Sundararajan:2010sr,Bernuzzi:2010ty,Bernuzzi:2010xj,Bernuzzi:2011aj,Barausse:2011kb}.

\begin{table}
\caption [State-of-the-art of {\rm PN} calculations in BH perturbation theory] {\label{Tab:PNem} State-of-the-art of {\rm PN} calculations in BH perturbation theory (i.e.,  
for an extreme mass-ratio compact binary) in the case of quasi-circular orbits. We list main references that 
contributed to the current accuracy. 
Unless otherwise specified, $n/2$-{\rm PN} refers to 
the {\rm PN} term of formal order ${\cal O}(1/c^n)$ relative to
the leading non-spinning term.}
\begin{tabular}{@{}c@{} c c c@{}}\hline \hline \\[-5pt]
& No Spin & Spin-Linear & Spin-Squared \\[5pt]\hline\\[-5pt]
{Energy Flux}  & {22PN} & {4PN} & {4PN}\\[2pt]
{at Infinity} & \cite{Poisson:1993vp,Sasaki:1994rw,Tagoshi:1994sm,Tanaka:1997dj,Fujita:2010xj,Fujita:2011zk,
Fujita:2012cm} 
& \cite{Poisson:1993zr,Shibata:1994jx,Tagoshi:1996gh} & \cite{Tanaka:1996ht,Tagoshi:1996gh} \\[5pt]\hline\\[-5pt]
{BH Horizon}&  {6PN} &  {6.5PN} & {6.5PN}\footnote{Spin couplings beyond the squared ones are also present.}\\[2pt]
Flux\footnote{We count the PN order with respect to 
the leading-order luminosity at infinity. BH horizon 
flux terms start at 4PN and 2.5PN orders in the non-spinning and 
spinning case, respectively.}  
& \cite{Poisson:1994yf,Mano:1996vt,Tagoshi:1997jy} & \cite{Mano:1996vt,Tagoshi:1997jy,Mino:1997bx} & 
\cite{Mano:1996vt,Tagoshi:1997jy,Mino:1997bx} \\[5pt]
\hline \hline
\end{tabular}
\end{table}
The perturbation sourced by the small test mass not only produces  
outgoing radiation in the wave zone that removes energy and angular momentum 
from the particle, but also produces a field in the near zone that acts 
on the test mass (i.e., the GSF) and gradually diverts it from its geodesic motion. 
Besides conservative terms, the GSF contains dissipative \index{gravitational self-force (GSF) formalism}
contributions that are responsible for the radiation-reaction force. The finite-mass corrections 
to the orbital motion due to the GSF are important for detection of extreme mass-ratio 
binaries and extraction of parameters. Although at any given time the GSF yields 
fractional corrections to the motion of the small body on the order of $q \ll 1$, 
these corrections accumulate over the very large number of cycles 
($\sim 1/q$), thus producing effects that cannot be neglected. \index{radiation reaction}
\index{extreme mass ratio inspiral (EMRI)}

The computation of the GSF is not an easy task because the field generated 
by the particle's motion diverges on the particle's worldline. Indeed, 
the gravitational field is infinite at the particle's position. Thus, one first needs 
to isolate the field's singular part. Quite interestingly, a careful analysis shows that the singular 
piece does not affect the motion of the particle, but only contributes to the particle's inertia and it 
renormalises its mass. The regular field is solely responsible for the GSF.

The case of a point electric charge moving in flat spacetime is well understood and dates 
back to work by Lorentz, Abrahams, Poincar\'e and Dirac~\cite{Dirac38}. The extension to curved spacetime 
has not been straightforward~\cite{DeWitt:1960fc,Mino:1996nk,Quinn:1996am}. The proper definitions 
of the singular and regular Green functions from the Hadamard elementary functions 
for the wave equation in curved spacetime were obtained only in 2003 
by Detweiler and Whiting~\cite{Detweiler:2002mi}. In curved spacetime the GSF is non-local in time.  
It is given by a tail integral describing radiation that is first emitted by the particle 
and then comes back to the particle after interacting with the spacetime curvature. Because 
the regular field that is fully responsible for the GSF  satisfies 
a homogeneous wave equation, it is a free radiation-field that interacts 
with the particle and carries information about its past. We now follow~\cite{Poisson:2011nh} 
and sketch the derivation of the GSF equation at first order. \index{gravitational self-force (GSF) formalism}

The gravitational perturbations produced by a point particle of mass $m$ can 
be described by the trace-reversed potentials $\gamma_{\mu \nu} = h_{\mu \nu} -1/2(g^{\rho \lambda} h_{\rho \lambda}) g_{\mu \nu}$. Imposing the Lorenz gauge $\gamma^{\mu \nu}_{\;\;\;;\mu}=0$, one finds 
that the trace-reversed potentials satisfy the equation 
$\Box \gamma^{\alpha \beta} + 2 R_{\mu\;\;\nu}^{\;\;\alpha\; \;\beta} \gamma^{\mu \nu} = 
-16 \pi G T^{\alpha \beta}/c^4$, where covariant differentiation 
uses the background metric $g_{\mu \nu}$, $\Box = g^{\mu \nu} \nabla_\mu \nabla_\nu$, $T^{\alpha \beta}$ being the 
point-mass's energy momentum tensor. The solutions for the potentials are obtained in terms of retarded 
Green functions $G_{+\;\mu \nu}^{\; \alpha \beta}$ as $\gamma^{\alpha \beta}(x) = 4 m \int_\gamma 
G_{+\;\mu \nu}^{\; \alpha \beta}(x,z) u^\mu u^\nu d \tau$ where the integral is done 
on the body worldline and $u^\mu = d z^\mu/d \tau$. 
The perturbations $h_{\mu \nu}$ are derived by inverting the equation $\gamma_{\mu \nu} = 
h_{\mu \nu} -1/2(g^{\rho \lambda} h_{\rho \lambda}) g_{\mu \nu}$. Furthermore, the equation of motion of 
the small body are obtained: (i) imposing that the body follows geodesics in the metric $\texttt{g}_{\mu \nu}$ 
of the perturbed spacetime, (ii) removing the singular part $h_{\mu \nu}^{\rm S}$ from the retarded 
perturbation and (iii) postulating that it is the regular part $h_{\mu \nu}^{\rm R}$ that acts on the 
small body. They read
\begin{equation}
a^\mu = -\frac{1}{2} (g^{\mu \nu} + u^\mu u^\nu) (2 h^{\rm tail}_{\nu \lambda \rho} - 
h^{\rm tail}_{\lambda \rho \nu})u^\lambda u^\rho \,,
\label{MiSaTaQuWa}
\end{equation}
with 
\begin{equation}
h^{\rm tail}_{\mu \nu \lambda }= 4 m\int_{-\infty}^{\tau^-} \nabla_\lambda (G_{+ \mu \nu \mu'\nu'}- \frac{1}{2} g_{\mu \nu}
G^{\;\;\rho}_{+\; \rho \mu' \nu'})(z(\tau),z(\tau')) u^{\mu'} u^{\nu'} d \tau',
\label{htail}
\end{equation}
where $\tau^- = \tau - \epsilon$ is introduced to avoid the singular behaviour when $\tau'=\tau$, $z(\tau)$ is the current 
position of the particle; all tensors with unprimed indices are evaluated at the current position, while 
tensors with primed indices are evaluated at prior positions $z(\tau')$. Finally, on the particle worldline the regular field 
is $h^{\rm R}_{\mu \nu;\lambda} = -4m(u_{(\mu}R_{\nu)\rho \lambda \xi} + u_\lambda R_{\mu \rho \nu \xi}) u^\rho u^\xi+h^{\rm tail}_{\mu \nu \lambda}$.

The equation of motion (\ref{MiSaTaQuWa}) was first derived in 1996 by Mino, Sasaki and Tanaka~\cite{Mino:1996nk}, and then by 
Quinn and Wald~\cite{Quinn:1996am}. It is known as the MiSaTaQuWa equation of motion. It is important to notice 
that whereas in the original derivation the MiSaTaQuWa equation appears as the geodesic 
equation in the metric $g_{\mu \nu} + h^{\rm tail}_{\mu \nu}$, in the interpretation by Detweiler and 
Whiting, it is a geodesic equation in the (physical) metric $g_{\mu \nu} + h^{\rm R}_{\mu \nu}$, which is regular 
on the worldline of the body and satisfies the Einstein equations in vacuum. The derivation 
in~\cite{Mino:1996nk,Quinn:1996am} is limited to point masses, but Gralla and Wald~\cite{Gralla:2008fg}, 
and Pound~\cite{Pound:2009sm} demonstrated that the MiSaTaQuWa equation applies to any compact 
object of arbitrary internal structure. The MiSaTaQuWa equation of motion is not gauge invariant~\cite{Barack:2001ph} 
and relies on the Lorenz gauge condition. 
To obtain physically meaningful results, one needs to combine the MiSaTaQuWa equation 
of motion with the metric perturbations $h_{\mu \nu}$ to obtain gauge invariant 
quantities that can be related to physical observables. 

Although considerable progress has been made in the last several years to develop methods to calculate the metric 
perturbation and GSF at first order~\cite{Barack:2001ph,Detweiler:2002gi}, the majority of the work has \index{gravitational self-force (GSF) formalism}
focused on computing the GSF on a particle that moves on a {\it fixed} worldline of the background 
spacetime --- for example for a static particle~\cite{Keidl:2006wk}, radial~\cite{Barack:2002ku}, 
circular~\cite{Barack:2007tm,Detweiler:2008ft} and eccentric~\cite{Barack:2009ey,Barack:2010tm} 
geodesics in Schwarzschild. Methods to compute the GSF on a particle orbiting a 
Kerr BH have been proposed (e.g., see~\cite{Barack:2002mh}) and actual implementations are underway.  
Recently, \cite{Warburton:2011fk} has carried out the first calculation, in Schwarzschild spacetime,  
that takes into account changes in the particle's worldline as the GSF acts on the particle. 
It is important to stress that it is computationally very intensive to integrate the Einstein equations 
for very long inspiraling orbits. For this reason approximation methods have also been developed. 
Assuming that secular effects associated with the GSF accumulate on time scales 
much longer than the orbital period, one can employ the adiabatic approximation~\cite{Mino:2003yg,
Sago:2005fn,Sago:2005gd,Hughes:2005qb,Drasco:2005kz}, which uses a field that 
is sourced by a geodesic and neglects periodic effects and the 
conservative portion of the GSF. For some choices of the binary parameters the adiabatic 
approximation can be sufficiently accurate for detection, 
but it has been shown~\cite{Hinderer:2008dm} to be generically inaccurate for extracting binary parameters. 

The GSF program is not yet complete.  \index{gravitational self-force (GSF) formalism} To obtain sufficiently accurate templates \index{extreme mass ratio inspiral (EMRI)}
produced by extreme mass-ratio binaries, the GSF needs to be computed at first order in $q$, but the 
gravitational energy flux at second order in $q$. This implies that the metric perturbations need to be 
computed at second order. The formalism to derive metric perturbations at second order has been 
developed~\cite{Rosenthal:2006iy,Gralla:2012db,Pound:2012nt}, calculations are underway and might 
be completed in a few years. Nevertheless, 
steady advances in the knowledge of the GSF have already been used to derive interesting, 
physical effects and higher-order PN terms, as we now discuss. 

Barack and Sago~\cite{Barack:2009ey} combined the conservative pieces of the GSF 
and metric perturbations to calculate the frequency shift in the innermost, stable circular orbit 
that originates from the GSF of the small body in Schwarzschild spacetime (see~\cite{Isoyama:2014mja} 
for the extension to the Kerr spacetime).  In \cite{Barack:2010ny} a similar shift was computed in the rate of periastron advance for eccentric orbits. In 2008  
Detweiler pointed out~\cite{Detweiler:2008ft} that the time component of the velocity vector, $u^t$, of a small test 
mass in the Schwarzschild spacetime is gauge invariant with respect to transformations that preserve the helical 
symmetry of the perturbed spacetime. 
The inverse of $u^t$ is an observable, as it is the gravitational redshift experienced by photons emitted by the orbiting body and observed 
at a large distance on the orbital axis. The redshift has been computed numerically through a 
GSF calculation as a function of the (gauge invariant) orbital frequency and it has been compared 
to analytical predictions~\cite{Blanchet:2010zd,Blanchet:2013txa,Bini:2013rfa} and used to extract 
yet unknown higher-order PN terms beyond the test-particle  \index{post-Newtonian (PN) formalism}
limit~\cite{Blanchet:2009sd,Blanchet:2010zd,LeTiec:2011ab,Blanchet:2012at,Shah:2013uya}. The latter 
was possible through the first-law of binary BH dynamics~\cite{Friedman2002,LeTiec:2011ab,Blanchet:2012at}. 
Quite importantly, the redshift factor has been shown to be simply related to the binding 
energy and angular momentum of circular-orbit binaries~\cite{LeTiec:2011ab}. Thus, the knowledge of the redshift 
can be employed to compute relativistic effects linear in $q$ in the (specific) binding energy and 
angular momentum~\cite{LeTiec:2011dp,Barausse:2011dq,Akcay:2012ea}. Other dynamical invariants have 
also been derived~\cite{Barack:2011ed,Shah:2012gu,Dolan:2013roa}. 

Lastly, in the absence of GSF results at second order in $q$ and of comparisons to NR 
simulations for intermediate mass-ratio binaries, it is difficult to assess the region covered 
by GSF results in Fig.~\ref{Fig:2bodymethods}.  \index{numerical relativity (NR)} It is generally believed that the knowledge of relativistic \index{intermediate mass black hole binaries}
effects at first order in $q$ in the conservative \index{gravitational self-force (GSF) formalism}
dynamics and second order in $q$ in the dissipative sector would be able 
to describe only waveforms from extreme-mass ratio inspirals having $q \,\laq \, 10^{-5}$. 
However, results at the interface between GSF, PN theory \index{gravitational waveform}
and NR, are suggesting that leading order GSF \index{post-Newtonian (PN) formalism}
results may have a much larger range of validity including intermediate mass-ratio 
binaries and perhaps even comparable mass binaries when $q$ is replaced by the 
symmetric mass ratio $m M/(m+M)^2$~\cite{LeTiec:2011bk,Tiec:2013twa,Nagar:2013sga}. 
Approximations to GR continue to be surprisingly successful.  

\subsection{The effective-one-body formalism}
\label{sec:EOB}

\index{effective-one-body (EOB)} In 1998-2000, motivated by the construction of LIGO and Virgo detectors and the absence of merger 
waveforms for comparable-mass binary BHs from NR, an analytical 
approach that combines the PN expansion and perturbation theory, known as the effective-one-body 
(EOB) approach~\cite{Buonanno:1998gg,Buonanno:2000ef}, was introduced. This novel approach was aimed \index{perturbation theory}
at modeling analytically both the motion and the radiation 
of coalescing binary systems over the entire binary evolution (i.e., from the inspiral to the plunge, 
then the merger and the final ringdown). Several predictions~\cite{Buonanno:2000ef,Buonanno:2005xu} of 
the EOB approach has been broadly confirmed by the results of NR simulations. \index{numerical relativity (NR)}
These include: (i) the blurred, adiabatic transition from the inspiral to a plunge, which is merely a continuation 
of the inspiral, (ii) the extremely short merger phase, (iii) the simplicity 
of the merger waveform (i.e., the absence of high-frequency features in it, with the burst of radiation produced at the merger being \index{gravitational waveform}
filtered by the potential barrier surrounding the newborn BH), (iv) estimates 
of the radiated energy during the last stages of inspiral, merger and ringdown ($0.6\%$ to $5\%$ of the binary total mass depending on BH spin magnitude) and spin of the final BH (e.g., $0.8 M_{\rm BH}^2$ for an 
equal-mass binary, $M_{\rm BH}$ being the final BH mass), and (v) prediction that a Kerr BH will 
promptly form at merger even when BHs carry spin close to extremal. Soon after its inception, the 
EOB model was extended to include leading-order spin effects~\cite{Damour:2001tu} and higher-order PN terms that 
meanwhile became available~\cite{Damour:2000we}. \index{post-Newtonian (PN) formalism}

We now describe how the EOB formalism is able, in principle, to predict the full waveform emitted by coalescing 
binary systems using the best information available from analytical relativity. In Sec.~\ref{sec:interface}, we 
shall show that the EOB approach can be made highly accurate by calibrating it to NR 
simulations, so that it can be used for detection and parameter estimation by ground- and space-based 
GW detectors. \index{effective-one-body (EOB)} \index{gravitational waveform}

There are three key ingredients that enter the EOB approach: (i) the 
conservative, two-body dynamics (or Hamiltonian), (ii) the radiation-reaction 
force and (iii) the gravitational waveform emitted during inspiral, merger and ringdown. 
In building these ingredients the EOB formalism relies on the assumption that the comparable-mass case is a smooth \index{radiation reaction}
deformation of the test-particle limit. Moreover, each ingredient is crafted through a resummation 
of the PN expansion to incorporate non-perturbative and strong-field effects that are lost when the dynamics 
and the waveforms are Taylor-expanded in PN orders. The construction of the three ingredients 
leveraged on previous results. The finding of the EOB Hamiltonian was inspired by results in the 1970's aimed at \index{post-Newtonian (PN) formalism}
describing the binding energy of a two-body system composed of comparable-mass, charged particles 
interacting electromagnetically~\cite{Brezin:1970zr}. The resummation of the radiation-reaction force was initially inspired 
by the Pad\'e resummation of the energy flux~\cite{Damour:1997ub}. The description of the merger-ringdown waveform was inspired 
by results in the 1970's on the radial infall of a test particle in a Schwarzschild BH~\cite{Davisetal72}, where it \index{gravitational waveform}
was found that the direct gravitational radiation from the particle is strongly 
filtered by the potential barrier once the test particle is inside it. The construction of 
the EOB merger-ringdown waveform was also inspired by results in the close-limit approximation~\cite{Price:1994pm}, in which 
one switches from the two-body to the one-body description close to the peak of the BH potential barrier. The 
recent description of the EOB inspiral-plunge waveform~\cite{Damour:2008gu} was inspired by the multiplicative (or factorised) 
structure of the waveform in the test-particle limit case. We now review the three basic ingredients. 

In the physical, {\it real} description, the centre-of-mass conservative dynamics of two particles of masses $m_1$ and $m_2$ and spins 
$\bm{S}_1$ and $\bm{S}_2$ is described by the PN-expanded Hamiltonian
$H_{\rm PN}(\bm{Q},\bm{P},\bm{S}_1,\bm{S}_2)$, where $\bm{Q}$ and $\bm{P}$ are the relative position and momentum 
vectors. The basic idea of the EOB approach is to construct an auxiliary, {\it effective} description 
of the real conservative dynamics in which an effective particle of mass 
$\mu =m_1\,m_2/(m_1+m_2)$ and effective spin $\bm{S}_*(\bm{S}_1,\bm{S}_2)$ 
moves in a deformed, Kerr-like geometry $g_{\mu \nu}^{\rm eff}(M,\bm{S}_{\rm Kerr};\nu)$, with mass $M =m_1+m_2$ and 
spin $\bm{S}_\text{Kerr}(\bm{S}_1,\bm{S}_2)$, such that the effective conservative dynamics is equivalent (when expanded 
in powers of $1/c$) to the original, PN-expanded dynamics. \index{post-Newtonian (PN) formalism}
The deformation parameter is 
the symmetric mass ratio $\nu = \mu/M$, ranging from $\nu = 0$ (test particle limit) to $\nu=1/4$ (equal masses). 
Exactly solving the problem of a spinning, effective particle in this deformed, Kerr-like geometry 
amounts to introducing a particular {\it non-perturbative} method for re-summing the PN-expanded 
equations of motion.

\begin{figure}
\begin{center}
\includegraphics*[width=0.55\textwidth, angle=0]{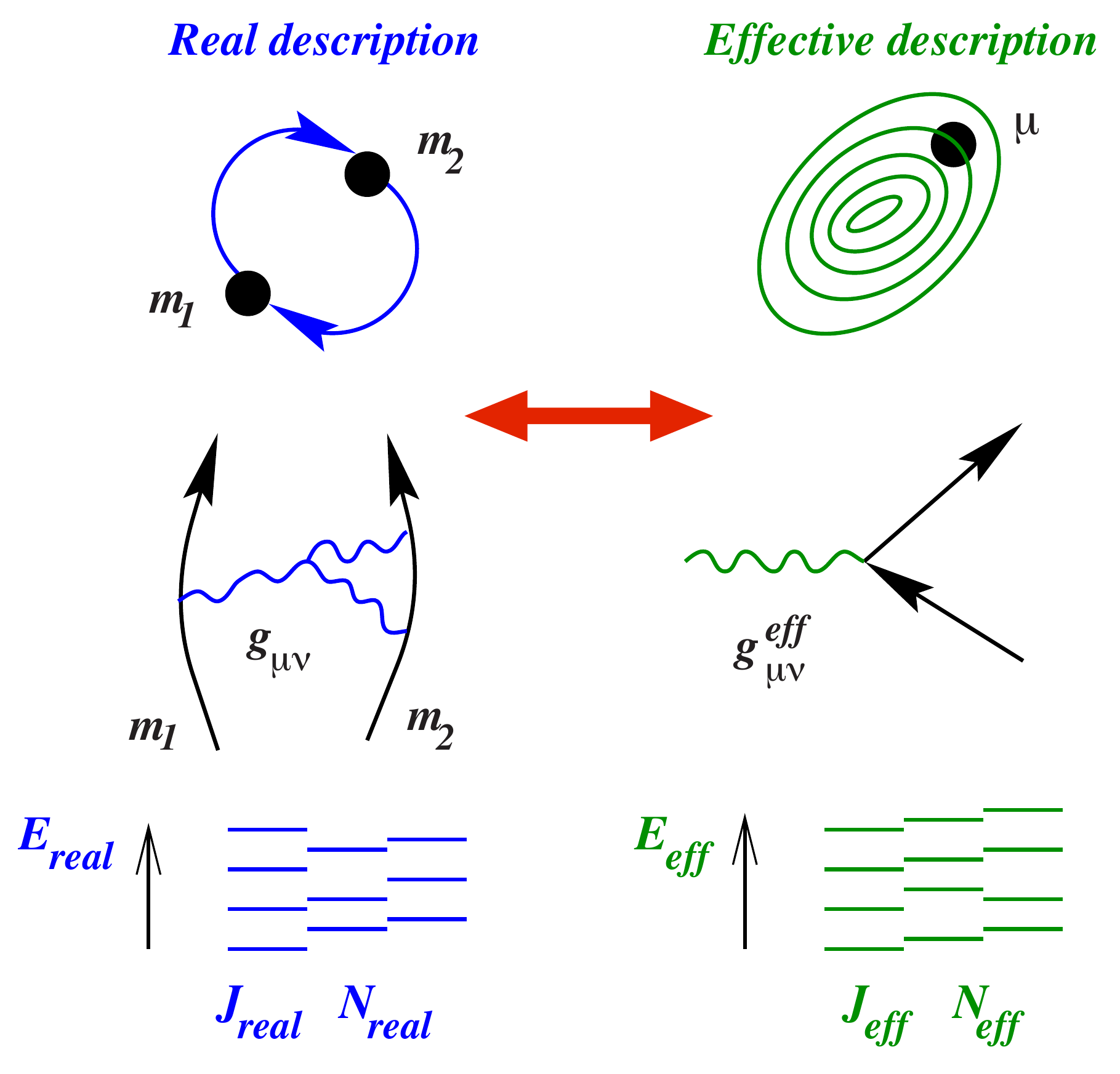}
\caption [The real and effective descriptions in the EOB formalism] {\label{mapEOB} The real and effective descriptions in the EOB formalism.}
\end{center}
\end{figure}
As done originally in \cite{Buonanno:1998gg}, even if the problem is purely classical, it 
is quite instructive to obtain such a mapping between the real and effective dynamics by
thinking quantum mechanically. For simplicity, let 
us restrict ourselves to non-spinning particles. Instead of considering the classical Hamiltonians 
$H_{\rm real}(\bm{Q},\bm{P})$ and $H_{\rm eff}(\bm{q},\bm{p})$ and their bounded orbits, we consider 
the energy levels ${E}_{\rm real}({N}_{\rm real},{J}_{\rm real})$ and 
${E}_{\rm eff}({ N}_{\rm eff},{ J}_{\rm eff})$ of the quantum bounded states associated 
with the Hamiltonian operators. The energy levels depend on 
the principal quantum number ${N}$ and the total-angular-momentum quantum number ${J}$, and they 
can be computed in a gauge-invariant manner within the Hamilton-Jacobi formalism, where ${ N}$ and ${ J}$ 
correspond to classical action variables. We sketch in Fig.~\ref{mapEOB} the real and effective descriptions. 
Because in quantum mechanics the action variables are quantised in integers, 
it is most natural to map the real and effective descriptions, requiring that 
the quantum numbers be the same (i.e., imposing ${ N}_{\rm real} = { N}_{\rm eff}$ and 
${ J}_{\rm real} = { J}_{\rm eff}$), but allowing the energy axis to change. 
Doing the mapping explicitly, \cite{Buonanno:1998gg} 
found the following, simple relation between the real and effective non-relativistic energies 
(${ E}^{\rm nr}_{\rm real} \equiv { E}_{\rm real} -M c^2$ and ${E}_{\rm eff}^{\rm nr} \equiv {E}_{\rm eff} - \mu c^2$) 
${ E}_{\rm eff}^{\rm nr} = { E}_{\rm real}^{\rm nr}[1 + \nu/2\, ({ E}_{\rm eff}^{\rm nr}/\mu c^2)]$, which can be re-written as
\begin{equation}
\frac{{ E}_{\rm eff}}{\mu c^2} = 
\frac{{ E}^2_{\rm real} -m_1^2 c^4 -m_2^2 c^4}{2 m_1 m_2 c^4}\,.
\label{EOBmap}
\end{equation}
Remarkably, the relation (\ref{EOBmap}) coincides with the one found in quantum electrodynamics~\cite{Brezin:1970zr},  
resumming part of the Feynman diagrams when mapping the one-body relativistic Balmer 
formula onto the two-body one, which describes charged particles of comparable masses 
interacting electromagnetically (e.g., positronium). (The mapping between the effective and 
real Hamiltonians can be also obtained through a suitable canonical transformation~\cite{Buonanno:1998gg}.) The 
improved resummed or EOB Hamiltonian, obtained by inverting the expression 
Eq.\,(\ref{EOBmap}), reads~\cite{Buonanno:1998gg} 
\begin{equation}
H_{\rm EOB} = M c^2\,\left [\sqrt{1+2 \nu\,\left (\frac{H_{\rm eff}}{\mu c^2}-1 \right )} -1 \right ]\,,
\label{EOBH}
\end{equation}
with 
\begin{equation}
\label{Heff}
H_{\rm eff}(\bm{r},\bm{p}) = \mu c^2\,\sqrt{A(r) \, \left [ 1 + \frac{\bm{p}^2}{\mu^2c^2} +
\left( B(r)^{-1} - 1 \right) \frac{(\bm{n}\cdot\bm{p})^2}{\mu^2c^2}+ \mathcal{Q}_4(\bm{p}) 
\right ] } \,,
\end{equation}
where $A(r)$ and $B(r)$ are the radial potentials of the effective metric 
$d s^2_\mathrm{eff} = - A(r) \, c^2d t^2 + B(r) \, d r^2 + r^2 d \Omega^2$ and 
$\mathcal{Q}_4(\bm{p})$ is a non-geodesic term quartic in the linear momentum 
that appears at 3PN order~\cite{Damour:2000we}. The metric potentials differ from the 
Schwarzschild ones by terms proportional to $\nu$. They can be computed in a PN 
series by matching the effective and real dynamics, thus  
$A_{k}(r) = \sum_{i=0}^{k+1} {a_i(\nu)}/{r^i}$ and $B_{k}(r) =\sum_{i=0}^k {b_i(\nu)}/{r^i}$. 
The EOB Hamilton equations read
\begin{equation}
\frac{d\bm{r}}{d t}=\frac{\partial H_{\text{real} }}{\partial \bm{p}}\,, \quad \quad 
\frac{d\bm{p}}{d t}=-\frac{\partial H_{\text{real}}}{\partial \bm{r}}
    +\bm{\mathcal{F}}\,,
\end{equation}
where $\bm{\mathcal{F}}$ denotes the radiation-reaction force that can be expressed, \index{radiation reaction}
assuming the energy balance equation and quasi-circular orbits, in terms of the 
GW energy flux at infinity~\cite{Damour:1997ub,Buonanno:2000ef} and through the BH 
horizons~\cite{Nagar:2011aa,Taracchini:2013wfa}. Using $-2$ spin-weighted 
spherical harmonics ${}_{-\!2}Y_{\ell m}(\theta,\phi)$, the gravitational polarisations 
can be written as  $h_+(\theta,\phi;t ) - i h_\times(\theta,\phi;t) = 
\sum_{\ell, m} {}_{-\!2}Y_{\ell m}(\theta,\phi)\, h_{\ell m}(t)$. The most recent 
description of the EOB inspiral-plunge modes proposed in \cite{Damour:2007xr,Damour:2008gu} 
read (symbolically)
\begin{equation}
\label{hlm}
h_{\ell m}^{\text{insp-plunge}}(t)=h_{\ell m}^{(N,\epsilon)}\,\hat{S}_\text{eff}^{(\epsilon)}\, T_{\ell m}\, e^{i\delta_{\ell m}}\,
f_{\ell m}\,N_{\ell m}\,,
\end{equation}
where the term $h_{\ell m}^{(N,\epsilon)}$ is the leading Newtonian mode, 
$\epsilon$ denotes the parity of the mode, the factor $T_{\ell m}$ 
resums the leading order logarithms of tail effects, the term $e^{i\delta_{\ell m}}$ 
is a phase correction due to sub-leading order logarithms, while the function 
$f_{\ell m}$ collects the remaining PN terms. Finally, the term $N_{\ell m}$ 
is a non-quasi-circular correction that models deviations from 
quasi-circular motion~\cite{Damour:2002vi}, which is assumed when 
deriving all the other factors in Eq.~(\ref{hlm}). 

Inspired by results in the 1970's~\cite{Davisetal72}, the EOB approach assumes that the merger 
\index{effective-one-body (EOB)}
is very short in time, although broad in frequency, and builds the merger-ringdown signal 
by attaching a superposition of quasi-normal modes (QNMs)~\cite{Buonanno:2000ef} to the 
plunge phase of the signal. Following the close-limit result~\cite{Price:1994pm}, 
in a first approximation the plunge and QNM signals are matched
at the light ring (i.e., at the unstable photon circular orbit), where the peak 
of the potential barrier around the newborn BH is located.  Thus, the EOB merger-ringdown 
waveform is built as a linear superposition of QNMs of the final Kerr \index{gravitational waveform}
BH~\cite{Buonanno:2000ef,Buonanno:2006ui}
\begin{equation}\label{ringdown}
  h_{\ell m}^{\text{merger-RD}}(t)=\sum_{n=0}^{N-1}
  A_{\ell m n}\,e^{-i\sigma_{\ell m n}(t-t_{\text{match}}^{\ell m})}\,,
\end{equation}
where $N$ is the number of overtones~\cite{Berti:2005ys,Berti2009}, $A_{\ell m n}$ is the complex amplitude of the $n$-th overtone, and 
$\sigma_{\ell m n}=\omega_{\ell m n}-i/\tau_{\ell m n}$ is the complex 
frequency of this overtone with positive (real) frequency $\omega_{\ell m n}$ and decay
time $\tau_{\ell m n}$. The complex QNM frequencies are known functions of
the mass and spin of the final Kerr BH. 

Finally, the full inspiral-plunge-merger-ringdown EOB \index{gravitational waveform}
waveform is obtained by joining the inspiral-plunge waveform
$h_{\ell m}^{\text{inspiral-plunge}}(t)$ and the merger-ringdown waveform
$h_{\ell m}^{\text{merger-RD}}(t)$ at the matching time
$t_{\text{match}}^{\ell m}$ as~\cite{Buonanno:2000ef}
\begin{equation}
 % \begin{split}
    h^{\text{EOB}}_{\ell m}(t) = h_{\ell m}^{\text{inspiral-plunge}}(t)\,
    \theta(t_{\text{match}}^{\ell m}-t) 
%\\&\quad 
+h_{\ell m}^{\text{merger-RD}}(t)\,
    \theta(t-t_{\text{match}}^{\ell m}) \,,
%&  \end{split}
\end{equation}
where $\theta(t)$ is the Heaviside step function. 
For $t > t_{\rm match}$ the GW emission is no-longer driven by the orbital motion,  
but by the ringing of spacetime itself and the production of QNMs. 

\begin{figure}
\begin{center}
\includegraphics*[width=0.73\textwidth, angle=0]{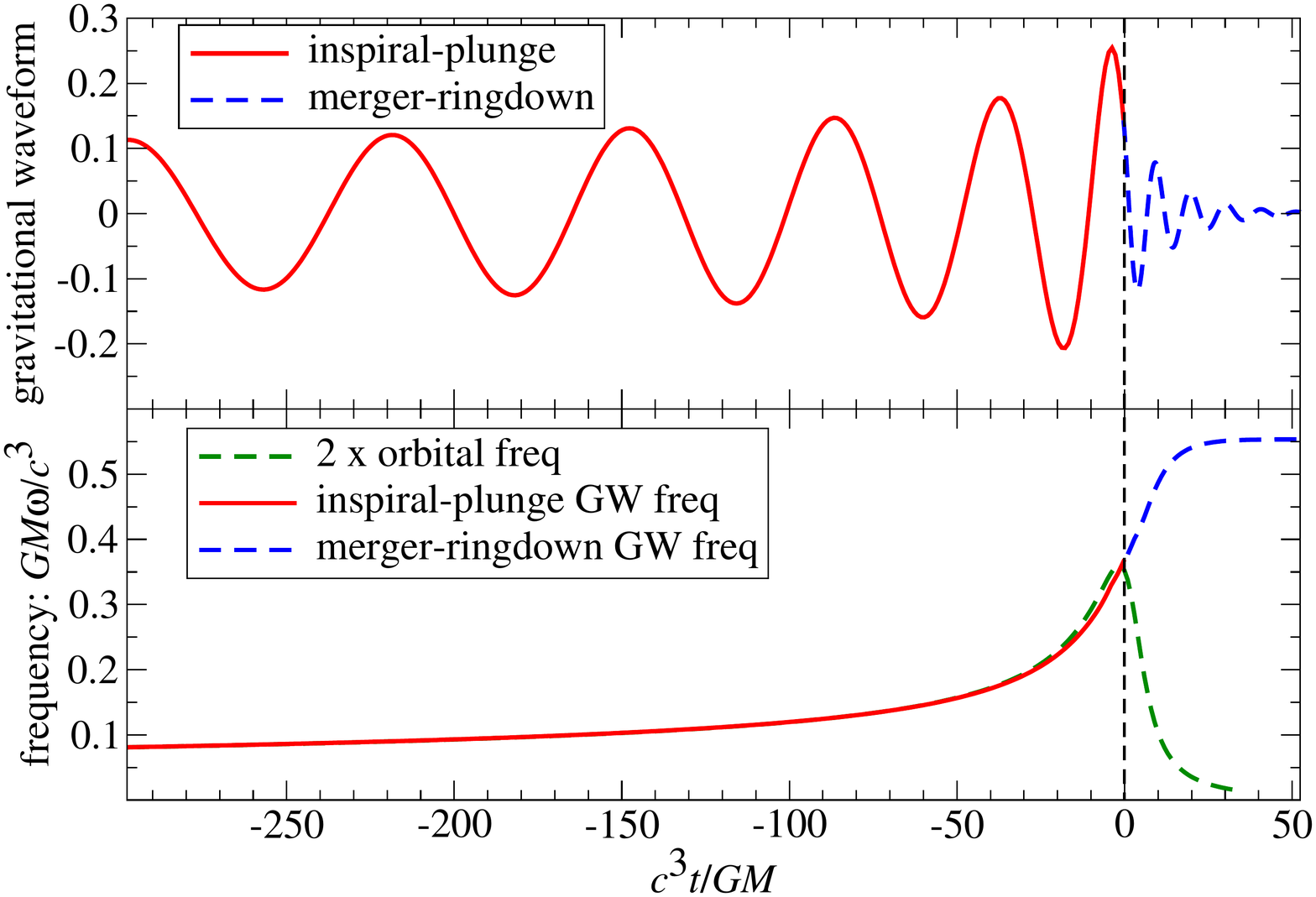}
\vskip-0.5truecm
\caption [Gravitational waveform and frequencies from in spiral, plunge, merger and ringdown] {\label{firstEOB} Gravitational waveform (upper panel), GW and twice orbital-angular frequencies (lower panel)  
from inspiral, plunge, merger and ringdown stages of a non-spinning, equal-mass binary BH as predicted in the EOB 
approach~\cite{Buonanno:1998gg,Buonanno:2000ef}.}
\vskip-0.5truecm
\end{center}
\end{figure}
In Fig.~\ref{firstEOB} we show in the top panel the first, full EOB waveform for a non-spinning, equal-mass binary BH \index{gravitational waveform}
obtained in \cite{Buonanno:2000ef} (see \cite{Buonanno:2005xu} for first, full EOB 
spinning, precessing waveforms). In the lower panel we show the EOB GW and twice orbital-angular frequencies, 
the former flattening at late times at the least damped QNM of the newborn BH, the latter having a peak around the 
EOB light ring. During  the inspiral and plunge stages the GW emission is driven by the motion of the effective particle. As the 
effective particle passes through the EOB light ring, the direct GW emission from the effective particle is filtered by the potential barrier around the newborn 
BH, and the GW radiation is driven by the perturbed spacetime geometry through the emission of QNMs.
Soon after the NR breakthrough, \cite{Buonanno:2006ui} compared 
EOB to NR waveforms finding very reasonable agreement for the late inspiral, plunge, merger and ringdown stages. \index{numerical relativity (NR)}
In particular, the EOB light ring was found to be located very close to the peak of the NR 
waveform and close to the location of the common apparent horizon, supporting the idea that GW radiation 
is quickly described by a superposition of QNMs as the two BHs merge. 
In Sec.~\ref{sec:interface} we shall discuss more recent, sophisticated comparisons and also calibrations 
of EOB waveforms. The EOB inspiraling dynamics has been compared directly to 
the one produced in NR simulations through the computation of the periastron 
advance and the binding energy. The agreements are  
remarkable~\cite{LeTiec:2011bk,Hinderer:2013uwa,Damour:2011fu}, even when 
no information from NR is used to improve the EOB model. \index{effective-one-body (EOB)}

The possibility of analytically modeling the merger waveform in the EOB approach stems from the waveform's simplicity. \index{gravitational waveform}
Does the simplicity imply that nonlinearities of GR do not play an important role? Not at all. Comparisons between 
numerical and analytical PN and EOB waveforms during the last $15$--$20$ orbits of evolution have 
demonstrated that the best agreement with NR results is obtained when corrections up to the 
highest PN order available today are included. \index{numerical relativity (NR)} Thus, as expected, non-linear 
effects are present and dominant in the strong-field regime. The waveform simplicity 
is the result of (i) the presence of only one characteristic scale close to merger, when 
radiation reaction, orbital and spin precession time scales \index{radiation reaction}
become of the same order of magnitude, (ii) the formation of a potential 
barrier around the newborn BH filtering the direct radiation from the 
merger {\it burst}, and (iii) the highly dissipative nature of 
disturbances in the BH spacetime because of QNMs. \index{post-Newtonian (PN) formalism}

The EOB conservative dynamics and waveforms have been extended to spinning BHs in 
\cite{Damour:2001tu,Damour:2008qf,Barausse:2009aa,Barausse:2009xi,Nagar:2011fx,Barausse:2011ys,Balmelli:2013zna} 
and \cite{Pan:2010hz}, respectively. In particular, motivated by the construction of a EOB Hamiltonian 
for spinning systems, that reduces to the Hamiltonian of a spinning particle in the extreme-mass ratio limit,  \index{extreme mass ratio inspiral (EMRI)}
\cite{Barausse:2009aa} worked out, for the first time, the Hamiltonian of a spinning particle in curved spacetime at all orders in PN 
theory and linear in the particle's spin. The EOB approach has also been extended to NS binary systems, incorporating \index{effective-one-body (EOB)}
tidal effects in the dynamics and waveforms~\cite{Damour:2009wj,Bini:2012gu}. 
 
To gain more insight and improve the transition from merger to ringdown 
\cite{Damour:2007xr,Bernuzzi:2010ty,Bernuzzi:2010xj,Bernuzzi:2011aj,Barausse:2011kb} 
combined the EOB approch to numerical studies in BH perturbation theory. 
Concretely, they used the EOB formalism to compute the trajectory 
followed by an object spiraling and plunging into a much larger BH, and then used that trajectory in the 
source term of either the time-domain RWZ~\cite{Regge:1957td,Zerilli:1971wd} or 
Teukolsky equation~\cite{Teukolsky:1973ha}. Solving those equations 
is significantly less expensive than evolving a BH binary in full numerical relativity. The possibility of 
using the test-particle limit to infer crucial information about the merger waveform of bodies of comparable masses \index{gravitational waveform}
follows from the universality of the merger process throughout the binary parameter space. 
The EOB approach has also been employed to generate quasi-circular, equatorial, very long, 
inspiraling waveforms in the extreme mass-ratio limit, with accuracy comparable 
to the ones produced by the Teukolsky-equation code~\cite{Yunes:2009ef,Yunes:2010zj}. 
Finally, the EOB  formalism has been improved by taking advantage of important developments 
in the GSF formalism and its interface with PN theory~\cite{LeTiec:2011ab}. \index{gravitational self-force (GSF) formalism}
In particular, using those results, the potentials entering the EOB metric have been derived at PN 
orders higher than previously known~\cite{Damour:2009sm,Barausse:2011dq,LeTiec:2011dp,Bini:2013zaa}.

\section{Compact-object binaries}
\label{sec:compact object binaries}

Binary systems of compact objects\footnote{Compactness of a body of mass $M$ and 
size $r_s$ is defined as ${\cal C} \equiv G M/(c^2 r_s).$ For BHs ${\cal C}_{\rm BH} 
= 0.5$, for NSs, depending on the EoS, ${\cal C}_{\rm NS} \sim 0.2 \mbox{--} 0.4$, 
while for the Sun ${\cal C}_{\odot} \ll 1$.} are the prime target for observation of almost 
all GW detectors. Loss of energy and angular momentum to GWs causes the companion 
stars of a binary system to spiral in toward each other, making the system more 
relativistic, in turn leading to a greater luminosity and a faster rate of inspiral. 
Indeed, a binary is a good example of a positive feedback 
system wherein radiation back-reaction makes the system more luminous. After
evolving adiabatically over millions of years, in the end, the two stars merge 
in a violent event, emitting extremely luminous gravitational radiation. 

The leading order expression for the GW luminosity of a system, 
Eq.\ (\ref{eq:leading order flux}), can be used to make an order of
magnitude estimate of how bright compact binaries can be and why 
they are the prime sources of GWs.  To leading order the luminosity
of a system composed of masses $m_1$ and $m_2$ (total mass $M=m_1+m_2$),
separated by a distance $d,$ on a circular orbit can be worked out from
Eq.\ (\ref{eq:leading order flux}) to be \cite{Blanchet2006}
${\cal L} = (32c^5\nu^2/5\,G) (v/c)^{10},$ where $\nu=m_1m_2/M^2$ is 
the symmetric mass ratio and $v=\sqrt{GM/d}$ is the orbital velocity. 
${\cal L}$ is a steep function of $v$ and depends quadratically on 
the mass ratio. Therefore, the source luminosity is greatest necessarily for 
relativistic, comparable mass systems.
Most binary systems in the Universe are asymmetric in mass and far 
from being relativistic: for the Jupiter-Sun system $v/c \sim 3\times 10^{-5}$ 
(and $\nu \sim 10^{-3}),$ for the Hulse-Taylor binary, J1913+16, $v/c\sim 10^{-3}$ 
\cite{WeisTay05} and for the AM CVn system RX J0806+1527 \cite{Israel:2002gq,Strohmayer:2005uc}
--- the most luminous source of GWs known today and expected to be observable 
in eLISA \cite{Roelofs:2010uv} --- it is $v/c \sim 4\times 10^{-3}$.

Neutron stars and BHs are the most compact objects in the Universe, with orbital velocities that can reach close to 
the speed of light; therefore, the most luminous sources are also the most 
compact and strongly-gravitating systems. Indeed, when BH binaries merge 
$v/c\sim 1/\sqrt{2}$ (see, e.g., Eq.\,(3.2) of \cite{Buonanno:2009zt})
and luminosities could reach the phenomenal levels of 
$\sim 4\times 10^{50}\,\rm W \sim 10^{24}\,L_{\odot},$ independent of the 
total mass of the binary.

\subsection{Characteristic evolution time scales and strain amplitude}
\label{sec:time scale and amplitude}

A binary predominantly emits GWs at twice its orbital 
frequency. As the system loses angular momentum and energy to gravitational
radiation its frequency increases. Noting that the luminosity in 
GWs is balanced by the loss in binding energy $E=-\nu M v^2/2,$ that is ${\cal L}=-\dot E$, at leading order 
in the PN expansion, the orbital angular frequency evolution reads
\begin{equation}
\dot \omega = \frac{\dot E}{dE/d\omega} = \frac{96}{5c^5} (G{\cal M})^{5/3}\omega^{11/3},
\label{eq:omega dot}
\end{equation} 
where we have substituted $v=(GM\omega)^{1/3}$. The quantity  ${\cal M}\equiv \nu^{3/5}M$ 
is called the {\em chirp mass} and  it turns out that at leading PN order a number 
of quantities depend on this specific combination of the component masses. 
Starting from a certain initial angular 
frequency $\omega_0,$ the time $\tau$ it takes for the system to merge (i.e.,
$\omega\rightarrow \infty$) can be estimated from $\tau=\int_{\omega_0}^\infty
d\omega/\dot\omega:$
\begin{equation}
\tau =  \frac{5c^5\,\omega_0^{-8/3}}{256\,(G{\cal M})^{5/3}}
\simeq {\rm 1000\,s} 
\left ( \frac{1.22 \,M_\odot}{\cal M}\right )^{5/3} 
\left (\frac{10\,\rm Hz}{f_0} \right)^{8/3},
\label{eq:chirp time}
\end{equation}
where $f_0 = \omega_0/\pi$ is the GW frequency, equal to twice 
the orbital frequency, and a chirp mass of $1.22\,M_\odot$ corresponds to a binary 
consisting of two 1.4\,$M_\odot$ NSs. 
The Hulse-Taylor binary, with two NSs of masses $\sim 1.4\,
M_\odot$ each and orbital period of 7.75 Hrs, will merge in about 300 
million years\footnote{The Hulse-Taylor binary has quite a large 
eccentricity $(e\simeq 0.62),$ which needs to be taken into account while computing
the merger time scale.}, its cousin J0737-3039 \cite{Burgay:2003jj}, with NSs 
of masses $1.35\,M_\odot$ and $1.25\,M_\odot$ and orbital period 
of 2.5 Hrs, will merge in 85 million years, while J0806+1527, with two white 
dwarfs of roughly $0.5\,M_\odot$ and orbital period of 321 s, will merge in 
$\sim 0.5\,\rm Myr.$ 

For a source at a distance $R$ from Earth, we can estimate the strain 
amplitude $h$ from its luminosity. Comparing Eqs.\ (\ref{quadr_form}) and 
(\ref{eq:leading order flux}) we can approximate ${\cal L} \sim 
c^3 (R\dot h)^2/20G.$ Writing $\dot h \sim \omega\, h,$ 
it follows that at leading order in the PN expansion
\begin{equation}
h \sim \frac{\sqrt{32}}{\pi c^4} \frac{(G{\cal M})^{5/3}\omega^{2/3}}{R}.
%4\times 10^{-23} \left(\frac{100\,\rm Mpc}{r}\right)
%\left(\frac{\cal M}{1.22\,M_\odot} \right)^{5/3} \left (\frac{f}{100\,\rm Hz}\right)^{2/3}
\label{eq:amplum}
\end{equation}
For a NS binary at 100 Mpc the amplitude of the waveform at 100 Hz is $h\sim 
4\times 10^{-23}$.  

\begin{figure}
\centering
\includegraphics[width=0.72\textwidth]{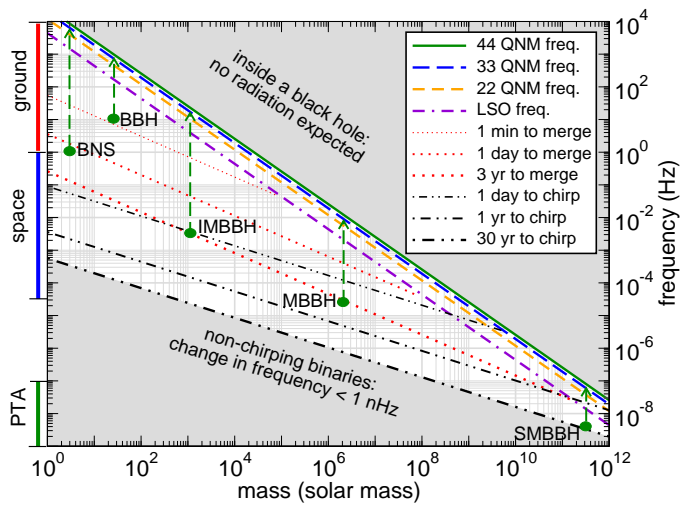}
\vskip-0.5cm
\caption [Frequency-mass diagram for equal-mass compact binaries] {Frequency-mass diagram for equal-mass compact binaries.}
\vskip-0.5cm
\label{fig:freq-mass}
\end{figure}
\subsection{Frequency-mass diagram}
\label{sec:freq-mass diagram}

In Fig.~\ref{fig:freq-mass} we plot several \index{frequency-mass diagram}
characteristic frequencies of equal-mass systems on (quasi-)circular
orbits and illustrate time scales over which they evolve.  The frequency range 
1 nHz--100 nHz is targeted by pulsar timing arrays (PTA), 30 $\mu$Hz--1 Hz by \index{pulsar timing array (PTA)}
space-based interferometers and 1 Hz--10 kHz by ground-based or underground 
detectors.  Dotted lines are frequencies starting from which a binary would last 
for 1 min, 1 day and 3 years until merger. As a binary evolves, its  orbital 
frequency changes.  For equal-mass sources that begin above the 
``30 yr (1 yr, 1 day) to chirp'' line, it will be possible to measure their chirp mass and 
luminosity distance after a 30-year (1-yr, 1-day) observational period; for systems
that begin below the ``30 yr (1 yr, 1 day) to chirp'' line, it will not be possible to 
infer the two quantities even after 30 years (respectively, 1 yr and 1 day).

During the final stages of the evolution, the component stars of a binary are no
longer able to stay on stable orbits. They plunge towards each other when the
orbital frequency is larger than a certain value, which we assume, for simplicity, 
to be that corresponding to the last stable orbit (LSO) of a Schwarzschild BH, i.e.\
$f_{\rm merge}\simeq c^3/(6^{3/2}\pi GM) \simeq (M/10 M_\odot)^{-1}\,440\,\rm Hz$. 
A binary's GW luminosity reaches its peak soon after it reaches the LSO. The merged object
in all cases, except for very low total-mass NS binaries, is a highly deformed 
BH that quickly settles down to a quiescent Kerr state, emitting a characteristic
spectrum of damped sinusoidal GWs --- the quasi-normal modes (QNMs) \cite{Berti:2009kk}. 
The complex frequencies of the QNMs depend on the mass and spin of the final BH.
Figure~\ref{fig:freq-mass} shows, for a non-spinning BH, frequency of the dominant quadrupole 
mode (i.e., $\ell=2,$ $m=2$ mode labeled `22 QNM freq.') and two of the higher-order 
modes (labelled ``33 QNM freq.'' and ``44 QNM freq.'', see \cite{Berti:2005ys}
for QNM frequencies) that are expected to carry a significant 
amount of energy. Although BHs have an infinite sequence of modes of higher frequencies, 
numerical simulations of BH mergers reveal that they are devoid of any appreciable 
energy in modes with 
$\ell>4$ \cite{PanEtAl:2011} and so we do not expect sources to radiate significantly
in the top shaded region.

\subsection{Zoo of compact-object binaries}
\label{sec:zoo of compact object binaries}

\index{compact-object binaries} Compact binaries occur in a very large range of masses and mass ratios.  In 
Fig.~\ref{fig:freq-mass} we show a few examples, but only for equal-mass, non-spinning
systems: (i) a NS binary of total mass $3\,M_\odot$ (BNS) 
that would be visible for about 15 minutes from 10 Hz in aLIGO/AdV and a few days 
from 1 Hz in ET, and it would chirp up in far less than a day from 1 Hz, 
(ii) a 20 $M_\odot$ BH binary (BBH) that lasts for almost 40 seconds from 10 Hz 
in aLIGO/AdV and $\sim 5$ hours from 1 Hz in ET, (iii) a $10^3\,M_\odot$ 
intermediate-mass BH binary (IMBBH) that would chirp up in just one day from 
3 mHz, but takes 3 years to merge, sweeping the bands of both eLISA and 
ground-based detectors, (iv) a $2\times 10^6\,M_\odot$ massive BH binary (MBBH) 
in the eLISA band that would chirp up in less than 
a year, but takes 3 years to merge, starting at 30 $\mu$Hz and (v) a $3\times 
10^{11}\,M_\odot$ supermassive BH binary in the PTA band that 
takes 100's of years to merge, but would just chirp up in 30 years. \index{supermassive black hole binaries}

\paragraph{Stellar-mass binaries:} 
Binaries of stellar-mass compact objects could contain two NSs, a NS and a BH \index{compact-object binaries} \index{stellar-mass binaries}
or two BHs. Advanced LIGO and Virgo are well positioned to observe all such systems.  
In the case of a binary composed of two NSs the merger dynamics can be quite 
complex and depends on the binary's total mass $M,$ the mass of the final remnant 
$M_{f}$ and the maximum NS mass $M_{\rm max}^{\rm NS}$ allowed by the (unknown)
NS EoS. For majority of mergers the final remnant is expected to be a BH 
\index{neutron star equation of state (EoS)}
with or without an accretion disk, except on rare occasions when $M_f<M_{\rm max}^{\rm NS},$
the final remnant can be a NS~\cite{Shibata:2006nm,Baiotti:2008ra,Duez:2009yz}.
A BH with an accretion disk might promptly form if $M_{\rm max}^{\rm NS} < M \lsim 3\,
M_\odot$ and the component masses are different from each other. However, if 
$M_{\rm max}^{\rm NS} < M \lsim 3\,M_\odot$ and the component masses are similar,  
a transient object called a {\em hypermassive} NS may form \cite{Shibata:2005ss,
Baiotti:2008ra}.  The hypermassive NS is expected to be a non-axisymmetric ellipsoid 
supported against collapse by a combination of thermal pressure and differential 
rotation \cite{Duez:2009yz} and can delay BH formation for 1 ms to 1 s 
\cite{Bartos:2012vd,Duez:2009yz}. This phase could witness quite a lot of 
rotational energy emitted as GWs, with a spectrum that is characteristic of the NS 
EoS~\cite{Shibata:2003yj,Baiotti:2006wn}.  A BH without an accretion 
disk is not a very likely outcome, but it can happen if $M\gsim 3\,M_\odot$ 
\cite{Shibata:2003ga,Shibata:2006nm, Baiotti:2008ra}.  The hypermassive NS 
phase and BHs with accretion disks could both be accompanied by significant 
emission of electromagnetic radiation \cite{Rosswog:2010ig}.
 
In the case of NS-BH binaries and binary BHs, the merger essentially produces a highly 
deformed BH that quickly settles down to a quiescent Kerr state by emitting QNM radiation.
In the case of NS-BH mergers with mild mass ratios (say, $m_{\rm NS}/m_{\rm BH}\,\gaq\, 1/3$ 
for a non-spinning BH), the NS is tidally disrupted before the LSO and forms an accretion disk, 
which can generate electromagnetic signals; for smaller mass ratios the NS directly plunges 
into the BH without forming an accretion disk \cite{Shibata:2011jka,Foucart:2012nc,Foucart:2012vn}.

It is apparent from Fig.~\ref{fig:freq-mass} that every binary in the band of ground-based 
detectors will merge within just a few days. How many such mergers might we expect each year within 
a given volume of the Universe?  From the small number of observed pulsar binaries 
it is not possible to reliably estimate the merger rate.  
The estimated median rate is about one event per year in 100 Mpc$^3,$ but it could 
be a factor 100 smaller or 10 greater due mainly to uncertainties in the distance 
to radio pulsars, their radio luminosity function, opening angle of the radio beam and
incompleteness of radio surveys \cite{Abadie:2010cfa}.
Rate predictions based on the evolution of populations of massive-star binaries
(called {\em population synthesis}) is also highly uncertain due to poor knowledge 
of the relevant astrophysics (e.g., supernova kick velocities and stellar metallicity). 
The upshot of these predictions is that aLIGO, with a horizon distance\footnote{The 
{\em horizon distance} of a detector is defined as the distance at which a 
face-on binary located directly above the plane of the detector produces a 
SNR of 8. The {\em reach} of a detector is the distance at \index{supernovae}
which a randomly oriented and located source produces the same SNR; 
the reach of a detector is a factor 2.26 smaller than its horizon distance
\cite{Finn1993}. A detector has roughly 50\% efficiency, i.e.\ is able to see half 
of all sources, within its reach \cite{Finn1993}.} of $\sim 450$ Mpc for binary 
NS mergers,  might observe between 0.4 to 400 mergers per year \cite{Abadie:2010cfa}. 

There is currently no evidence for compact binaries in which one or both of the components is 
a BH.  Population synthesis models predict a median rate of 5 binary BH \index{compact-object binaries}
mergers and 30 NS-BH mergers per Gpc$^{3}$/year; also in this case the 
uncertainties are large. Advanced LIGO,
with a horizon distance of 2.2 Gpc and 930 Mpc for these sources, could
detect 10 and 20 mergers per year, respectively \cite{Abadie:2010cfa}.  Metallicity of 
stars plays a key role in the evolution of massive stars. Black holes could be 
far more common in the Universe for low metallicities because stars would lose far 
less of their mass by stellar wind due to lower opacities and lead to more massive 
remnants towards the end of main sequence evolution. 
Expected binary BH detection rates in aLIGO for low metallicities are a factor 10 larger 
\cite{Belczynski:2010ApJ} and binary NS rates a factor 10 smaller.  
More recently, it has been noted that high mass X-ray binaries, such as IC10 X-1 
and NGC300 X-1, could be progenitors of BH binaries, in which case their merger 
rate could be far higher \cite{Bulik:2008ab}. 

\paragraph{Supermassive and intermediate-mass black-hole binaries:}
\index{supermassive black hole binaries} \index{intermediate mass black hole binaries}
There is growing evidence that certain galactic nuclei contain binary
supermassive BHs \cite{Komossa:2003wz} and eLISA would
observe binary mergers if their total mass is in the range 
$10^4$--$10^7\, M_\odot$ (see Fig.\,\ref{fig:freq-mass}). 
The merger rate of binaries of interest to eLISA is highly uncertain.  This is 
because there are only a handful of such candidate binaries that would merge within
the Hubble time \cite{Dotti:2011um}.  eLISA could observe
$\sim 10$-$100$ mergers per year, depending on the model that is used for the 
formation and growth of massive BHs \cite{Sesana:2010wy,Seoane:2013qna}. 
PTAs are expected to detect in five years or more the background produced by a population 
of $>10^7\,M_\odot$ supermassive BH binaries~\cite{Siemens:2013zla}; 
while they are not likely to observe mergers, they could detect individual 
systems if binaries of appropriate masses exist at redshifts $z\sim 0.1$-1, 
with orbital periods of $\sim 1$-30 years \cite{Sesana:2013cta}.

\begin{figure*}
\centering
\includegraphics[width=0.75\textwidth]{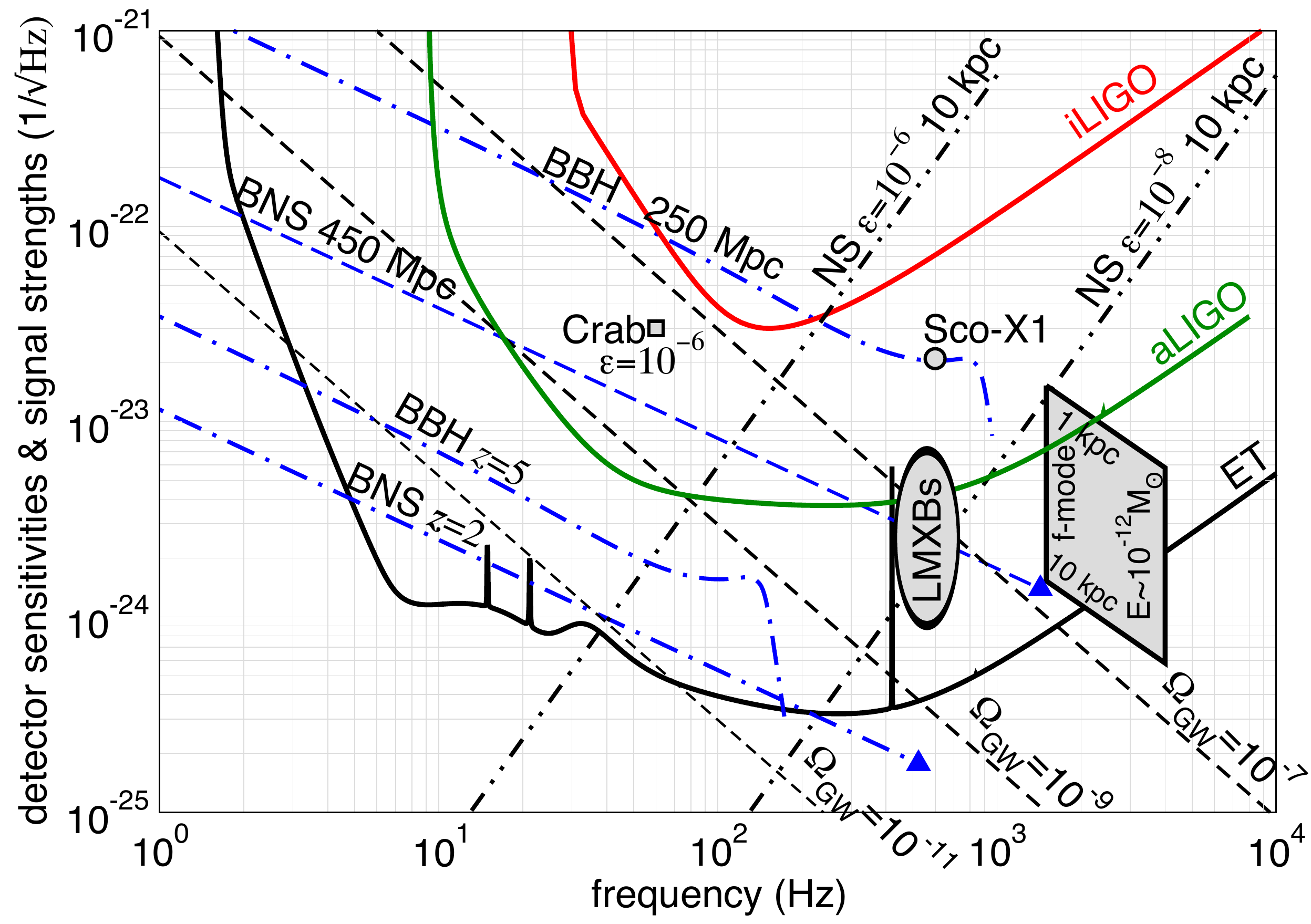}
\caption [ Sources of gravitational waves for ground-based detectors] {Sources of GWs for the ground-based detectors iLIGO, aLIGO and ET. For
continuous waves and stochastic backgrounds the characteristic amplitude of the
signals are plotted assuming an integration time of 1 year.} 
\label{fig:sources-g}
\end{figure*}
Although there is strong observational evidence for the existence of 
stellar-mass (5-30 $M_\odot$) and supermassive ($\gsim 10^6\,M_\odot$) BHs, 
little is known about BHs of intermediate mass $\sim 10^2$--$10^5\,M_\odot,$ 
not to mention their binaries.  However, there are hints that certain ultra-luminous X-ray
sources (e.g., HLX-1 in ESO 243-49 \cite{Farrell:2009ab}) might host 
intermediate mass BHs.  If a population of such objects exists and \index{supermassive black hole binaries} \index{intermediate mass black hole binaries}
they grow by merger of smaller BHs, then, depending on their masses,  
ET and eLISA will be able to detect them (see Fig.\,\ref{fig:freq-mass}). Their 
merger rates are highly speculative and range from 10 to 100
per year \cite{Gair:2010dx,AmaroSeoane:2009ui,Seoane:2013qna}. 

\paragraph{Extreme mass-ratio binaries:}
When one of the companion masses is far smaller than the other (e.g., a $10\,M_\odot$ 
\index{extreme mass ratio inspiral (EMRI)}
BH orbiting a $10^6\,M_\odot$ BH), we have the problem of a test body in {\em near} 
geodesic motion in BH geometry. Such binaries are called {\em extreme mass-ratio 
binaries,} as the mass ratio could be stupendously small $\sim 10^{-6}$--$10^{-4}$. 
eLISA would be best suited to observe the inspiral of stellar and intermediate-mass 
BHs into massive $10^4$--$10^7\,M_\odot$ BHs (see Fig.\ \ref{fig:freq-mass}).
Supermassive BHs in galactic nuclei are believed to grow \index{supermassive black hole binaries}
by the infall of stellar mass and intermediate-mass BHs. Such events could 
be observed by eLISA at cosmological distances. For instance, the inspiral of a 
$10\,M_\odot$ BH into a $10^6\,M_\odot$ supermassive BH at 
1 Gpc would be visible in eLISA. The rates in this case too are highly uncertain 
and range from a few to several hundreds \index{intermediate mass black hole binaries}
per year \cite{Sesana:2009b,Gair:2008bx,Seoane:2013qna}. 

\begin{figure*}
\centering
\includegraphics[width=0.75\textwidth]{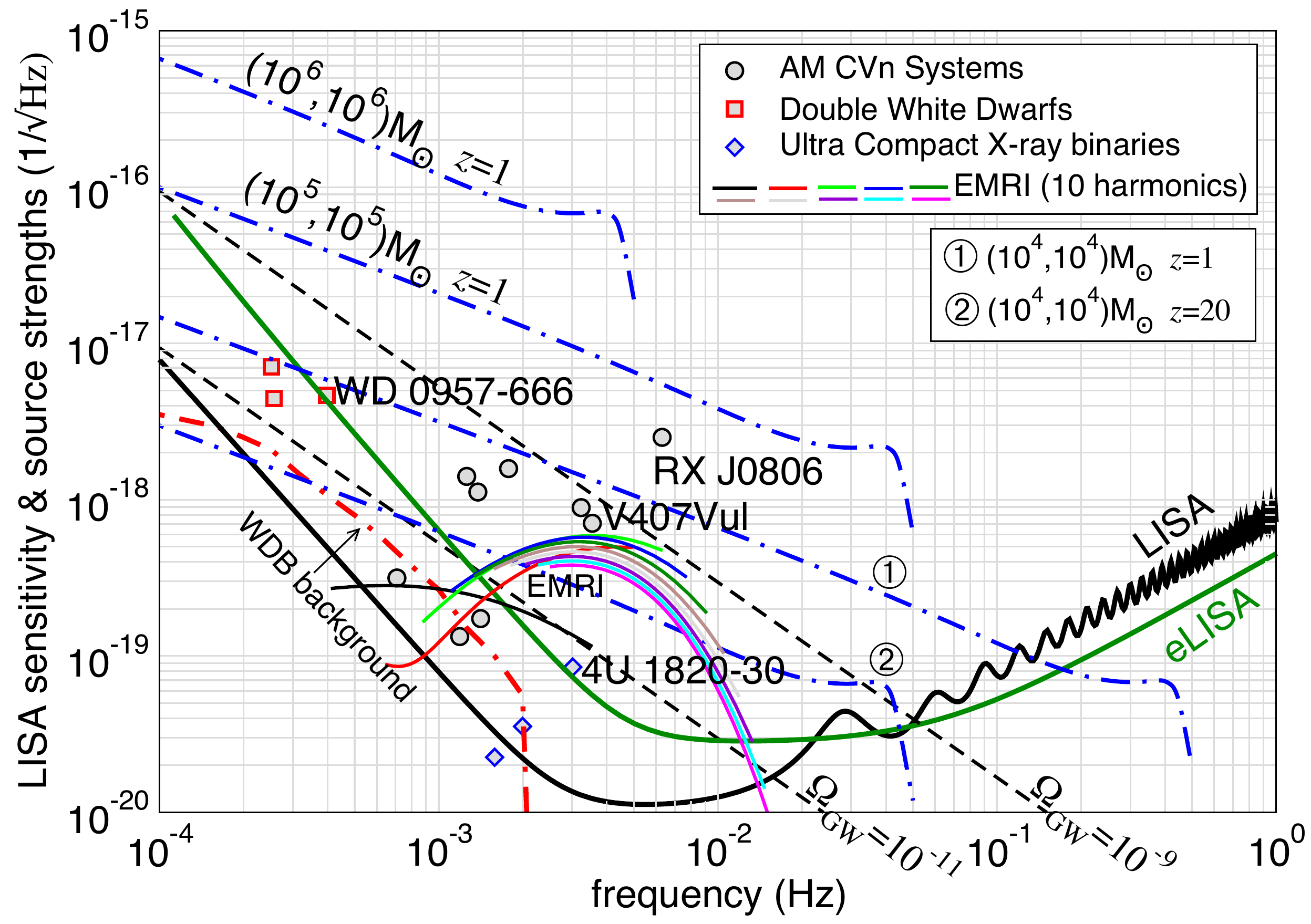}
\caption [Sources of gravitational waves for space-borne detectors] {Sources of GWs for the space-borne detectors 
LISA and eLISA. For double white dwarfs, AM~CVn systems, 
X-ray binaries and the stochastic background, the characteristic 
amplitude is computed assuming an integration time of 1 year.}
\label{fig:sources-s}
\end{figure*}

Figures \ref{fig:sources-g} and \ref{fig:sources-s} plot the characteristic amplitude 
$h_c\equiv \sqrt{f}|H(f)|,$ where $H(f)$ is the Fourier transform of the signal, for 
several non-spinning, equal-mass binaries with random orientation and 
sky position; for BH-BH binaries we use the inspiral-merger-ringdown 
signal, for BNS systems we have plotted only the inspiral part of the signal 
terminated at the LSO frequency because the spectrum of the merger signal is known 
only numerically and it varies largely, depending on the microphysics and NS EoS.
\index{neutron star equation of state (EoS)}
Also plotted are the detector noise amplitude spectra of three generations of 
ground-based interferometers in Fig.\,\ref{fig:sources-g} and two versions of LISA in 
Fig.\,\ref{fig:sources-s}.  

\begin{figure}
\includegraphics[width=0.7\textwidth]{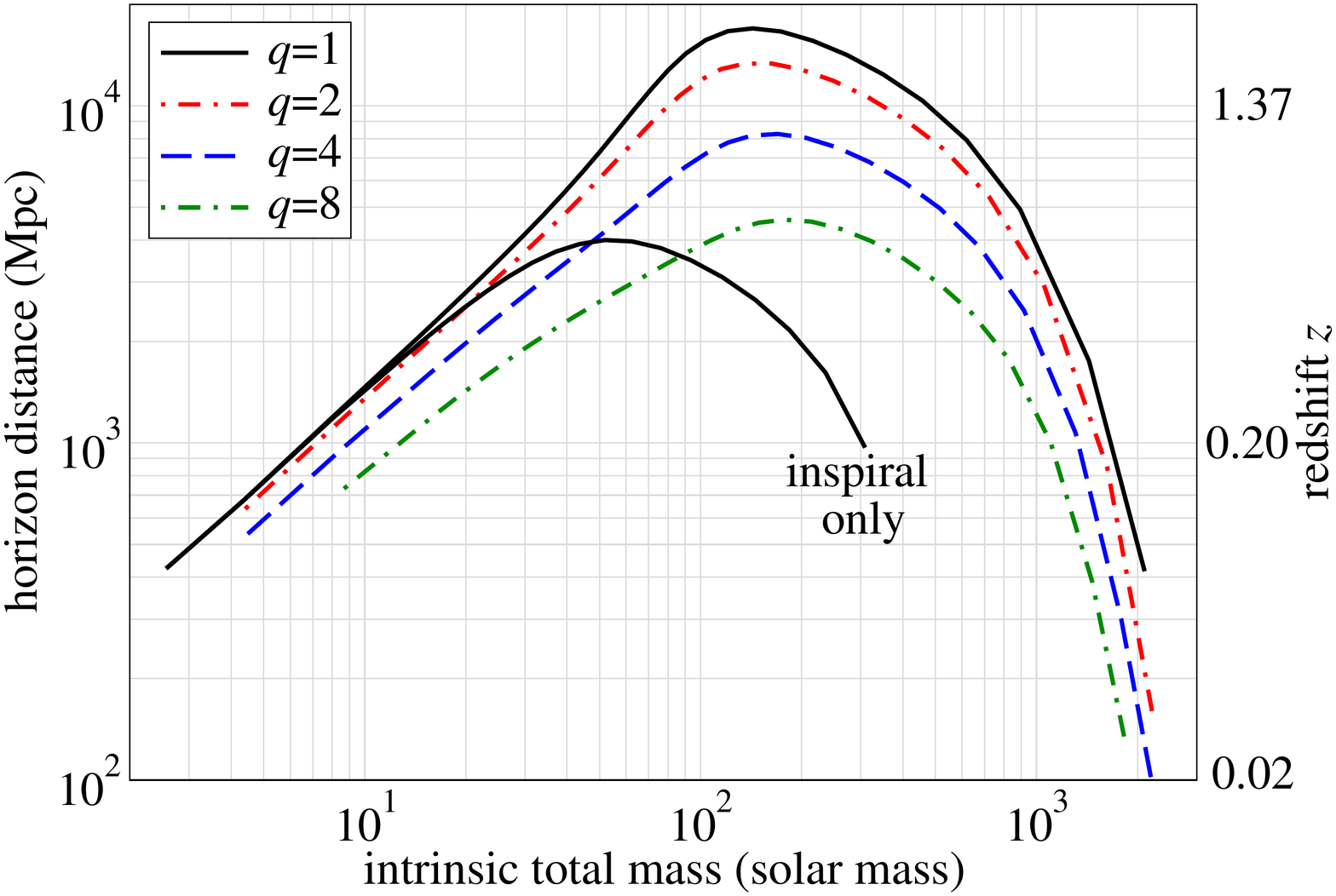}
\caption{Plots of the horizon distance of aLIGO as a function of the intrinsic 
total mass of a binary using (non-spinning) inspiral-merger-ringdown EOB waveforms calibrated to 
NR simulations~\cite{PanEtAl:2011}. We also plot the horizon-distance when we only include the inspiral 
phase terminated at the LSO of Schwarzschild.}
\label{fig:SNR aLIGO}
\end{figure}

Figure~\ref{fig:SNR aLIGO} plots the horizon distance, computed using the full 
inspiral-merger-ringdown EOB waveforms, for aLIGO as a function of the intrinsic mass 
of the binary, for several mass ratios $q=m_1/m_2 \geq 1.$ In the equal-mass case we 
also show the horizon distance using only the inspiral phase of the signal. We see that
the inclusion of the merger and ringdown portions of the signal has a significant 
effect in the horizon distance for $\gsim \,40\,M_\odot$ binaries \cite{Flanagan1998a,Damour:2000zb,Buonanno:2005xu}. 

Black-hole binaries of mass $50$--$2000\,M_\odot$ can be detected by aLIGO/AdV
at redshifts $z\sim 0.3\mbox{--}1.4$. The largest confirmed BH mass in the stellar range is $\sim 15\,M_\odot$ 
\cite{Orosz:2007ng}, but there are hints of even heavier BHs of $23\text{--}34 M_\odot$ 
\cite{Prestwich:2007mj}. Theoretically, low metallicity massive stars could lead to 
BHs of $50 M_\odot$ or higher \cite{Belczynski:2010ApJ}.  For binary systems of total 
mass $50$--$100$ $M_\odot$, detectors are sensitive to the final moments of merger, 
when the strong field dynamics dominates the evolution.  Comprehensive studies have 
shown that depending on the mass ratio, full inspiral-merger-ringdown waveforms should be used 
as matched filters when $M \,\gsim\, 10\mbox{--}15\,M_\odot$, if one wishes a loss in detection 
rate of no more than $10\%$ \cite{Buonanno:2009zt}.  Space-based detectors would observe 
the merger dynamics from $\sim 10^5$--$10^7\,M_\odot$ binaries, with SNRs 
$\sim 300$ for sources at $z\sim 3$ \cite{Seoane:2013qna,AmaroSeoane:2012km}. 
These SNRs are so large that templates would need to be improved beyond their 
current status, so as not to bias the estimation of the system's masses and its 
position on the sky. \index{matched filtering template}

\subsection{Interface between theory and observations}
\label{sec:interface}
A search for GWs from sources with known amplitude and phase evolution can 
make use of {\em matched filtering}, which involves \index{matched filtering template}
cross-correlating the detector output with a copy of our best guess of
the expected signal called a {\em template.}  If the template 
matches the signal well, then the correlation between the noisy signal 
and the template builds up with time, giving rise, on the average, to a positive output.
Matched filtering, however, is very sensitive to the signal's phase evolution; even
tiny phasing errors in the template can destroy the cross correlation. It is critical
to have accurate templates so that the SNR lost due to incorrect templates is negligible.

For low-mass, inspiraling binary systems, carrying very mild spins, 
i.e., for NS-NS binaries, any 3.5PN approximant is accurate enough for detection~\cite{Buonanno:2009zt},  
a remarkable result of the PN formalism.  
\index{post-Newtonian (PN) formalism} Until what time in the binary evolution (or, equivalently, for which total mass of the binary) 
is it safe to employ PN approximants in GW searches? Is there one 
particular PN approximant that is more accurate than others for any mass ratio and spin? Such questions 
remained unsolved for many years and were among the motivations of the EOB formalism in late 1990's.  \index{effective-one-body (EOB)}
To cope with those uncertainties, for a few years before the NR breakthrough,  
{\it detection-template families} were developed~\cite{Buonanno:2002ft,Buonanno:2002fy,Damour:2002vi,Buonanno:2005pt} 
and some of them were used in LIGO searches~\cite{Abbott:2005kq,Abbott:2007ai}. 
To incorporate possible systematics present in PN approximants, \index{post-Newtonian (PN) formalism}
those template families either extended the binary parameter space to unphysical regions or incorporated \index{matched filtering template}
higher-order physical effects, so that they could reach higher overlaps with both PN approximants 
and EOB waveforms. Eventually, after the NR breakthrough in 2005, PN approximants started 
to be compared to highly accurate NR waveforms~\cite{Boyle:2007ft,Lovelace:2011nu} and also to 
EOB waveforms calibrated to NR waveforms~\cite{Buonanno:2009zt}. \index{numerical relativity (NR)}
It was found that for $ M \,\gsim \, 10\mbox{--}15 M_\odot$, non-quasi-circular effects 
cannot be neglected and templates that include 
inspiral, merger and ringdown should be employed to avoid a large loss in the detection
rate~\cite{Buonanno:2009zt}. It was also found that PN \index{gravitational waveform}
approximants did not perform very well for large mass ratios, i.e., for 
NS-BH binaries or IMBHs. This is because in the PN approach, exact, known results in the 
test-particle limit are expanded in a PN series, washing out crucial 
non-perturbative information --- a drawback that was another motivation for the EOB formalism. 

In recent years, a variety of studies have been carried out at the interface between 
analytical and numerical relativity.  \index{numerical relativity (NR)} The results have indicated that 
the best way to provide accurate templates for a successful detection and extraction of binary 
parameters is to combine the knowledge from all the available methods: PN, GSF 
and NR. \index{gravitational self-force (GSF) formalism} One could try to directly combine PN-computed waveforms with NR \index{gravitational waveform} \index{matched filtering template}
waveforms, thus building a hybrid waveform. However, if the goal is to produce {\it highly} accurate 
templates, this method would still require high computational cost, because the different PN 
approximants agree sufficiently well with each other {\em only} at large separations, thus the hybridisation \index{post-Newtonian (PN) formalism} \index{matched filtering template}
should start hundreds of GW cycles before merger~\cite{Damour:2010zb,MacDonald:2012mp,
Pan:2013tva}. An alternative avenue is provided by the EOB approach. \index{effective-one-body (EOB)}

\begin{figure}
\includegraphics[width=0.75\textwidth]{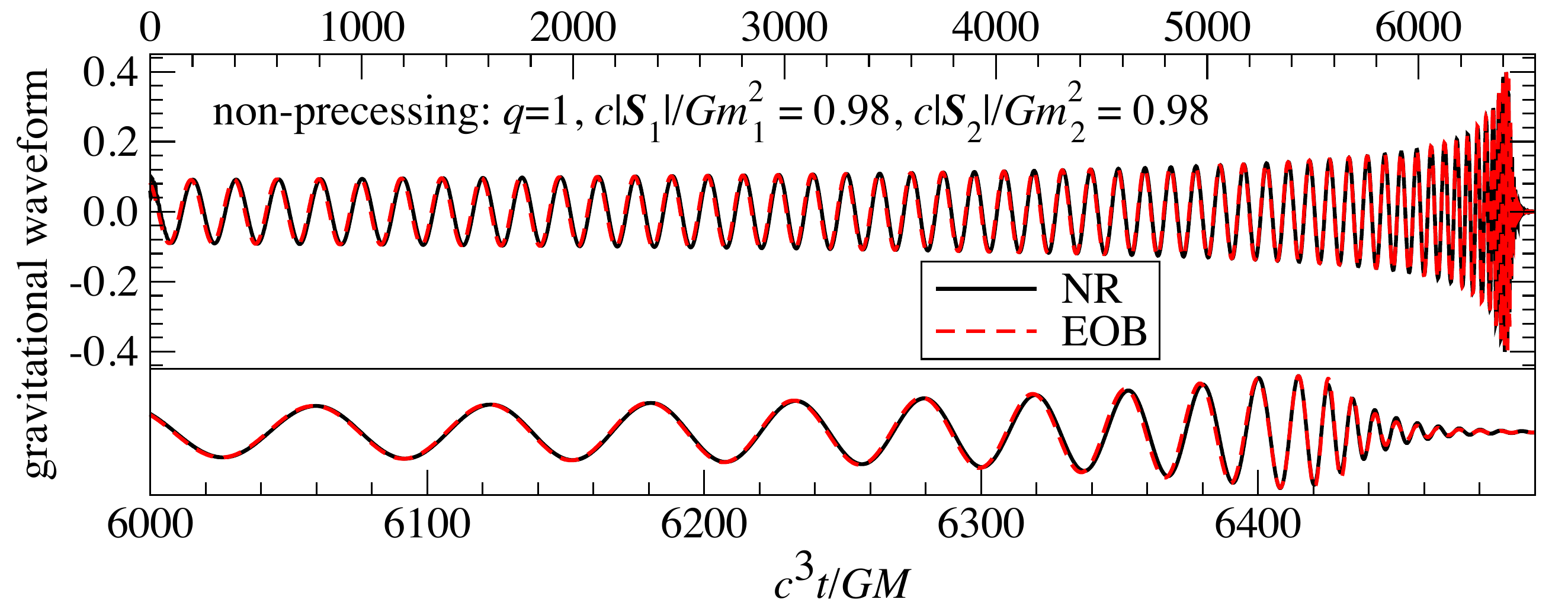}
\includegraphics[width=0.75\textwidth]{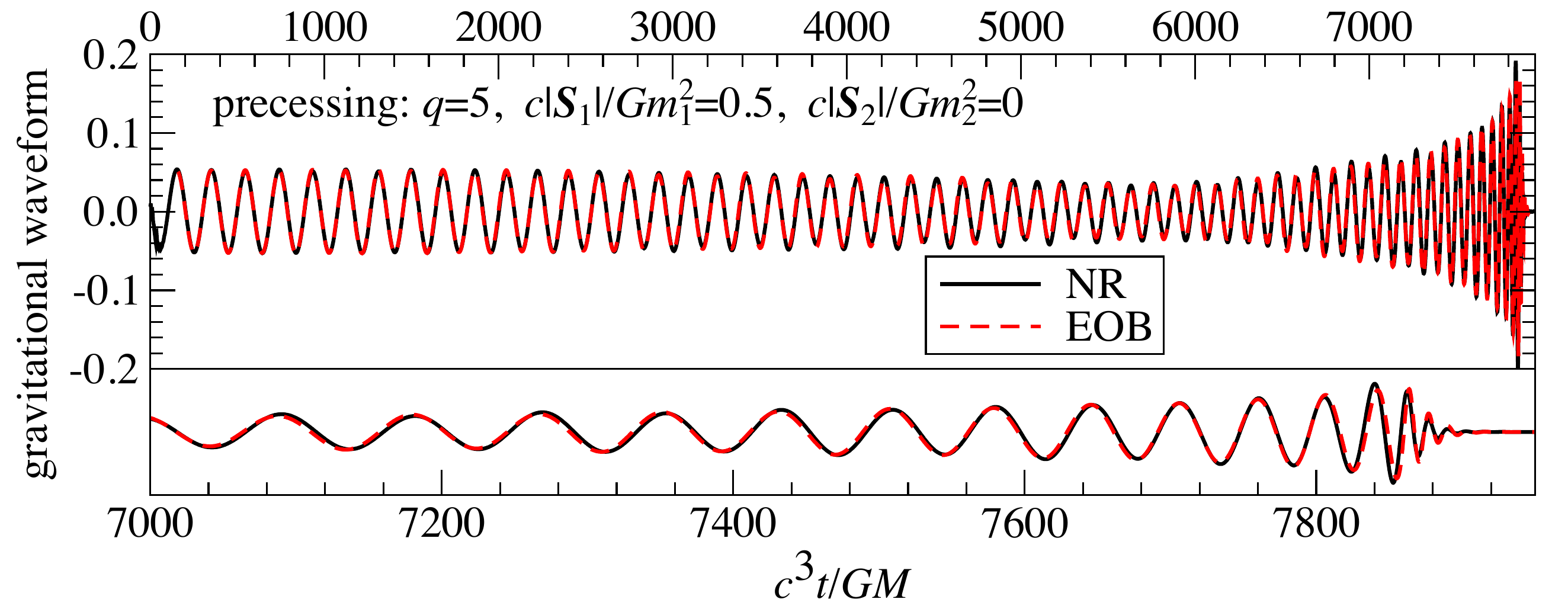}
\caption [State-of-the-art comparison between EOB and NR waveforms] {State-of-the-art comparison~\cite{Taracchini:2013rva} between (calibrated) EOB  and NR waveforms 
for quasi-extremal, non-precessing spins (top panel) and precessing spins (bottom panel); lower parts
show final few cycles.}
\label{fig:EOBNR}
\end{figure}

\paragraph{Analytical vis-a-vis numerical relativity:}
\index{numerical relativity (NR)} As earliest comparisons with NR waveforms demonstrated~\cite{Buonanno:2006ui,
Pan2007} (cf.\, Sec.~\ref{sec:EOB}), the EOB formalism is able to describe 
waveforms emitted during the inspiral, plunge, merger and ringdown stages using {\it only} analytical information.  \index{effective-one-body (EOB)}
Those first comparisons employed the 3.5PN EOB dynamics and leading-order PN waveforms. Subsequent studies 
carried out with highly accurate NR waveforms revealed the necessity of 
including higher-order PN terms in the EOB dynamics, energy flux and waveforms if the goal 
is to develop highly accurate templates for aLIGO/AdV searches. As a consequence, 
higher-order PN terms (in particular, the test-particle limit terms) are included in the gravitational 
modes $h_{\ell m}$~\cite{Damour:2008gu,Pan:2010hz,Taracchini:2012}. Since PN corrections 
are not yet fully known in the two-body dynamics, higher-order PN terms are included in the EOB dynamics \index{post-Newtonian (PN) formalism} \index{matched filtering template}
with arbitrary coefficients~\cite{Buonanno2007,DN2007b,DN2008,Buonanno:2009qa,Damour2009a,Pan:2009wj,
PanEtAl:2011,Taracchini:2012,Damour:2012ky,Taracchini:2013rva}, which are then calibrated by minimising the phase and 
amplitude difference between EOB and NR waveforms aligned at low frequency. Those coefficients have been denoted {\it adjustable or flexible 
parameters}. In particular, EOB non-spinning waveforms (including the first four subdominant modes) have been developed with any mass ratio 
and shown to be indistinguishable from highly-accurate NR waveforms with mass ratios $1$--$6$ 
up to SNRs of $\sim 50$~\cite{Littenberg:2012uj}. Note, however, that current NR waveforms cover the 
full detector bandwidth only for binaries with total mass larger than $M \,\gsim\,100 M_\odot$, thus those results 
are not yet conclusive. EOB waveforms are also stable with respect to the length of the numerical 
waveforms~\cite{Pan:2013tva}.  \index{gravitational waveform}
EOB waveforms for non-precessing systems with any mass ratio and spin have also
been developed and calibrated to existing, highly accurate numerical waveforms, which, however, do 
not yet span the overall parameter space~\cite{Taracchini:2013rva}. EOB waveforms for precessing systems
can be built from those for non-precessing ones~\cite{Pan:2013rra}; they capture remarkably well the spin-induced modulations in 
the long inspiral of NR waveforms and will be calibrated and improved in the near future.  
In Fig.~\ref{fig:EOBNR} we show the agreement between state-of-the-art EOB~\cite{Taracchini:2013wfa} \index{numerical relativity (NR)}
and NR waveforms for an equal-mass BH-BH binary with both spins aligned with the orbital-angular 
momentum and quasi-extremal (top panel), and a single-spin binary BH with mass ratio $5$, 
precessing with mild spin magnitude (bottom panel). 

Starting with \cite{Pan2007,Ajith-Babak-Chen-etal:2007}, a more phenomenological 
avenue has also been followed to produce inspiral-merger-ringdown waveforms. In this case, the 
original motivation was to provide aLIGO/AdV detectors with
inspiral, merger and ringdown waveforms that could be computed \index{compact-object binaries}
efficiently during searches and be used to detect high-mass coalescing 
compact binaries. The full waveforms are constructed by first matching inspiral 
PN templates and NR waveforms in either the time or frequency domain,
and then fitting this hybrid waveform in the frequency domain to a \index{matched filtering template}
stationary-phase-approximation based template, augmented by a Lorentzian 
function for the ringdown stage. As NR waveforms started 
spanning larger regions of the parameter space, the phenomenological 
waveforms have been improved~\cite{Santamaria:2010yb}, and extended to non-precessing~\cite{Ajith2009} \index{post-Newtonian (PN) formalism}
and precessing~\cite{Hannam:2013oca} binary BHs. 

Today, highly accurate NR waveforms, having several tens ($\sim 70$) of GW cycles and  
generic BH--spin orientations, can be produced by the pseudo-spectral Einstein code (SpEC) of the 
Simulating eXtreme Spacetime (SXS) collaboration~\cite{Mroue:2013xna}. 
Although such waveforms are not long enough and do not span the entire parameter space
to be employed as search templates, they do allow testing of the stability of calibrated analytical waveforms with respect \index{gravitational waveform} 
to the length of the simulations, improvement of the accuracy of analytical templates and also discrimination between different \index{matched filtering template}
PN/EOB approximants. We expect that longer and more accurate numerical waveforms will be produced even 
more efficiently in the near future. Furthermore, the production of many short numerical waveforms 
with finite-difference codes~\cite{Pekowsky:2013ska} 
will continue to help with extracting interesting information about the characteristics 
of the merger signal. Eventually, all those advances will reduce systematics in the analytical waveforms, 
so that they can be used not only to observe GWs with advanced detectors,  
but can also be employed in the future by space-based detectors to extract binary parameters and to test 
general relativity at high SNRs ($\sim 10^3$). 

\subsection{Results from LIGO and Virgo}
\label{sec:compact binary results}

\index{gravitational wave detection results} Data from several science runs of iLIGO and Virgo have been analysed to search
for compact binary coalescences. No GW signals \index{compact-object binaries}
were found, but the results were used to set upper limits on merger rates and
exclusion distances to short-hard gamma ray bursts (GRBs). \index{gamma ray bursts} The \index{gravitational waveform}
searches have used PN templates for ``low-mass" binaries \cite{Abadie:2012ac} 
with total mass $<25\,M_\odot$ and component masses $>1\,M_\odot$ and EOB 
templates calibrated to NR waveforms~\cite{Buonanno2007} 
for ``high-mass" binaries \cite{Aasi:2013ab} with total mass in the range 
$[25,\,100]\,M_\odot$ and mass ratio in the range $1\le m_1/m_2\le 6.$ The 
horizon distance to low-mass systems during iLIGO Science Run S6 and Virgo 
Science Run VSR3 was 40 Mpc, 80 Mpc and 90 Mpc, for binaries with $(1.35 + 1.35) 
M_\odot,$ $(1.35+5.0) M_\odot$ and $(5.0+5.0) M_\odot$, respectively.
The corresponding upper limits, in units of $\rm Mpc^{-3}\,Myr^{-1},$ were 
$130,$ $31$ and $6.4$ \cite{Abadie:2012ac}. These limits were derived for binaries with 
non-spinning components; if spins are included the upper limits are 15\% \index{matched filtering template}
higher for binaries containing one or more BHs. The horizon distance to high-mass 
binaries \cite{Aasi:2013ab} ranged from 230 Mpc for a (14+14) $M_\odot$ binary to 
nearly 600 Mpc for a (50+50) $M_\odot$ binary; the corresponding upper limits, 
in units of $\rm Mpc^{-3}\,Myr^{-1},$ were 0.87 and 0.07. Additionally, searches
for intermediate mass BHs with component masses $50\,M_\odot$ to $350\,
M_\odot$ were carried out \cite{Abadie:2012ab} using an excess-power algorithm in the
time-frequency plane, setting upper limits in the range 0.14--13 $\rm Mpc^{-3}\, Myr^{-1}$ for 
equal-mass binaries. \index{intermediate mass black hole binaries}

The upper limits from the low-mass searches are roughly two orders of magnitude
away from the expected ``realistic'' merger rates \cite{Abadie:2010cfa}. 
Advanced detectors, at their design sensitivity \cite{Aasi:2013ac}, will improve 
the distance reach by a factor of $\sim 10,$ and an increase in search volume 
by a factor of 1000. We can, therefore, expect the network of aLIGO/AdV/KAGRA 
to be making detections once they reach their design sensitivities.  The current plan \cite{Aasi:2013ac}
is to collect data intermittently for periods of 3 to 6 months each year, when 
the detectors are commissioned from their initial sensitivity in late 2015 to 
their design sensitivity by the end of the decade; detectors are expected to 
be in continuous operation after 2020. There is a fair chance that the first
detections might happen around 2017, when the  distance reach for binary NSs 
is expected to reach $\sim 100$ Mpc (or horizon distance of 250 Mpc). 
Initial detections, with two or three
detectors, might only have a moderate SNR ($\sim 12$-15) and it might only be 
possible to localise events to within hundreds of square degrees. The addition
of KAGRA and a new detector in India \cite{Iyer:2012ab} will help improve the
angular resolution of the network to tens of square degrees and facilitate 
easier EM follow-up of mergers. The analysis methods deployed in the searches have proven their ability to make 
use of the predicted waveforms to identify events and measure their parameters and 
compute the false alarm probability and statistical significance of detected events. 
The best example of this is the GW100916 event (popularly called the {\em Big Dog} event) 
that was secretly injected into the iLIGO-Virgo data streams as part of the {\em Detection 
Challenge} (\texttt{http://www.ligo.org/news/blind-injection.php}). It was successfully 
identified by the on-line and off-line analysis pipelines, attributing a very high 
significance to the detected event \cite{Abadie:2012ac}. \index{gravitational waveform}

Finally, new collaborative efforts have been established to coordinate activities between 
GW data analysts (or astronomers), numerical relativists and analytical relativists 
which have enabled testing of data analysis pipelines, production of a variety of 
NR simulations of binary BHs, and building of more robust models for use in searches and 
in extracting astrophysical information from the data~\cite{Aylott:2009ya,Ajith:2012tt,Hinder:2013oqa,Aasi:2014tra}.

\subsection{Science targets and  challenges}
\label{sec:compact binary science and challenges}

Gravitational waves from compact sources will unravel many unsolved
problems in astronomy, fundamental physics and cosmology. In this section
we will discuss science targeted by GW observations \index{compact-object binaries}
and the challenges that must be addressed in achieving those targets. 

\vspace{0.25cm}
\centerline{\emph{Science targets}}
\vspace{0.25cm}

\paragraph{Formation and evolution of stellar mass compact binaries:}
Coalescing compact binaries form from massive stars. Their formation \index{compact-object binaries}
involves a number of stellar processes. These include the evolution of the
two stars through the main sequence; gravitational collapse; supernova of 
the more massive star and associated kick that could disrupt the binary; 
evolution of the binary through the common envelope phase, wherein the larger 
giant star transfers mass to its compact companion; the second supernova and 
associated kick \cite{Postnov:2007jv}. The property of the final compact binary remnant depends on
all these factors, as well as on the metallicity, chemical composition, masses and
spins of the progenitors and the initial separation of the stars \cite{Belczynski:2010ApJ}.
Understanding the formation and evolution of compact object stars is an open problem in 
astrophysics as many of the mechanisms mentioned above are poorly understood, 
not accessible from observations and difficult to model. \index{supernovae}

Different evolutionary models of compact binaries predict different \index{compact-object binaries}
coalescence rates for the three types of compact binary mergers. By 
determining the rates for these different populations it will be possible 
to discriminate amongst the many competing models that are currently 
prevalent. Determining the mass function and spin distributions (i.e., spin
magnitudes and relative orientations) and mass ratios of companion stars,
will add more discriminatory power to single out the correct model that
describes the formation and evolutionary mechanisms \cite{OShaughnessy2008}. 
Some models also predict a gap in the largest NS mass allowed by the EoS 
and the smallest mass of a BH formed by stellar evolution. Gravitational-wave 
observations could help verify the existence of such a mass gap \cite{Hannam:2013uu}.

\paragraph{Gamma-ray bursts:} Observing GWs in coincidence with GRBs will have a tremendous 
impact on understanding the progenitors of GRBs and how they are powered. Moreover, coincident 
observation of GRBs and GWs will help identify the host galaxy \index{gamma ray bursts}
of a GW source and measure its redshift. If progenitors of short GRBs (shGRBs) are binary 
NS mergers~\cite{Fong2013}, then this would help measure both the luminosity distance 
and redshift to the source, {\em without} the use of the cosmic distance ladder 
\cite{Schutz:1986,DelPozzo:2012}. Clearly, such observations will have a great potential for 
precision cosmography \cite{Dalal:2006ab,Sathyaprakash:2009xt,Nissanke:2009kt}. 

If binary NS mergers are progenitors of shGRBs then in addition 
to beamed emission of gamma-rays, they could also emit isotropic radiation 
at optical and infra-red wavelengths. The neutron-rich material that is
ejected in the process of merger could produce heavy nuclei that decay by 
r-process radioactivity, powering a transient optical and infra-red source 
called \emph{kilonova} \cite{Rosswog:2010ig}. Follow-up observations with 
the Hubble Space Telescope of the shGRB 130603B at a redshift of $z=0.356$ is 
the first evidence of a kilonova \cite{Berger:2013,Tanvir:2013}, with apparent
magnitude of 25.8. At 200 Mpc, the distance reach of ground-based advanced detectors to binary 
NS mergers, such transients would have an apparent magnitude of about 23 and be
observable by some of the ground-based optical and near-IR telescopes, but sky 
localisation of GW events will be a challenge. \index{gamma ray bursts}

\paragraph{Cosmology:} 
Compact binary sources are quite unique for cosmology as they are standard candles 
\cite{Schutz:1986}: for binaries that chirp up during the course of observation \index{compact-object binaries}
one can measure the source's luminosity distance from GW observations alone. 
Weak gravitational lensing would bias distance measurements of individual 
sources \cite{Holz:2005df}, but a large population of events, as might be expected in the case of ET, 
can average out lensing biases \cite{Sathyaprakash:2009xt}. If the host galaxy of a 
merger event is identified and its redshift measured, then we can use a population of 
binary coalescence events to infer cosmological parameters.  In fact, GW observations 
might also measure the redshift through galaxy clustering using wide-field galaxy 
surveys~\cite{Schutz:1986,DelPozzo:2012}. Moreover, tidal effects can be used to 
determine the source's redshift provided the NS EoS is known \cite{Messenger:2011gi}. 

There is strong observational evidence that galactic centres,
including the Milky Way \cite{Ghez:2003qj}, host supermassive 
BHs of $10^6$--$10^{12}\,M_\odot.$ When and how did such BHs form? 
Did the BHs precede the galaxies or did they form after the galaxies 
were assembled? What were their initial masses and how did they grow? These are 
among the most pressing unsolved questions in cosmology. Gravitational-wave   
observations by PTAs, eLISA and ground-based interferometers will
together cover the entire spectrum of BH binaries up to $z\sim 5$-20, 
redshifts so large that the Universe was in its infancy assembling the first stars 
and galaxies. \index{supermassive black hole binaries}

\paragraph{Neutron-star equation of state:}
\index{neutron star equation of state (EoS)}
By determining the EoS of NSs we can infer the composition and structure of NS 
cores, which has remained largely unknown nearly half-a-century after the discovery 
of the first pulsar, although astronomical observations have begun to indicate hints of 
neutron superfluidity in the core \cite{Shternin:2010qi,Page:2010aw}.  Tidal 
interaction in compact binaries, where one or both of the companions is a NS, depends 
on the EoS. Tidal effects are imprinted in the inspiral phasing starting at 5PN order 
beyond the leading term and in the merger and post-merger dynamical phases, \index{compact-object binaries}
as well. Advanced LIGO/AdV could distinguish between extreme models of equations 
of state by observing $\sim 25$ NS binary inspirals \cite{DelPozzo:2013}, while ET 
should measure the EoS quite accurately with a single loud event \cite{Markakis:2011vd}.

\paragraph{Testing gravitational dynamics:}
Signals from coalescing compact binaries can be used to probe the strong-field 
dynamics of gravity and as such facilitate tests of GR and its alternatives.
Proposed tests either assume that GR is correct and look for small 
deviations from GR \cite{Mishra:2010tp,Li:2012}, or begin \index{compact-object binaries}
with an alternative theory of gravity and determine the degree to which 
observations favour the alternative \cite{Will:2005va,Yunes:2013dva}. 
Orbits of small BHs plunging into massive BHs are very complex
and capture the non-linear dynamics of gravity. By observing the emitted 
radiation one can map out the geometry of the massive objects that reside 
in galactic nuclei and check if it agrees with the Kerr geometry 
or if BHs have extra ``hair'' \cite{Ryan97,Barack:2006pq,Gair:2012nm}.  
An alternative approach checks for consistency of GWs from QNMs emitted 
during the ringdown phase of a BH binary merger \cite{Gossan:2011ha}.

\vspace{0.25cm}
\centerline{\emph{Challenges}}
\vspace{0.25cm}

\paragraph{Critical problems in numerical relativity:} It is important to continue to test the accuracy 
of analytical inspiral-merger-ringdown waveforms by comparing them to long NR waveforms and to characterise \index{numerical relativity (NR)}
any systematic biases in parameter estimation that 
might result due to inaccurate modeling. If BHs carry large spins, waveforms emitted by NS-BH and BH-BH 
binaries with mass ratios $\,\gaq\,4$ will have modulations due to spin-induced precession, which
accumulate mostly during the long inspiral. Thus, more comprehensive 
studies are needed to understand the dynamics as a function of mass ratio, BH spins and EoSs of NSs. \index{neutron star equation of state (EoS)}
Moreover, at present, accurate NR waveforms span the entire aLIGO/AdV bandwidth only if their total 
mass is larger than $\sim 100\, M_\odot$ and their mass ratio is $\,\laq \, 10$. 
Longer waveforms for larger mass ratios and generic \index{gravitational waveform}
spin configurations would be highly challenging, but they will be invaluable in validating analytical models, 
even if they sparsely sample the full parameter space.

Binary neutron star and NS-BH merger simulations that use realistic EoSs and include 
neutrino transport and other microphysics will be necessary to extract the best science 
from the data. Such simulations will also provide more accurate templates for the merger 
and bar-mode instability phases to infer redshifts from GW observations alone. 
Analytical and semi-analytical models of NS-NS and NS-BH mergers will be crucial for
parameter estimation and Bayesian hypothesis testing.  This will be a huge challenge 
for the bar-mode instability regime where the signal does not seem to have any phase 
coherence (see, e.g., simulations in \cite{Shibata:2003yj,Baiotti:2006wn}). Even so, 
accurate modeling  \index{matched filtering template}
of the expected spectrum or time-frequency content of the signal can be useful in 
understanding the physics of dynamical instabilities.  Several related questions remain 
open such as the role of large NS magnetic fields and spins in the merger dynamics of 
binary NSs and NS-BHs \cite{Rosswog:2010ig}.

\paragraph{Critical problems in analytical relativity:} As discussed in Sec.~\ref{sec:PN}, 
advances in PN theory over the last thirty years will 
enable the detection of BNS inspirals (if NSs carry mild spins) with negligible loss in 
detection rates and in extraction of binary parameters. When spins 
are present, the PN phasing and amplitude are not known as accurately as in 
the non-spinning case, causing PN approximants to differ substantially even during 
the inspiral phase~\cite{Nitz:2013mxa}. Thus, spin couplings through at least 
4.5PN order beyond the leading order are required in the conservative dynamics and 
gravitational flux at infinity, and through a similar PN order in the BH-absorbed 
horizon flux. Once available, spin couplings will also be employed in the EOB formalism 
to further improve inspiral-merger-ringdown waveforms. Considering that today 
the two-body, non-spinning conservative dynamics is known at 4PN order, it will be relevant 
to derive the energy flux at  4PN order to allow the computation of
the phase evolution at 4PN order, further validating analytical templates against 
NR waveforms. \index{effective-one-body (EOB)}

Recent results that take advantage of GSF calculations have allowed \index{gravitational self-force (GSF) formalism}
the computation of new terms in the conservative dynamics at PN orders higher than 4PN. 
However, if we consider that to fully complete the computation of the dynamics 
at 3PN and 4PN orders, it was necessary to overcome novel, specific subtleties that appeared
at those PN orders (notably, at 3PN order one had to replace the Hadamard regularisation with dimensional regularisation 
and at 4PN order it was necessary to introduce a non-local in time action to properly include tail effects), 
it is difficult to imagine that, in the future, PN calculations could be systematically and automatically extended 
to an {\it arbitrarily high} PN order by simply using algebraic computer programs. This limitation 
does not depend on the particular technique that is used (MPM-PN or DIRE formalisms, 
EFT or Hamiltonian canonical formalism), but it seems simply a consequence of 
quibbles and complexities of the nonlinearities of GR. It is worth noticing that at present
it does not also seem necessary to have Taylor-expanded PN results at arbitrarily high PN 
orders to detect the signals and to extract the best science. A combination of PN, perturbative, 
and NR results and resummation techniques can effectively work around the problem.  \index{direct integration of relaxed Einstein equation (DIRE)}  \index{effective field theory (EFT)}

As discussed in Sec.~\ref{sec:GSF}, to obtain sufficiently accurate templates for extreme 
mass-ratio binaries, the metric perturbations need to be computed at second order. The 
formalism has been developed, calculations are underway and hopefully will be completed in the 
next few years. \index{extreme mass ratio inspiral (EMRI)} \index{matched filtering template}

\paragraph{Refining analytic source models and waveforms:}
Current searches in ground-based detectors assume that compact binaries are on 
quasi-circular orbits when they enter the detector sensitivity band \index{compact-object binaries}
(see, e.g., \cite{Aasi:2013ab}). This is very
likely a good approximation for binaries formed in fields as radiation back reaction circularises a binary much \index{radiation reaction}
faster than the orbit decays \cite{PetersMathews1963,Kowalska:2010qg}.  
However, eccentric binaries may form through Kozai mechanism or 
dynamical capture in dense stellar environments~\cite{Miller-Hamilton2002,Wen2003,Kocsis:2011jy,Naoz:2012bx} and 
in this case the eccentricity can be large at merger.  
Eccentric-binary event rates are very uncertain. Eccentric-binary waveforms are not required for detection 
in aLIGO/AdV/KAGRA, unless the eccentricity at 20 Hz is larger than 0.1. 
Faithful models that take into account eccentricity would be needed to detect 
GWs from highly eccentric binaries or to accurately estimate parameters of 
mildly eccentric binaries.

Except for \cite{Abbott:2003pj}, past analysis pipelines have mostly searched 
for binaries composed of non-spinning objects in the two-dimensional space of component 
masses~\cite{Aasi:2013ab}. It is important to explore the relevance of including spins 
for detection. Several studies have indicated that due to degeneracies in the binary 
parameter space, template families containing a single effective spin could be 
sufficient to detect waveforms emitted by double-spin binary systems~\cite{Buonanno:2002fy,Pan:2003qt,Buonanno2004,
Buonanno:2005pt,Ajith:2011ec}. Moreover, precession-induced modulations can be incorporated in template waveforms in \index{matched filtering template}
an efficient way, reducing also the dimensionality of the parameter space~\cite{Buonanno:2002fy,Buonanno:2005pt,
Schmidt:2012rh,Pan:2013rra,Hannam:2013uu}. However, we still miss a comprehensive study that spans the entire parameter 
space and determines, after taking into account how the improvements in sensitivity can be set aside by increases in  
false alarm probability, in which region of the parameter space single- and/or double-spin, 
non-precessing or precessing searches, are needed (for first steps in this direction see 
\cite{Harry:2011qh,Privitera:2013xza}). Furthermore, studies are needed on the systematic biases 
in the estimation of parameters due to the use of incomplete waveform models, especially when spin-induced 
modulations in NS-BH and BH-BH are included. At present EOB waveforms for spinning BH binaries are 
computationally expensive to generate (although far faster than doing NR simulations)
and hence not suitable for use in Markov Chain Monte Carlo-based parameter estimation methods. 
Quite importantly, accelerated waveform generation techniques that use reduced-order 
algorithms or singular-value decomposition techniques have been proposed to address such 
problems~\cite{Cannon:2010qh,Cannon:2011xk,Field:2011mf,Field:2013cfa,Purrer:2014fza}.

Finally, using PN results and NR simulations, analytical templates that extend up to merger 
and include tidal effects are under development~\cite{Bernuzzi:2012ci,Damour:2012yf}. 
They will be needed to extract the best information on tidal effects and EoS in NS-NS 
and NS-BH coalescences. Including tidal effects in {\it point-particle} templates calibrated 
to highly accurate BH-BH waveforms from NR (e.g., 
EOB waveforms) is the best way of controlling systematic errors due to lack of knowledge of higher-order \index{gravitational waveform} \index{matched filtering template} \index{neutron star equation of state (EoS)}
point-particle terms in the PN expansion. In fact, employing inspiraling, point-particle PN templates augmented 
by tidal effects, leads to large systematic biases and limits the extraction of 
tidal information~\cite{Favata:2013rwa,Yagi:2013baa}.

Lastly, all sources of systematic effects in GR waveforms  need to be under control if one wants to 
measure possible deviation from GR. Current waveforms are likely not to satisfy this requirement in the majority of 
the parameter space for ground-based and especially space-based detectors. 
A comprehensive study that also employs accurate waveforms from alternative theories of gravity 
both for the inspiral, but also the plunge and merger stages, is needed to understand how much one should 
to reduce systematic errors in GR waveforms and to be able to observe 
deviations from GR. \index{gravitational waveform}

\paragraph{Synergy between EM and GW observations:}
Following up GW events using EM telescopes, and likewise
analysing GW data at the time of EM transients, will be invaluable in 
enhancing the scientific returns of observations.  Since the EM sky is full
of transient events, understanding which EM transients to follow-up in
GW data is important as otherwise coincidences will lose significance.
While many EM transients are easily identifiable, challenges remain in
unraveling the nature of astronomical transients \cite{Djorgovski:2011rv}. A study 
of the fraction of EM transients that might look like GW progenitors, and 
hence contribute to false coincidences, is desirable.

EM followups of GW events rely on accurate estimation of source position 
\cite{Fairhurst:2010is}. What are the methods by which we might be able to 
improve sky localisation?  For example, sub-dominant harmonics in systems with 
large mass ratios \cite{VanDenBroeck07b,Capano:2013raa} and binaries with 
rapidly spinning components \cite{Raymond:2008im}, could 
both enhance sky resolution and galaxy surveys could help target specific sky 
patches \cite{Hanna:2013yda}.  A proper understanding of 
biases in the estimation of sky position due to inaccurate waveform models 
or the use of galaxy catalogues is necessary.  

Pulsar timing arrays are approaching astrophysically relevant sensitivity levels
and setting limits on the detectability of binary supermassive BH binaries \cite{Arzoumanian:2014gja}.
SKA could observe continuous waves from an isolated supermassive BH binary, if one
\index{isolated black holes} \index{pulsar timing array (PTA)}
exists in the relevant frequency range, within $z\lsim 1$ \cite{Burke-Spolaor:2013aba}. 
The challenge here would be to control systematics in pulsar timing noise and to
discover a large number of ($\sim 100$) stable (timing noise $\lsim 20$~ns) millisecond 
pulsars. \index{supermassive black hole binaries}

\section{Isolated compact objects}
\label{sec:sources:neutron stars}

Over the past 30 years, astronomy has made great strides in observing 
compact objects and their environments and equally raised new puzzles 
about the interior structure of NSs. In particular, X- and gamma-ray 
observations have identified new potential sources of GWs. On the 
theoretical front, there is now a vast amount of literature aiming
to understand the structure and composition of NS cores and observational 
signatures expected of them.  Supernova simulations have become
more sophisticated, with predictions of GW amplitudes that are far more 
pessimistic now than it was thought three decades ago, but challenges
have remained in producing realistic simulations that resolve all
the relevant scales and include all the macro- and micro-physics. 
Gravitational-wave observations with initial interferometers at
design sensitivity have broken new ground, setting the best \index{supernovae}
upper limits on the strength of GWs from known pulsars. \index{gamma ray bursts}
These observations are already constraining theoretical models and 
advanced interferometers will greatly improve upon them.
In this section we will take a census of the most important
GW sources of isolated compact objects, current observational status
and science targets and challenges for future observations.

\subsection{A menagerie of neutron-star sources}

Neutron stars in isolation with a time-varying quadrupole moment are 
potential sources of GWs. The birth of a NS
in a supernova, a non-axisymmetric spinning NS, a NS
accreting from a companion in a low-mass X-ray binary, differentially 
rotating NSs, could all produce GWs. \index{supernovae}

For an isolated body the energy available for radiation is in the form of 
its gravitational binding energy, rotational energy or energy stored in its  
magnetic field.  Most of the available energy might be emitted 
in a burst of GWs, resulting in a source with a large amplitude,
or else the energy might leak out slowly over a long period of time, giving a 
continuous, but low-amplitude, source of radiation.

{\noindent \em Supernovae:} 
Neutron stars are born in the aftermath of the gravitational collapse of a massive star 
of $\sim 8$--$100\,M_\odot$ or when the core of a white dwarf becomes more massive 
than the Chandrasekhar limit of $1.4\,M_\odot.$  \index{supernovae}
Supernovae were the prime targets for the first GW detectors  
and they are still among the most important sources. The Galactic supernova
rate is uncertain and is thought to be between 0.01--0.1 per year,
but the rate within about 5 Mpc could be one per few years \cite{ando:05}. 

In supernovae, GWs are emitted at the expense of the gravitational 
binding energy. The time scale over which the radiation is emitted is the 
dynamical free-fall time $\tau_{\rm FF}\sim \sqrt{4\pi/G\rho_{\rm NS}},$ 
where $\rho_{\rm NS}\sim 5\times 10^{17}\,\rm kg\,m^{-3}$ is the mean
density of a NS\footnote{The density of the pre-collapse star is not 
relevant as most of the energy in GWs is emitted in \index{supernovae}
the final moments of the collapse and core bounce.}. Thus the time scale 
for collapse is $\tau_{\rm FF} \sim 2\,\rm ms$ and the frequency of GWs
would be $f\sim \tau_{\rm FF}^{-1} \sim 500\,\rm Hz.$ 
Numerical simulations also reveal that the time domain
waveform is a short burst and the energy in the burst is spread
over a frequency range of 200~Hz to 1~kHz, with the peak of the
radiation at $f_{\rm peak} \simeq 500\,\rm Hz$ \cite{Dimmelmeier:2008iq,ott:09b}. 
If a fraction $\epsilon \sim 10^{-8}$ of the rest mass energy of the star 
is converted to GWs then the characteristic amplitude of the signal 
would be $h_c \sim 2\times 10^{-22}\,\rm Hz^{-1/2}.$ From Fig.\, \ref{fig:sources-g} we see 
that a Galactic supernova from a random sky position would be easily observable in 
aLIGO, producing an SNR\footnote{Laser interferometers have the best 
response to burst sources that occur directly above their plane, but
for an event at a random position on the sky and for waves of arbitrary 
polarisation the response and the SNR are a factor $\sqrt{2/5}$ smaller.}
of $\sim \sqrt{2/5}\, h_c/\sqrt{S_h(f_{\rm peak})} \sim 30.$ At 4 Mpc, the 
distance at which the rate could be one per few years, the characteristic amplitude 
would be $h_c \sim 5 \times 10^{-24}\, \rm Hz^{-1/2},$ which would be observable 
in ET with a similar SNR if $\epsilon \sim 10^{-6}.$ \index{gravitational waveform}

{\noindent \em Spinning neutron stars:}
A NS that is perfectly spherically symmetric or spinning about its 
symmetry axis emits no radiation since its quadrupole moment would not vary 
with time.  Non-axisymmetric NSs would produce radiation at
twice the spin frequency.  The amplitude of GWs for a NS 
at a distance $R,$ is \cite{Maggiore2008} $h_0 = {4\pi^2G}\epsilon\,I_{zz}\,f^2/(c^4 R)$, 
where $f$ is the frequency of GWs and  
$\epsilon\equiv(I_{xx}-I_{yy})/I_{zz}$ is the NS ellipticity given in 
terms of the principal moments of inertia with respect to the rotation axis, 
$I_{xx},$ $I_{yy}$ and $I_{zz}.$ Typical NS moments of inertia are 
$I\sim 3\times 10^{38} \,\rm kg\,m^2,$ so a NS at 10 kpc, spinning at 50 Hz 
(GW frequency of 100 Hz) and an ellipticity of $10^{-6},$ has an amplitude 
of $h_0 \simeq 3\times 10^{-27}.$ 

Since GWs from spinning NSs are essentially 
continuous waves (CW), Fourier transforming the signal would focus all 
its power into one frequency bin. The SNR grows as the 
square-root of the integration period. Thus, the characteristic strain 
amplitude $h_c$ of a signal integrated over a time $T$ is $h_c=h_0\sqrt{T}$, 
which for the example considered above is $h_c\sim 1.7\times 10^{-23}\,\rm Hz^{-1/2},$  
for $T=1$~yr. For the Crab (B0531+21), the youngest known pulsar, at a
distance of $2\,\rm kpc$ and spin frequency of 30 Hz, $h_c\sim 3\times 10^{-23}
\,\rm Hz^{-1/2},$ for the same ellipticity. Figure~\ref{fig:sources-g} 
plots the characteristic amplitude as a 
function of GW frequency for two choices of ellipticities, $\epsilon = 10^{-6}$ and
$10^{-8},$ and for NSs at 10 kpc, located and oriented randomly with respect 
to the detector and for an integration time of 1 year. 
Neutron stars of spin frequencies in the range 20--400 Hz (GW frequencies 
of 40--800 Hz) would be accessible to advanced detectors if their 
ellipticities are $\epsilon \gsim 1.6 \times 10^{-5} 
(f/100\,{\rm Hz})^{-2}(R/{10\,\rm kpc}).$ 

Nearly 2000 pulsars are currently known\footnote{ATNF pulsar catalogue: 
\texttt{http://www.atnf.csiro.au/people/pulsar/psrcat/}} and it is estimated 
that our galaxy is host to $\sim 10^9$ NSs. The fraction of 
%The only decent reference for the number of NSs I could find was
%http://www.nasa.gov/mission_pages/GLAST/science/neutron_stars.html
NSs accessible to the gravitational window is uncertain.
The biggest uncertainty is the ellipticity that can be sustained in a 
NS, with largest estimates of $\epsilon \sim 10^{-4}$ \cite{Owen05}, but more 
typically $\epsilon \sim 10^{-6}$ or smaller \cite{Andersson:2009yt}.  
Statistical arguments suggest that a NS with ellipticity 
$\epsilon = 10^{-6}$ could be close enough to have an amplitude of 
$h_{\rm max}\simeq 1.6\times 10^{-24}$ in the frequency range 250--680 Hz 
\cite{Knispel:2008ue}.  

Figure~\ref{fig:sources-s} shows the expected
amplitude for several known AM CVn systems, white dwarf binaries, and X-ray 
binaries in the eLISA band. The latter sources are often referred to as 
{\em calibration sources,} because eLISA should see them at these amplitudes.

{\noindent \em Pulsar Glitches and Magnetar Flares:}
Radio pulsars have very stable spins and their periods ($P$) change very slowly 
over time.  Their small spin-down rate ($\dot P \lsim 10^{-12}),$ is occasionally 
marked by a sudden increase in angular frequency $\Omega,$ an event that is called a 
{\em glitch} \cite{Chamel:2008}.  To date more than 300 glitches have been observed 
in about 100 pulsars \cite{Espinoza:2011}.
Vela (B0833-45) is a nearby ($R \sim 300 \,\rm pc$) pulsar in which 16 glitches
have been observed since its discovery in 1969. The magnitude of a glitch is
measured in terms of the fractional change in the angular velocity, which is found 
to be in the range $\Delta \Omega /\Omega \sim 10^{-5}$--$10^{-11}.$ 
Some time after a glitch, the pulsar returns to its regular spin-down evolution. 
Pulsar glitches are not the only transient phenomena observed in NSs. 
Sources of giant X- and gamma-ray flashes are thought to arise in highly 
\index{gamma ray bursts}
magnetised NSs, called {\em magnetars}, with B-fields $\sim 10^{15}$--$10^{16}$~Gauss.
The source of high-energy radiation is believed to be powered by the decay of 
the magnetic field associated with stellar quakes \cite{Kaspi:2010jq}.  

Pulsar glitches and magnetar flares could excite a spectrum of normal mode oscillations 
of the ultra dense NS core, the characteristic mode frequencies varying over a range of 
1.5--6 kHz and damping times $\tau\sim \rm ms,$ depending on the mode in question and 
the NS EoS \cite{Andersson:1997rn}.  \index{neutron star equation of state (EoS)}
The energy in normal modes could be emitted as a narrow-band burst of exponentially 
damped sinusoidal GWs.  Figure \,\ref{fig:sources-g} shows 
plausible characteristic amplitudes produced by normal modes of energy 
$10^{-12} M_\odot,$ for mode frequencies in the range
of 1.5--4 kHz and NS distances in the range 1 kpc to 10 kpc.
Third generation detectors like ET should be able to detect such amplitudes in 
coincidence with radio observations.  

{\noindent \em Low Mass X-ray Binaries:}
Low-Mass X-ray Binaries (LMXBs) are accreting NSs or BHs that emit 
bursts of X-ray flashes lasting for about 10 s and repeat once every few hours or days,
with  millisecond oscillations in burst intensity \cite{Chakrabarty:2005}.  
X-ray bursts are believed to be caused by thermonuclear burning
of infalling matter, while oscillations are suspected to be caused by the 
NS spin.  About 100 galactic LMXBs are known to-date as well as many 
extra-galactic ones. Inferred spin frequencies of NSs in LMXBs 
seem to have an upper limit of about 700 Hz \cite{Chakrabarty:2005},
although this is nowhere close to the value at which centrifugal break-up would
limit the star's spin frequency. It has been proposed that GWs
might be responsible for limiting the spin frequencies of NSs in LMXBs 
\cite{Wag84,Bildsten:1998,Chakrabarty:2003kt}. 

The expected characteristic amplitude of gravitational radiation is shown 
in Fig.~\ref{fig:sources-g} for the well-known LMXB Sco X-1 and for the 
known Galactic population of LMXBs.  Advanced detectors could detect 
Sco X-1 if it is losing all of its accreted angular momentum to GWs (which
is unlikely to be the case), while ET targets the full galactic population 
\cite{Watts:2008}.

\subsection{Results from LIGO and Virgo} 
{\noindent \em Searches for burst signals:} Searches for bursts of GWs
essentially fall into one of two classes: all sky, blind searches and 
astrophysically triggered searches. \index{gravitational wave detection results}
In the first approach there is no a priori information about what and when to 
look for. The goal of this approach is to detect radiation from unmodeled, or
poorly modeled, transient sources, as well as hitherto unknown sources, that 
last for less than $\sim 1$ second. Since no assumption about the nature of 
GWs is made, this approach has the greatest serendipitous discovery potential. 
The search algorithm uses wavelet transforms to look for excess power 
\cite{Klimenko:2008} and the sensitivity of the search is characterised
in terms of the root-sum-square strain amplitude $h_{\rm rss}$ of the 
signal\footnote{The strain amplitude $h_{\rm rss}$ is defined as: 
$h_{\rm rss} = \sqrt{\int \left [\, |h_+(t)|^2 + |h_\times(t)|^2\,\right ] dt}.$}.
Analysing data from the various science runs has determined that the rate of  
strong GW bursts (i.e., bursts with $h_{\rm rss}>10^{-19}
\rm \,Hz^{-1/2}$ in the frequency region from 70 Hz to 3 kHz) reaching the 
Earth is less than 1.3 events per year at 90\% confidence \cite{Abadie:2012ad}.
For hypothesised standard candle sources that emit 1 $M_\odot$ equivalent of
energy in GWs as sine-Gaussian waveforms, the inferred rate density
of events in the local Universe, in units of $\rm Mpc^{-3}\,yr^{-1},$ is 
less than $10^{-6}$ in the frequency range 100--200 Hz and less than $10^{-2}$
in the frequency range 1--2 kHz. Alternatively, generic burst 
sources within 10 kpc emitted less than $\simeq 2\times 10^{-8}\,M_\odot$ in 
GWs in the frequency range 100--200 Hz; that limit increases to 
$\simeq 10^{-5}\,M_\odot$ at 1 kHz \cite{Abadie:2012ad}.

In the second approach, analysis is carried out around the time of an astrophysical 
transient, such as a supernova or a magnetar flare. Knowledge of the epoch and sky
position of the event helps reduce the amount of data that needs to be searched for,
which in turn decreases the false alarm probability and improves the search 
sensitivity. Searches have been carried out at the time of pulsar glitches
\cite{Abadie:2011md}, magnetar flares \cite{Abadie:2011ac} and GRBs
\cite{Abadie:2012ae}. Of particular significance is the search for GWs
\index{supernovae}
around the time of GRB070201 \cite{Abbott:2008ab}. The event in this case was a
\index{gamma ray bursts}
shGRB that is believed to have followed either from giant quakes in
highly magnetised NSs or from merging binary NSs. Location 
of GRB070201 coincides with the spiral arms of the Andromeda galaxy (M31) at 780 kpc.
LIGO detectors, which were taking data at the time of this event, would have
quite easily detected signals from a merging NS binary at this distance,
but not bursts associated with a magnetar flare. The analysis found no plausible
GW candidates within a 180 s window around the time of the GRB
and in particular excluded binary NS-NS and NS-BH 
mergers at M31 with more than 99\% confidence \cite{Abbott:2008ab}. The analysis 
also concluded that isotropic energy in GWs from 
the source, if it were at M31, was most likely less than $4.4\times 10^{-4}\, 
M_\odot,$ lending support for the possibility that this was the first Soft
Gamma Repeater flare observed outside the Milky Way. \index{gamma ray bursts}

More recently, searches have also been performed around the times of 128 
long GRBs and 26 shGRBs \cite{Abadie:2012ae}; no GW
candidates of any significance were found, which meant that the bursts could
not have occurred closer than a certain distance determined by the horizon 
distance of the detectors in the direction of the GRBs. The maximum exclusion 
distance for the population of shGRBs was 80 Mpc, which is not 
surprising since the closest known shGRB is at a distance of $\sim 500$ 
Mpc. However, extrapolating current results to advanced detectors, it 
seems quite plausible that GWs coincident with GRBs could be
detected within 2.5--5 years of observing \cite{Metzger,Abadie:2012ae}, 
or place upper bounds on the number of GRBs arising from merging binaries. 
\index{gravitational wave detection results} \index{gamma ray bursts}

{\noindent \em Searches for continuous waves:}
Most CW signals are monochromatic in the rest frame of their sources.
Their detection is complicated by the fact that the signal received at 
the detector is modulated due to the Earth's motion. Because of 
Doppler modulation in frequency, the spectral lines of fixed frequency sources spread power 
into many Fourier bins about some mean frequency.  Although the modulation of the signal 
makes the search prohibitively expensive, imprinted in the modulation is the source's
position on the sky; it will be possible to resolve the source's location
subject to Rayleigh criterion, $\delta \theta = 2\pi \lambda/L,$ where $\delta\theta$
is the angular resolution, $\lambda$ is the wavelength of the radiation and
$L=2\,\rm AU$ is the baseline for an observation period of 1 year.
At a frequency of 100 Hz, $\delta\theta \sim 2^{\prime\prime}.$ In the case of
CW sources, two different types of searches have been performed: searches
for known pulsars (with precisely known sky position and frequency evolution) and 
blind searches (sources with unknown sky position and spin frequency).

Searches for CW signals from known,
isolated pulsars, not being limited by computational resources, have achieved 
the best possible sensitivity \cite{Aasi:2013cw}.  In particular, upper limits 
\index{isolated neutron stars}
on the strength of GWs from the Crab and Vela pulsars are now determined by 
these observations to be well below the level expected from the observed rate 
at which these pulsars are spinning down. The loss in energy to GWs in Crab 
is less than 1\% of the rotational energy lost due to the observed 
spin down \cite{Aasi:2013cw}; the corresponding number for Vela is 10\% 
\cite{Abadie:2011md}. A search for CW signals in the frequency range of 100--300 Hz
from the compact central object, believed to be a NS, in the supernova remnant 
Cassiopea A at 3 kpc, has set best upper limit on the strain amplitude of 
$\sim 3\times 10^{-24}$ and equatorial ellipticity of 0.4--$4\times 10^{-4},$ as well
as setting the first ever limit on the amplitude of r-modes in this young
NS \cite{Abadie:2010hv}.  \index{gravitational wave detection results} \index{supernovae}

Given the possibility that the strongest CW sources may be electromagnetically quiet or 
previously undiscovered, an all sky, all frequency search for such unknown sources is 
very important, though computationally formidable. Clever and computationally efficient
algorithms and distributed volunteer-computing \texttt{Einstein@Home} \cite{Abbott:2008uq} 
have made the searches ever more sensitive, and have been successful in discovering new
radio pulsars in old radio data (see, e.g., \cite{Knispel:2013da,Allen:2013sua}). 
A blind GW search using \texttt{Einstein@Home} excluded signals in the 50
Hz to 1.2 kHz band, with upper limits on strain amplitudes $\sim 10^{-24}\,$--$10^{-23}$
depending on the frequency of the source. For example, strain amplitudes greater than 
$7.6\times 10^{-25}$ were excluded at 152.5 Hz the (frequency where LIGO S5 run had the 
best sensitivity), over a 0.5 Hz-wide band \cite{Aasi:2013ein}. This means there are no NSs
at this frequency within 4 kpc and spinning down faster than 2 nHz\,s$^{-1}$ and 
ellipticities greater than $2\times 10^{-4}.$ Targeted searches for sources within 8
pc of the Galactic centre Sag A$^*$, in the frequency range of 78--496 Hz, and maximum
spin down rates of $\sim -8 \times 10^{-8}\,\rm Hz\,s^{-1}$ have achieved the best
sensitivities for blind searches, ruling out NSs with GW amplitude 
$\sim 3\times 10^{-25}$ around 150 Hz in this region of the sky (for 
details and caveats see \cite{Aasi:2013dir}). \index{gravitational wave detection results}

Advanced detectors will beat the spin down limit of 
several pulsars \cite{Aasi:2013cw}. For the fastest pulsars in the frequency 
range ~200--400 Hz, advanced detectors will reach ellipticity limits of 
$\sim 10^{-8}$ (or a differential radius of 100 microns in 10 km!),
significantly below the spin-down limits; ET will be sensitive to ellipticities
as low as $10^{-9}$ \cite{ET-DSD}. \index{gravitational wave detection results}

\subsection{Science targets and challenges}
Neutron stars are the most compact objects with matter known today. They have
strong surface gravity that is responsible for very intense sources of X-rays 
and gamma-rays.  Their dense cores could \index{gamma ray bursts}
be superfluid and might consist of hyperons, quark-gluon plasma or other 
exotica \cite{Chakrabarty:2005} and are, therefore, laboratories of high energy 
nuclear physics.  Observing a representative sample of the galactic population 
of NSs could transform astrophysical studies of compact objects, but there are
still some challenges in theoretical modeling of NSs and analysis of data. 

\vspace{0.25cm}
\centerline{\emph{Science Targets}}
\vspace{0.25cm}

\paragraph{Physics of low-mass X-ray binaries:}
Detecting GWs from LMXBs should help to understand the mechanism is behind 
limiting spin frequencies in LMXBs.  The centrifugal breakup of NS spins for 
most EoS is $\sim 1500$ Hz, far greater than the maximum spin of $\sim 700$ 
Hz inferred from X-ray observations \cite{Benacquista:2011kv}.  It has, 
therefore, been a puzzle as to why NS spin frequencies are stalled.  One 
reason for this could be that some mechanism operating in the NS is emitting 
GWs and the resulting loss in angular momentum explains why NSs cannot be spun 
up beyond a certain frequency.  The exact mechanism \index{neutron star equation of state (EoS)}
causing the emission of GWs can account for this amplitude if NS can support 
an effective ellipticity of $\epsilon \sim 10^{-8}$.  This ellipticity could 
be produced by a time-varying, accretion-induced quadrupole moment 
\cite{Bildsten:1998}, or by relativistic instabilities (e.g. r-modes) 
\cite{Andersson:1999}, or by large toroidal magnetic fields \cite{Cutler:2000}. 
Targeted observations of known LMXBs could confirm or rule out astrophysical
models of such systems.

\paragraph{Understanding supernovae:}
Supernovae produce the Universe's dust and some of its heavy elements; 
their cores are laboratories of complex physical phenomena requiring general 
relativity, nuclear physics, magneto-hydrodyna\-mics, neutrino viscosity and 
transport and turbulence to model them.  Much of the physics of supernovae 
is poorly understood: How non-axisymmetric is the collapse? How much energy is 
converted to GWs and over what time scale? What causes shock revival 
in supernovae that form a NS: neutrino, acoustic and/or magneto-rotational 
mechanisms? Depending on the supernova \index{supernovae}
mechanism, the predicted energy in GWs from supernovae varies by large 
factors (see, e.g., \cite{Mueller:2010nf,Ott:2012mr}), indicating 
the complexity of the problem in numerical simulations.
Until we know the mechanism that revives the stalled shock it will not
be possible to correctly predict the amplitude of the emitted gravitational 
radiation or its time-frequency structure. Gravitational-wave observations 
could provide some of the clues for solving these questions \cite{ott:09b,Logue:2012zw}. 
Moreover, GWs could also be produced by neutrino emission during the supernova explosion 
and the signal spectrum could extend down to $\sim 10$ Hz \cite{Mueller:2003fs}. 
More realistic studies are needed to quantify this signal. 

\paragraph{Testing neutron-star models:}
Models of NSs are mostly able to compute their maximum ellipticity by subjecting 
the crust to breaking strains with predicted ellipticities ranging from $\epsilon 
\sim 10^{-4}$ (for exotic EoS) \cite{Owen05} to $10^{-7}$ for conventional crustal 
shear \cite{Ushomirsky:2000}. Large toroidal magnetic fields of order $10^{15}\,\rm G$ 
could sustain ellipticities of order $10^{-6}$ \cite{Cutler:2002} and accretion along 
magnetic fields might produce similar, or a factor 10 larger, deformations 
\cite{Payne:2003}. The large range in possible ellipticities shows that GW 
observations could have a potentially high impact and science return in this area. 
Confirmed detections of NSs with known distances will severely constrain models
of the crustal strengths and a catalogue of CW sources would help understand the 
galactic supernova rate and their demographics could lead to insights on the 
evolutionary scenarios of compact objects. \index{neutron star equation of state (EoS)} \index{supernovae}

\vspace{0.25cm}
\centerline{\emph{Challenges}}
\vspace{0.25cm}

\paragraph{Interfacing theory with searches:} 
Models that can accurately predict the spectrum and complex mode frequencies 
will be very useful in searches for GW signals at the time of glitches in 
pulsars and flares in magnetars \cite{Andersson:1997rn}. In fact, robust 
predictions can also help to tune detectors to a narrower band, where signals 
are expected, with greater sensitivity. To take advantage of such techniques, 
which will become feasible in the era of routine observations, models would need 
to become realistic and reliable. 

Supernova simulation is one area where a breakthrough in understanding the
core bounce that produces the explosion could be critical to produce reliable
models. At present, it is not clear if models will ever be able to produce
waveforms that can be deployed as matched filters. A catalogue of predicted
waveforms are routinely used to calibrate the sensitivity of a search 
(see, e.g., \cite{Abadie:2012ad}). \index{gravitational waveform} \index{supernovae}
Accurate models of the frequency range and spectral features of the emitted
radiation will obviously aid in better quantification of the search sensitivity.

\paragraph{The problem of blind searches:} 
Looking for CWs in GW data is a computationally formidable problem 
\cite{Schutz:1987uh,Brady:1997ji,Jaranowski:1998qm}. 
Blind searches have to deal with many search parameters, such as the sky 
position of the source, its spin frequency and one or more derivatives 
of the spin frequency \cite{Brady:1997ji}.  The number of floating point operations 
grows as the 5th power of the integration time $T$ for a blind search with
unknown sky position, unknown spin frequency and one spin-down parameter
\cite{Creighton:2011}.  Most blind searches are able to coherently integrate 
the data for about a few hours to days (depending on the number of spin-down 
parameters searched for) \cite{Aasi:2013ein,Aasi:2013dir} and the 
sensitivity of searches will always be limited by the available 
computing resources.  Algorithms that can integrate for longer 
periods are desirable, as are multi-step hierarchical searches that could 
achieve optimal sensitivity given the computational power.

\section{Gravitational radiation from the early Universe}
\label{sec:primordial}

During the past 30 years, several new predictions for GW signals from the 
primordial Universe have been made, greatly stimulated by the construction and 
operation of the first GW detectors and the planning of future experiments.

The epoch of big-bang nucleosynthesis (BBN), when light elements first formed, is the earliest epoch 
of the Universe that we understand today with any confidence. The Universe was only a second old
at this epoch, it was radiation dominated and had a temperature of $\sim 1\, {\rm MeV.}$ In contrast,
the Universe was much older (age of $\sim 10^{5}\mbox{--}10^{6}$ years) and cooler (temperature of 
$\sim 1\, {\rm eV}),$ when the CMB radiation, measured today with amazing accuracy, 
was emitted. It is expected that the Universe is filled with cosmic neutrinos 
produced when the Universe's temperature was $1\, {\rm MeV}$, but this has not been observed, yet.
In fact, no primordial background, of radiation or particles, produced before the epoch of CMB has
ever been detected.

\subsection{Primordial sources and expected strengths}

Gravitational waves emitted prior to BBN in the so-called {\it dark age} 
would travel unscathed, due to their weak interaction with matter, and provide us
with a view of the Universe at that time.

A rapidly varying gravitational field during inflation can produce a stochastic
background of GWs by parametric ``amplification" of quantum, vacuum fluctuations~\cite{Starobinsky:1979ty,Grishchuk:1974ny}. Today this background would span the frequency range of $10^{-16}$ --
$10^{10}$ Hz, which covers the frequency band of current and future detectors on 
the ground and in space (see Figs.~\ref{fig:sources-g} and \ref{fig:sources-s}).  \index{stochastic background}
This is the same mechanism that is believed to have produced the scalar density perturbations that led 
to the formation of large scale structures in the Universe. 
Single-field, slow-roll models of inflation predict that the GW background today 
slightly decreases as the frequency increases, 
$\Omega_{\rm GW}=\Omega_0 (f/f_{\rm eq})^{n_T}$, for $f>f_{\rm eq}$ with $n_T\,\laq\, 0$, 
while it rises as a power-law, $\Omega_{\rm GW}$ = $\Omega_0  ({f}/f_{\rm eq})^{-2}$, 
for $f<f_{\rm eq}$~\cite{Allen1988,Maggiore:2000}, where $\Omega_{\rm GW}(f) =
{d \rho_{\rm GW}(f)}/{d\log f}/{\rho_C}$~\cite{Allen:1997ad}, with   
$\rho_C=3H_0^2/(8\pi G)$ and $H_0$ the present value of the Hubble parameter.
The transition frequency $f_{\rm eq}\simeq 10^{-16} {\,\rm Hz}$ corresponds to the Hubble radius at the time 
of matter-radiation equality, redshifted to the current epoch. The value of $\Omega_0$ 
is not known, but  the current upper limit on the tensor-to-scalar ratio 
from the CMB~\cite{Ade:2013zuv} implies $\Omega_0 \,\lsim\, 10^{-15}.$  A cosmological 
GW background would leave an imprint in the CMB polarisation 
map~\cite{Kamionkowski:1996ks,Seljak:1996gy} and, as mentioned 
in Sec.~\ref{sec:history}, the BICEP2~\cite{Ade:2014xna} experiment has claimed a detection 
of this signature. However, further scrutiny suggests that at this time BICEP2 result cannot be 
excluded from being of astrophysical origin~\cite{Mortonson:2014bja, Flauger:2014qra}.
Concurrently with the construction of ground-based detectors and the planning of
next generation of experiments, studies of the GW background from inflation have 
been refined and several physical effects that may impact the high-frequency portion 
of the spectrum have been predicted~\cite{Weinberg:2003ur,Pritchard:2004qp,Smith:2005mm,Boyle:2005se}.

The preheating phase, which occurs at the end of inflation, is a highly non-thermal phase
that creates transient density inhomogeneities with time-varying mass multipoles, which
would generate a stochastic background of GWs~\cite{Khlebnikov:1997di}. Symmetry breaking 
phase transitions or the ending stages of brane inflation might witness the creation of cosmic 
(super)strings~\cite{Kibble:1976sj,Vilenkin:1984ib,Sarangi:2002yt}. Due to their 
large tension these strings undergo relativistic oscillations and thereby
produce GWs, which causes them to shrink in size and disappear. However, they could be constantly 
replaced by smaller loops that brake off from loops of size larger than the Hubble radius. 
In this way, a network of cosmic (super)strings could generate a stochastic GW background 
\index{stochastic background} but they could also produce bursts of gravitational radiation 
when cusps and kinks form along strings~\cite{Berezinsky:2000vn,Damour:2000wa,Copeland:2003bj,
Damour:2004kw,Olmez:2010bi}. Likewise, a strong first order phase transition could create
bubbles of true vacuum, which collide with each other and produce a GW 
background~\cite{Turner:1990rc,Kamionkowski:1993fg}. 
For more details on the mechanisms responsible for gravitational radiation in the 
early Universe, we refer the reader to the following 
reviews~\cite{Allen1988,Maggiore:2000,Buonanno:2003th,Binetruy:2012ze} and references therein.

\subsection{Results from LIGO and Virgo}
\label{sec:analysis-stochastic}
\index{gravitational wave detection results} \index{stochastic background}
Stochastic signals are a type of continuous waves, but with two important differences. 
In general they do not arrive from any particular direction and, by definition, 
have no predictable phase evolution. Therefore, conventional matched filtering would not 
\index{matched filtering template}
work and sliding data of one detector relative to another (to account for the difference
in arrival time) has no particular advantage. Even so, data from one detector could 
serve as a ``template'' to detect the same stochastic signal present in another 
detector. If the detectors are located next to each other and have the same orientation
then a simple cross-correlation of their outputs weighted by their noise spectral
densities would result in optimal SNR~\cite{Thorne:1987af}. \index{matched filtering template}
The problem with two nearby detectors is that they will have a common noise background
that would contaminate the correlated output. By placing detectors far apart, one
could mitigate the effect of common noise, but in that case wavelengths smaller than 
the distance between the detectors will not all be coherent in the two detectors, which
effectively reduces the sensitivity bandwidth. This is appropriately taken into account
in the cross-correlation statistic of a pair of detectors by using what is called the 
{\em overlap reduction function}~\cite{Flanagan:1993}, which is a function of frequency that accounts for 
the lack of coherence in stochastic signals in detectors of different orientations and
separated by a given distance~\cite{Allen:1997ad}.  Additionally, since the template 
is essentially noisy, the amplitude SNR grows as the {\em fourth-root} of the product 
of the effective bandwidth $\Delta f$ of the detector and the duration $T$ over which 
the data is integrated~\cite{Thorne:1987af}. In the case of PTAs, the detection technique 
is similar; instead of a pair of detectors one constructs the correlation between the 
timing residuals of many stable millisecond pulsars. \index{stochastic background} \index{pulsar timing array (PTA)}

The energy density in GWs is related to the strain power spectrum $S_{\rm GW}(f)$ 
by \cite{Allen:1997ad} $S_{\rm GW}(f) = 3H_0^2\,\Omega_{\rm GW}(f)/(10 \pi^2f^3)$. 
The {\em characteristic amplitude} $h_c$ of a stochastic background, i.e.\ the strain
amplitude produced by a background after integrating for a time $T$ over a bandwidth
$\Delta f$, is given by $h^2_c= \sqrt{T\,\Delta f}\, S_{\rm GW}$.  
In Figs.\,\ref{fig:sources-g} and \ref{fig:sources-s} we plot (in dotted lines) $h_c(f)$ 
for several values of $\Omega_{\rm GW}$ assumed to be independent of $f$, 
setting $T=1\,\rm yr$ and $\Delta f=100\,\rm Hz$. Due to the overlap 
reduction function, the SNR is built up mostly from the low-frequency part of the signal
such that $\lambda_{\rm GW} \gsim d/2,$ where $d$ is distance between detectors. A stochastic
signal would be detectable if it stays above the noise curve roughly over a frequency 
band $\Delta f\simeq f.$ Advanced detectors should detect $\Omega_{\rm GW} \ge 10^{-9}$ 
at tens of Hz, while ET, due to its much improved low-frequency sensitivity and collocated
detectors, could detect $\Omega_{\rm GW}\sim 10^{-11}$, while eLISA could detect $\Omega_{\rm GW} \ge 10^{-12}$ 
at mHz frequencies.  For an integration time of $T=5\,\rm yr$ and bandwidth of
$\Delta f=6\,\rm nHz,$ a stochastic background with $\Omega_{\rm GW} =2.5 \times 10^{-10}$
would be detectable in PTA with an SNR of 5, assuming 20 millisecond pulsars that have an rms stability \index{stochastic background}
of 100 ns~\cite{Sesana:2008,Sesana:2012ak}. SKA will improve the sensitivity of PTAs
by two orders of magnitude to $\Omega_{\rm GW} \sim {\rm few} \times 10^{-13}$~\cite{Kramer:2004rwa}.

Gravitational radiation from the early Universe does not only generate stochastic backgrounds: 
bursts from cusps and kinks along cosmic (super)strings produce power-law signals in the frequency 
domain~\cite{Damour:2004kw} that can be searched for using matched-filtering techniques. 
\index{matched filtering template} \index{stochastic background}
Data from several science runs of iLIGO and Virgo detectors have been analysed to 
search for signals from the early Universe. In those science runs, the detectors' sensitivity 
has passed the BBN bound~\cite{Copi:1996pi,Abbott:2009ws} in the frequency band around
100 Hz, $\Omega_{\rm GW} < 6.9 \times 10^{-6},$ but not yet the CMB bound~\cite{Smith:2006nka}.
The results have started to exclude 
\index{gravitational wave detection results}
regions of the parameter space of expected signals from cosmic (super)strings
~\cite{Abbott:2009rr,Abbott:2009ws,Aasi:2013vna}, and have constrained the equation of state of 
the Universe during the dark age~\cite{Boyle:2007zx,Abbott:2009ws}. Moreover, 
pulsar timing observations have set physically meaningful upper limits for the 
supermassive BH binary background ($\Omega_{\rm GW} < 1.3\times 10^{-9}$ at 2.8 nHz) 
\cite{Shannon:2013wma} and cosmic (super)strings~\cite{Jenet:2006sv}. 
\index{supermassive black hole binaries} \index{pulsar timing array (PTA)}

\subsection{Science targets and challenges}

Stochastic GW signals might carry signature of unexplored physics in the energy range
$\sim 10^9$ GeV to $\sim 10^{16} $ GeV. The detection of GWs 
from the dark age could therefore be revolutionary.
The spectrum of the detected radiation could reveal phase transitions that might have occurred
in the Universe's early history, unearth exotic remnants like cosmic
(super)strings, prove that a cosmic inflationary phase existed and 
that gravity can be reconciled with quantum mechanics. 
No other observation can ever take us closer to the origin of our Universe and hence the science 
potential of discovering primordial GWs will be immense. 
However, the challenges in this area are equally formidable. 

As mentioned before, one of the biggest problems in identifying a stochastic
GW background is how to disentangle it from the environmental \index{stochastic background}
and instrumental noise backgrounds. At present, data from two or more detectors
are cross correlated to see if there is any statistical excess. If the detectors
are geographically widely separated then one could reasonably hope for the
environmental noise backgrounds in different instruments not to correlate,
although correlations could exist due to large scale magnetic fields, 
cosmic rays and anthropogenic noise and the like. When the number of detectors 
grows, noise correlations go down. However, because of the overlap reduction 
function, the sensitivity to a stochastic background diminishes quickly 
with geographically separated detectors. Due to 
environmental and instrumental noise, searching for 
stochastic backgrounds in collocated detectors like ET or eLISA will be a 
real challenge.

In the case of PTAs, the problem is less severe as one is looking for correlations 
in the residuals of the arrival times of radio pulses from an array of millisecond 
pulsars, after subtracting the model of the pulsar from the original data. 
In principle, any systematics in timing residuals could be mitigated by 
integrating the correlation over long periods, but the problem here is that the 
time scale for integration tends to be large (i.e., tens of years). \index{pulsar timing array (PTA)}

In the case of the GW background from inflation, CMB bounds on inflationary potentials 
are not very informative on the value of $\Omega_0$ except for placing an upper limit.  
In the absence of probes from epochs prior to BBN, it is very hard to infer the 
equation of state of the Universe between the end of inflation 
to the epoch when the radiation era started. Thus, it is difficult to predict the spectral 
slope of the relic GW background from BBN (when the Universe was certainly radiation 
dominated) to the scales where PTA, space-based and ground-based detectors are sensitive~\cite{Boyle:2007zx}. 
It is customary and (perhaps) natural to assume that the slope is the same 
over the huge range of frequency, spanning twenty orders of magnitude, and 
that it can be determined by CMB observations. 
However, as we look backward past BBN, a stiff energy component might overtake radiation 
as the dominant component in the cosmic energy budget~\cite{Peebles:1998qn,Giovannini:1998bp,
Giovannini:1999bh,Riazuelo:2000fc,Tashiro:2003qp}, 
without coming into conflict with any current observational constraints. The detection of the 
B-mode polarisation in the CMB will certainly have a huge impact, determining $\Omega_0$, but 
we will still not know whether the slope remains the same for twenty orders of magnitude.

Predictions for the GW background from preheating at the end of 
inflation~\cite{Khlebnikov:1997di,Binetruy:2012ze} 
lie typically (except for some choices of parameters in hybrid inflation) 
in the MHz frequency range where no GW detectors exist or are currently planned. 
Experimental proposals would need to be made, but since, when holding $\Omega_{\rm GW}$ fixed,  
the noise spectral density decreases as the frequency increases, 
the requirements on the detector sensitivity can be very hard to achieve. Finally, depending on 
string parameters, both the stochastic background and single powerful bursts from 
cusps and kinks of cosmic (super)strings could be observed by PTAs, eLISA and the aLIGO/AdV/KAGRA network. \index{stochastic background}
More robust predictions of (super)string loop sizes (large versus small loop sizes 
at birth) will be important to restrict regions of parameter space that are searched 
over (see~\cite{Binetruy:2012ze} and references therein).

\section*{Acknowledgements}
We are greateful to Luc Blanchet, Eric Poisson and Riccardo Sturani for carefully reading and providing comments 
on the manuscript.  We have benefited from useful discussions 
with Leor Barack (who supplied a Mathematica code to compute EMRI spectra in Figure
\ref{fig:sources-s}),  Chris Messenger, Bernard Schutz and Patrick Sutton.
A.B. acknowledges partial support from NSF Grant No. PHY-1208881 and NASA Grant NNX09AI81G. 
B.S.S. acknowledges support from STFC (UK) grant ST/L000962/1, ST/L000342/1 and ST/J000345/1.

\bibliographystyle{cambridgeauthordate}
\vskip-3truecm
\bibliography{References}

\end{document}